\documentstyle[fleqn,psfig,run2col,amstex]{article}
\input epsf
\def\slashchar#1{\setbox0=\hbox{$#1$}           
   \dimen0=\wd0                                 
   \setbox1=\hbox{/} \dimen1=\wd1               
   \ifdim\dimen0>\dimen1                        
      \rlap{\hbox to \dimen0{\hfil/\hfil}}      
      #1                                        
   \else                                        
      \rlap{\hbox to \dimen1{\hfil$#1$\hfil}}   
      /                                         
   \fi}                                         %

\def\pythia{{\sc PYTHIA}}
\def\herwig{{\sc HERWIG}}
\def\isajet{{\sc ISAJET}}

\def\that{\hat{t}}
\def\uhat{\hat{u}}
\def\q2{Q^2}
\def\qt{$Q_T$}

\def\beq{\begin{equation}}
\def\eeq{\end{equation}}
\def\beqa{\begin{eqnarray}}
\def\eeqa{\end{eqnarray}}

\def\ie{{\it i.e.,\ }}
\def\etal{{\it et al.\ }}

\def\slfrac#1/#2{\leavevmode\kern.1em\raise.5ex\hbox{\the\scriptfont0
         #1}\kern-.1em/\kern-.15em\lower.25ex\hbox{\the\scriptfont0 #2}}

\def\gev{{\rm\,GeV}}
\def\tev{{\rm\,TeV}}

\newcommand{\lessim}{\raisebox{-0.8mm}%
{\hspace{1mm}$\stackrel{<}{\sim}$\hspace{1mm}}}

\newenvironment{Itemize}{\begin{list}{$\bullet$}%
{\setlength{\topsep}{0.2mm}\setlength{\partopsep}{0.2mm}%
\setlength{\itemsep}{0.2mm}\setlength{\parsep}{0.2mm}}}%
{\end{list}}
\newcounter{enumct}
\newenvironment{Enumerate}{\begin{list}{\arabic{enumct}.}%
{\usecounter{enumct}\setlength{\topsep}{0.2mm}%
\setlength{\partopsep}{0.2mm}\setlength{\itemsep}{0.2mm}%
\setlength{\parsep}{0.2mm}}}{\end{list}}
\def\bea{\begin{eqnarray}}
\def\eea{\end{eqnarray}}
\def\bes{\begin{eqnarray*}}
\def\ees{\end{eqnarray*}}

\begin{document}

\title{\bf Report of the QCD Tools Working Group}

\author{{\bf Convenors:}
Keith Ellis (FNAL), Rick Field (Florida), Stephen Mrenna (Davis) and 
Greg Snow (Nebraska)\\
~\\
{\bf Working Group Members:} 
     C. Bal\'azs (Hawaii), E. Boos (Moscow), 
     J. Campbell (FNAL), 
     R. Demina (Kansas State), 
     J. Huston (MSU), C-Y.P. Ngan (MIT), A. Petrelli (ANL), 
     I. Puljak (LNPHE), 
     T. Sj\"ostrand (Lund), 
     J. Smith (Stony Brook), 
     D. Stuart (FNAL),
K. Sumorok (MIT)}

\begin{abstract}
We report on the activities of the ``QCD Tools for heavy flavors and
new physics searches'' working group of the Run II Workshop on QCD
and Weak Bosons.  The contributions cover the topics of
improved parton showering and
comparisons of Monte Carlo programs and resummation calculations,
recent developments in \pythia,
the methodology of measuring backgrounds to new physics searches,
variable flavor number schemes for heavy quark electro-production,
the underlying event in hard scattering processes, and
the Monte Carlo MCFM for NLO processes.
\end{abstract}   

\maketitle    

\section{Overview}
The task of the ``QCD Tools for heavy flavors and new physics searches 
working group'' 
was to evaluate the status of the tools -- invariably computer programs that
simulate physics processes at colliders -- that are being used to estimate
signal and background rates at the Tevatron, and to isolate areas of concern.
The contributions presented here cover several topics related to that
endeavor.  It is hoped that the next period of data-taking at the
Tevatron will reveal indirect or direct evidence of physics beyond
the Standard Model.  
The precise measurement of the $W$ boson mass
$M_W$ and its correlation with the top quark mass $m_t$ 
is one example of an indirect probe of the Standard Model.  
The production of a light Higgs boson in association with a $W$ or $Z$
boson is an obvious example of a direct one.  While both measurements
are related to electroweak symmetry breaking, it requires a
quantitative understanding of perturbative and non--perturbative QCD
to interpret data.

Because of the importance of the $M_W$ measurement, and since 
gauge boson production in association with jets is a serious background
in many new physics searches, much effort was devoted to understanding
gauge boson production processes.  It is well known that the
emission of many soft gluons has a profound effect on the kinematics
of gauge boson production.  Two calculational methods have been
used to compare ``theory'' with data:  (1) analytic resummation
of several series of important logarithms, and (2) parton showering
based on DGLAP--evolved parton distribution functions.  Here, there
are reports on our understanding of both, and improvements.
Note also that diboson production is often a background too.

In the Standard Model, and its minimal supersymmetric extension,
the mechanism that generates mass for the electroweak gauge bosons
also generates fermion mass.  From an agnostic point of view,
the fact that the $W$ and $Z$ bosons and the top quark 
have roughly similar masses, and these masses are quite disparate
from, say, the electron or neutrino masses, is some evidence
that heavy flavor is related to electroweak symmetry breaking.
Many of the search strategies for Run II rely on tagging
$c$ and $b$ quarks or $\tau$ leptons.  For this reason, there
are several contributions regarding issues of determining
backgrounds in Run II.

\section{Performing parton showering at Next-to-Leading-Order
Accuracy}

\centerline{\it by S. Mrenna}\vskip 2.mm

\subsection{Introduction}
\label{sec:intro}

In the near future, experiments at the Tevatron
will search for evidence of physics that supersedes the standard
model.  Important among the tools that will be used in these
searches are showering event generators or showering Monte Carlos
(SMC's).  Among the most versatile and popular of these are the
Monte Carlos \herwig \cite{herwig}, \isajet \cite{isajet}, 
and \pythia \cite{pythia}.  SMC's are useful because they accurately
describe the emission of multiple soft gluons, which is, in effect,
an all orders problem in QCD.  However, they only predict
total cross sections to a leading order accuracy, and, thus,
can demonstrate a sizeable dependence on the choice of scale
used for the parton distribution functions (PDF's) or 
coupling constants (particularly $\alpha_s$).  Also, in general,
they do not translate smoothly into kinematic configurations
where only one, hard parton is emitted.  In distinction
to SMC's are certain analytic calculations which account for
multiple soft gluon emission and higher order corrections to the
hard scattering.  These resummation calculations, however,
integrate out the kinematics of the soft gluons, and, thus,
are limited in their predictive power.  They can, for example,
describe the kinematics of a heavy gauge boson produced in
hadron collision, but cannot predict the number or distribution
of jets that accompany it.  
However, searches for new physics, either directly or indirectly through
measurements of precision electroweak observables, often
demand detailed knowledge of kinematic distributions and
jet activity.
Furthermore, $W$+jets (and $Z$+jets) processes are often backgrounds to
SUSY or technicolor signatures, and we demand a reliable
prediction of their properties.  Here, we report on recent
progress in improving the predictive power of showering
Monte Carlos by incorporating
the positive features of the analytic resummation calculations
into the showering algorithms.
In the ensuing discussion, we focus on the specific example 
of $W$ boson production at a hadron collider, when the
$W$ decays leptonically.  The results apply equally well
to $\gamma^*, Z$ and Higgs bosons (or any heavy, color--singlet
particle) produced in hadron collisions.

\subsection{Parton Showers}

SMC's are based on the factorization theorem \cite{factor}, which,
roughly, states that physical observables in QCD
are the product of short--distance functions and long--distance
functions.  The short--distance functions are calculable in
perturbation theory.  The long--distance functions are fit
at a scale, but their evolution to any other scale is also
calculable in perturbation theory.  

A standard application of the factorization theorem is
to describe $W$ boson production at a $p\bar p$ collider
at a fixed order in $\alpha_s$.
The production cross section is obtained by
convoluting the partonic subprocesses evaluated at the
scale $Q$ with the PDF's evaluated at $Q$.  The partons
involved in the hard collision must be sufficiently virtual
to be resolved inside the proton, and a natural choice for
the scale $Q$ is $Q=M_W$ \cite{ISR}.
However, the valence quarks in the proton
have virtualities at a much lower scale $Q_0$ of the order
of 1 GeV.  The connection between the partons at the
low scale $Q_0$ and those at the high scale $Q$ is described by
the DGLAP evolution equations \cite{DGLAP}.  The DGLAP
equations include the most important
kinematic configurations of the splittings $a \to b c$,
where $a,b$ and $c$ represent different types of partons
in the hadron ($q,g$, etc.).  Starting from a measurement
of the PDF's at a low scale $Q_0$, a solution of the DGLAP
equations yields the PDF's at the hard scale $Q$.
Equivalently, starting with a parton $c$ involved
in a hard collision, it is also possible to 
determine probabilistically which splittings generated $c$.
In the process of evolving parton $c$ back to the valence
quarks in the proton, a number of 
spectator partons (e.g. parton $b$ in the branching $a\to bc$) 
are resolved.  These partons
constitute a shower of soft and/or collinear 
jets that accompany the $W$--boson,
and influence its kinematics.  

The shower described above occurs with unit probability and does not change
the total cross section for $W$--boson production calculated
at the scale $Q$ \cite{odorico}.  The showering can be attached to the 
hard--scattering
process based on a probability distribution {\it after} the
hard scattering has been selected.
Once kinematic cuts are applied, the
transverse momentum and rapidity of the $W$--boson populate
regions never accessed by the differential 
partonic cross section calculated at a fixed order.
This is consistent, since the fixed--order calculation was
inclusive (i.e., $p\bar p\to W+X$) and was never intended to
describe the detailed kinematics of the $W$--boson in isolation.  
The parton shower,
in effect, resolves the structure of the inclusive state
of partons denoted as $X$.
In practice, the fixed order
partonic cross section (without showering) 
can still be used to describe properties of the decay
leptons as long as the observable is well defined 
(e.g., the number of leptons with central rapidity 
and high transverse momentum, but not the distribution
of transverse momentum of the $W$).  

Here, we focus on the case of initial state gluon radiation.
More details can be found in Ref. \cite{kbook}, for example.
Showering of the parton $b$ with momentum fraction $x$ resolved at
the scale $Q^2=e^t$ is driven by a Sudakov form factor $\exp(-S)$, such 
as\cite{backwards}
\begin{multline}
\exp\!\!\left( -\!\!\int_{t'}^{t}\!\int_{x \over 1-\epsilon}^{x \over x + 
\epsilon} 
dt'' dz {\alpha_{s}(z,t'') \over 2\pi}
\hat{P}_{a\to bc}(z) {x' f_a(x',t')\over x f_b(x,t') } \!\right), \\
x'=x/z,
\label{sjostrand_shower}
\end{multline}
which is implemented in \pythia, and the formally equivalent 
expression\cite{back2}
\begin{multline}
{\Delta(t') \over f_b(x,t')} {f_a(x,t) \over \Delta(t) },\nonumber \\
\Delta(t') = \exp\!\!\left( -\!\!\int_{t_0}^{t'}\!\int_{\epsilon}^{1-\epsilon}
dt'' dz {\alpha_{s}(z,t'') \over 2\pi}
\hat{P}_{a\to bc}(z) \right),  
\label{webber_shower}
\end{multline}
which is implemented in \herwig.  In the above expressions, $t_0$ is 
a cutoff scale for the showering,
$\hat{P}$ is a DGLAP splitting function,
and $f_i$ is a parton distribution function.
The Sudakov form factor presented here is a solution of the DGLAP equation, and
gives the probability of evolving from the scale ${Q}^2=e^{t}$ to 
${Q'}^2=e^{t'}$ with
no resolvable branching.
The Sudakov form factor contains all the information necessary
to reconstruct a shower, since it encodes the change in virtuality of
a parton until a resolvable showering occurs.
A parton shower is then an iterative solution
of the equation $r=\exp(-S)$, where $r$ is a random number
uniformly distributed in the interval  $[0,1]$, until a solution for $Q'$ 
is found which is below a cutoff.  For consistency, the cutoff
should represent the lowest scale of resolvable emission $Q_0$.
The evolution proceeds 
backwards from
a large, negative scale $-|Q^2|$ to a small, negative cutoff scale
$-|Q_0^2|$.  

After choosing the change in virtuality, a particular backwards branching
and the splitting variable $z$
are selected from the probability function based on
their relative weights (a summation over all possible branchings
$a\to bc$ is implied in these expressions).
The details of how a full shower
is reconstructed in the \pythia~Monte Carlo, for example, can be found 
in Ref.~\cite{pythia}. The structure of the shower can be complex:
the transverse momentum of the $W$--boson
is built up from a whole series of splittings and boosts, and
is known only at the end of the shower, after the final boost.

The SMC formulation
outlined above is fairly independent of the hard scattering
process considered.  Only the initial choice of partons
and possibly the high scale differs.  Therefore, this
formalism can be applied universally to many different
scattering problems.  In effect, soft and collinear
gluons are not sensitive to
the specifics of the hard scattering, only the color charge of
the incoming partons.

\subsection{Analytic Resummation}

At hadron colliders, the partonic cross sections can
receive substantial corrections at higher orders in
$\alpha_s$.  This affects not only the total production
rate, but also the kinematics of the $W$ boson.
At leading order ($\alpha_s^0$), the $W$--boson has a
$\delta(Q_T^2)$ distribution in $Q_T^2$.
At next--to--leading order, the
real emission of a single gluon generates a contribution
to $d\sigma/dQ_T^2$ that behaves as 
$Q_T^{-2}\alpha_s(Q_T^2)$ and $Q_T^{-2}\alpha_s(Q_T^2)\ln(Q^2/Q_T^2)$
while the leading order, soft, and virtual corrections are proportional
to $-\delta(Q_T^2)$.  At higher orders, the most singular terms
follow the pattern of $\alpha_s(Q_T^2)^n\sum_{m=0}^{2n-1}\ln^m(Q^2/Q_T^2)$
$=\alpha_s^n L^m\equiv V^n$.
The logarithms arise from the incomplete cancellation of the virtual
and real QCD corrections, but this cancellation becomes complete for
the integrated spectrum, where the real gluon can become arbitrarily
soft and/or collinear to other partons.  The pattern of singular terms
suggest that perturbation theory should be performed in 
powers of $V^n$ instead of $\alpha_s^n$.
This reorganization of the perturbative series is called resummation.

The first studies of soft gluon emission resummed the leading
logarithms \cite{DDT,Parisi}, leading to a suppression of the cross section
at small $Q_T$.  The suppression
underlies the importance of including sub--leading logarithms \cite{logs}.
The most rigorous approach to the problem of
multiple gluon emission is the Collins--Soper--Sterman
(CSS) formalism for transverse momentum resummation \cite{backtoback},
which resums all of the important logarithms.
This is achieved 
after a Fourier transformation with respect to $Q_T$ in 
the variable $b$, so that the series involving the delta
function and terms $V^n$
simplifies to the form of an exponential.  Hence, the soft gluon
emission is resummed or exponentiated in this $b$--space formalism.  
Despite the successes of the $b$--space formalism,
there are drawbacks: the soft gluon dynamics are integrated out, and
the Sudakov form factor is a Fourier transform.

The CSS formalism was used by its authors to predict both the
total cross section to NLO and the kinematic distributions
of the $W$--boson to all orders \cite{cssWZ} at
hadron colliders.
A similar treatment was presented using the AEGM formalism \cite{AEGM},
that does not involve a Fourier transform, but is evaluated directly
in transverse momentum $Q_T$ space.  
When evaluated at NLO, the two formalisms are equivalent to NNNL order
in $\alpha_s$, and agree with the fixed order calculation
of the total cross section \cite{AEM85}.  A more detailed numerical 
comparison of the
two predictions can be found in Ref. \cite{Arnold-Kauffman}.

Recently, the AEGM formalism has been re-investigated,
and an approximation to the $b$--space formalism has been developed
in $Q_T$--space
which retains its predictive features 
 \cite{resum2} (see also the recent 
eprint \cite{stirling}).
This formulation {\it does} have a simple, physical interpretation,
and can be used to develop an alternate algorithm for 
parton showering which {\it includes} higher--order corrections
to the hard scattering.  For this reason, we focus on the $Q_T$--space
formalism.
To NNNL accuracy, the $Q_T$ space expression agrees
exactly with the $b$--space expression, and has the form \cite{resum2}:
\begin{equation}
\begin{split}
{d\sigma(h_1 h_2 \to V^{(*)}X) \over dQ^2\,dQ^2_T\,dy} = 
  {d\over dQ_T^2}\widetilde{W}(Q_T,Q,x_1,x_2) \\ + Y(Q_T,Q,x_1,x_2). 
\label{eq:qt_space}
\end{split}
\end{equation}
In this expression, $Q$, $Q_T$ and $y$ describe the kinematics of the
boson $V$, the function $Y$ is regular as $Q_T\to 0$
and corrects for the soft gluon 
approximation, and the function $\widetilde W$ has the form:
\begin{equation}
\begin{split}
{\widetilde W} = & e^{-S(Q_T,Q)} H(Q,y) \times \\
&\Big(C \otimes f\Big)(x_1,Q_T) 
\Big(C \otimes f\Big)(x_2,Q_T), 
\end{split}
\end{equation}
where
\begin{equation}
\begin{split}
  S(Q_T,Q) = & \\ 
  \int_{Q_T^2}^{Q^2}
  {d {\bar \mu}^2\over {\bar \mu}^2} &
       \left[ \ln{Q^2\over {\bar \mu}^2}
        A\big(\alpha_s({\bar \mu})\big) + 
        B\big(\alpha_s({\bar \mu})\big) \right],
\end{split}
\end{equation}
and
\begin{multline}
  \left( C_{jl} \otimes f_{l/h_1} \right) (x_1,\mu) = \\[1.mm]
  \int_{x_1}^{1} {d \xi_1 \over \xi_1} \, 
  C_{jl}( {x_1 \over \xi_1}, Q_T)
  f_{l/h_1}(\xi_1, Q_T).
\end{multline}
$H$ is a function that describes the hard
scattering, and $A$, $B$, and $C$ are calculated perturbatively in
powers of $\alpha_s$:
\bes
(A,B,C)=\sum_{n=0}^{\infty} \left({\alpha_s(\mu)\over\pi}\right)^n 
(A,B,C)^{(n)}
\ees
(the first non--zero terms in the expansion of $A$ and $B$ are for $n=1$).
The functions $C^{(n)}$ are the Wilson coefficients, and are responsible for
the change in the total production cross section at higher orders.
In fact, $\left( C\otimes f\right)$ is simply a
redefinition of the parton distribution function obtained by
convoluting the standard ones with an ultraviolet--safe function.

Ignoring $Y$ and other kinematical dependence, Eq.~(\ref{eq:qt_space}) 
can be rewritten as:
\bea
{d\sigma(h_1 h_2 \to WX) \over dQ_T^2 } = \sigma_1
  \left({d\over dQ_T^2}  \left[e^{-S(Q_T,Q)}~R
\right]\right),
\eea
where
\begin{equation}
R=  {\left(C\otimes f\right)(x_1,Q_T) \left(C\otimes f\right)(x_2,Q_T) 
   \over \left(C\otimes f\right)(x_1,Q) \left(C\otimes f\right)(x_2,Q)} 
\end{equation}
and 
\bes
  \sigma_1 = \kappa \int_{}^{} {dx_1 \over x_1} \left(C\otimes 
f\right)(x_1,Q) \left(C\otimes f\right)(x_2,Q).
\ees 
The factor $\sigma_1$ is the total cross section to a fixed order, 
while the rest of 
the function yields the probability that the $W$--boson has a transverse 
momentum $Q_T$.

At leading order, the expression for the 
production of an on--shell $W$--boson simplifies
considerably to:
\begin{multline}
 {d\sigma(h_1 h_2 \to WX) \over dQ^2_T} =  \\[1.mm]
\sigma_0 
  {d \over dQ^2_T} 
  \left(e^{-S(Q_T,Q)} {f(x_1,Q_T) f(x_2,Q_T) \over f(x_1,Q) f(x_2,Q)}\right),
\end{multline}
\bes
 \sigma_0 =  \kappa \int_{}^{} {dx_1 \over x_1} f(x_1,Q) f(x_2,Q), 
\ees  
where $\kappa$ contains physical constants and we ignore the function $Y$
for now.  The expression contains
two factors, the total cross section at leading order $\sigma_0$,
and a cumulative probability function in $Q_T^2$ that describes the
transverse momentum of the $W$--boson. The term $e^{-S/2} f(x,Q_T)/f(x,Q)$
is of the same form as the Sudakov form factor of 
Eq.~(\ref{webber_shower}) and, hence,
to that of Eq.~(\ref{sjostrand_shower}).

\subsection{A modified showering algorithm}

The primary result of this analysis is to exploit the
expression for the differential cross section, which
has the form of a leading order cross section times a 
backwards evolution, to incorporate NLO corrections to the parton shower.  We 
generalize the function $\Delta(t)/f(x,t)\times f(x,t')/\Delta(t')$ of the
standard backwards showering algorithm
to $\sqrt{\widetilde{W}}$ (the square root
appears because we are considering the evolution of each parton line 
individually).

To implement this modification in a numerical program, like \pythia,
we need to provide the new, modified PDF (mPDF) based on the Wilson 
coefficients.  At leading
order, the only Wilson coefficient is $C_{ij}^{(0)} = \delta_{ij}\delta(1-z)$,
and we reproduce exactly the standard showering formulation.
For $W$--boson production at NLO,
the Wilson coefficients $C$ are:
\begin{eqnarray}
C_{jk}^{(1)} & = &\delta_{jk} \left\{{2\over 3}(1-z) + {1\over 
3}(\pi^2-8)\delta(1-z)\right\},\\[1.mm]
C_{jg}^{(1)} & = & {1\over 2}z(1-z).
\end{eqnarray}
To NLO, the convolution integrals become:
\begin{multline}
\left(C\otimes f_i\right)(x,\mu) = f_i(x,\mu) 
\left(1+{\alpha_s(\mu)\over\pi} {1\over 3}(\pi^2-8)\right) \nonumber \\[1.mm]
 + {\alpha_s(\mu)\over\pi} \int_{x}^{1} {dz\over z}\left[{2\over 
3}(1-z)f_i(x/z,\mu) \right .\\[1.mm]
\left . + {1\over 2}(1-z)f_g(x/z,\mu)\right],
\end{multline}
and $f_g(x,\mu)$ is unchanged.
The first term gives the contribution of an unevolved parton to the hard
scattering, while the other two contain contributions from
quarks and gluons with higher momentum fractions that split
$q\to q g$ and $g \to q \bar q$, respectively.  

We are assuming that
the Sudakov form factor used in the analytic expressions and in the SMC 
are equivalent.  In fact,
the integration over the quark splitting function in $\Delta(Q)$ 
yields an expression similar to the analytic Sudakov:
\begin{multline}
\int_{z_m}^{1-z_m} dz C_F \left( {1+z^2\over 1-z} \right) = \\[1.mm]
C_F \left( \ln\left[{1-z_m\over z_m}\right]^2
-3/2(1-2z_m) \right) \\[1.mm]
\simeq A^{(1)}\ln(Q^2/Q_T^2) + B^{(1)},
\label{eq:analytic_sud}
\end{multline}
where $z_m = {Q_T \over (Q+Q_T)}$ is an infrared cutoff, terms of order 
$z_m$ and higher are neglected, and the $z$ dependence of the
running coupling has been ignored \cite{soft}.
Note that the coefficients $A^{(1)}$ ($C_F$) and $B^{(1)}$ ($-3/2 C_F$)
are universal to 
$q\bar q$ annihilation into a color singlet object, 
just as the showering Sudakov form
factor only knows about the partons and not the details of the hard scattering.
For $gg$ fusion, only the coefficient $A^{(1)}$ (3) is universal.
In general, at higher orders,
the analytic Sudakov is sensitive to the exact hard scattering process.

While the Sudakov form factors are similar, there is no one--to--one
correspondence.  First, the \qt--space Sudakov form factor is expressed
directly in terms of the $Q_T$ of the heavy boson, while, in the SMC's, 
the final $Q_T$ is built
up from a series of branchings.  
Secondly, the integral on the left of Eq.~(\ref{eq:analytic_sud}) is 
positive (provided that
$z_m<\frac{1}{2}$), while the analytic expression on the right
can become negative.
This is disturbing, since it means sub-leading logarithms (proportional
to $B$) are dominating leading ones.  In the exact SMC Sudakov,
the kinematic constraints guarantee that $\Delta(Q)<1$.
In this sense, the Sudakov in the SMC is a more exact implementation
of the analytic one.
Nonetheless, the agreement apparent between the analytic and parton
shower expressions
is compelling enough to
proceed assuming the two Sudakov form factors are equivalent.

\subsection{Hard Emission Corrections}

The SMC and resummation formalisms are optimized to deal with kinematic
configurations that have logarithmic enhancements $L$.
For large $Q_T\simeq Q$, there are no such enhancements, and
a fixed order calculation yields the most accurate predictions.
The region of medium $Q_T$, however, is not suited to either
particular expansion, in $\alpha_s^nL^m$ or $\alpha_s^n$.

The problem becomes acute for SMC's.
In the standard implementation of SMC's, the highest $Q_T$ 
is set by the maximum virtuality allowed, $Q=M_W$ in our
example, so that the region $Q_T\ge Q$ is never accessed.
However, at $Q_T \ge Q$, the fixed order calculation is
preferred and yields a non--zero result, 
so there is a discontinuity between the two
predictions.  This behavior does not occur in the analytic
resummation calculations, because
contributions to the cross section that are not
logarithmically enhanced as $Q_T\to 0$ are added back order--by--order
in $\alpha_s$.  This procedure corrects for the 
approximations made in deriving the exponentiation of soft gluon
emission.  The correction is denoted $Y$.
If the coefficients
$A$ and $B$ are calculated to high--enough accuracy, one sees a
relatively smooth transition between Eq.~(\ref{eq:qt_space}) and the NLO 
prediction
at $Q_T=Q$.  In the $Q_T$--space calculation,
this matching between the two calculations at $Q_T=Q$ is guaranteed at 
any order. The function $Y$ has the form
\begin{multline}
Y(Q_T,Q,x_1,x_2)=\int_{x_1}^1{\frac{d\xi _1}{\xi _1}}
\int_{x_2}^1{\frac{d\xi_2}{\xi _2}}\sum_{n=1}^\infty \left[ {\frac{\alpha
_s(Q)}\pi }\right] ^n  \\[1.mm]
f_{a}(\xi _1,Q)\,R_{ab}^{(n)}(Q_T,Q,\frac{x_1}{\xi _1},
\frac{x_2}{\xi _2})\,f_{b}(\xi _2,Q).
\label{Ypiece}
\end{multline}
For $W$ or $Z$ boson production, the $a=q,b=\bar q$ component of $R$  at 
first order in $\alpha_s$ is
\begin{multline}
R^{(1)}_{q\bar q} = C_F{ {(\that-\q2)^2 + (\uhat-\q2)^2} \over 
{\hat{t}\hat{u}} }\delta(\hat{s}+\hat{t}+\hat{u}-Q^2) \\[1.mm]
 - {1\over Q_T^2}\hat{P}_{q\to q}(z_B)\delta(1-z_A) - (A \leftrightarrow B).
\end{multline}
The invariants $\hat{s},\hat{t}$ and $\hat{u}$ are defined in terms of 
$z, Q, Q_T$:
\begin{equation}
\begin{split}
&\hat{t}/Q^2 = 1 - 1/z_B\sqrt{1+Q_T^2/Q^2}, \\
&\hat{u}/Q^2 = 1 - 1/z_A\sqrt{1+Q_T^2/Q^2}.
\end{split}
\end{equation}
The term in $R$ proportional to the delta function is simply the
squared matrix element for the hard emission, while the
terms proportional to $Q_T^{-2}$ are the asymptotic
pieces from $\widetilde{W}$.

We would like to include similar corrections into the SMC.  However, this 
is not entirely straightforward.  Though it is not obvious from 
Eq.~(\ref{Ypiece}),
the ($a=g,b=q$+permutations) components are negative for $Q_T<Q$, 
though the sum $Y$ is positive.  
Retaining negative weights in an intermediate part of the calculation is
not a problem in principle.  We can artificially force the negative 
weights to
be positive, and then include the correct sign of the weight when filling
histograms, for example.  However, this would involve some modification
to the~\pythia~code used in this study.  

A pragmatic approach is to ignore the negative weights entirely, and 
multiply the exact $W+$parton 
cross sections by a factor so that their sum reproduces the $Q_T$ 
distribution and normalization
of the analytic $Y$ piece.  For the Tevatron in Run I, we find that 
the multiplicative
factor $f_{\rm COR}={1\over 2}(Q_T/50)^2\times (1+Q_T/25)$ reproduces 
the correct behavior
for $Q_T<50$ GeV.  For $Q_T\ge 50$ GeV, the uncorrected $W+$ parton 
cross sections
are employed.  Since the matching between the ``resummed'' and ``fixed order''
calculations is now occurring at $Q_T=50$ GeV instead of $Q_T=M_W$, we 
further limit the maximum
virtuality of showering to $50$ GeV.  This is in accord with the fact that the 
``resummed'' part of the analytic calculation becomes negative around 
$Q_T=50$ GeV.
This choice does have some effect on the overall normalization of the 
parton showering component.

At this point, it is useful to compare the scheme outlined above to
other approaches at improving the showering algorithm.
One scheme is
based on phase--space splitting of 
a NLO matrix element into a piece with LO kinematics and another 
with exclusive NLO kinematics \cite{matching,matching2}.  
The separation depends on an adjustable parameter
that splits the phase space.
In the approach of Ref. \cite{matching},
the separation parameter is tuned so that the contribution with
LO kinematics vanishes.  The resultant showering of the term
with exclusive NLO kinematics can generate emissions which are
harder than the first ``hard'' emission, which is not consistent.
More seriously, physical observables are
sensitive to the exact choice of the separation parameter
(see the discussion in Ref. \cite{balazs}
regarding $Q_T^{sep}$).  
Furthermore, the separation parameter must be retuned for different
processes and different colliders.
This scheme is guaranteed to give
the NLO cross section before cuts, but does not necessarily
generate the correct kinematics.

The other scheme is to modify
the showering to reproduce the hard emission
limit \cite{merging,merging2}.  While this can be accomplished,
it does so at the expense of transferring events from
low $Q_T$ to high $Q_T$.  There is no attempt to predict
the absolute event rate, but only to generate the correct
event shapes.  In some implementations, the
high scale of the showering is increased to the maximum
virtuality allowed by the collider energy.  This is contrary
to the analytic calculations, where the scale $Q=M_W$, for example,
appears naturally (in the choice of constants $C_1, C_2$ and $C_3$
which eliminate potentially large logarithms).  This scheme
will generate the correct hard limit, but will not generate the
correct cross section in the soft limit.
\subsection{Numerical results}

For our numerical results, 
we predict the $Q_T$ distribution of $W$ and $Z$ bosons
produced at the Tevatron in Run I.
The modified PDF (mPDF) was calculated using CTEQ4M PDF's.
These distributions are in good
agreement with analytic calculations, 
but the shape and overall normalization cannot be predicted accurately 
by the standard showering algorithm.  Some of the alternative
showering algorithms reproduce the shape, but not the overall
normalization.
Secondly, we discuss jet properties for the same processes,
which are not significantly altered from the predictions of
the standard showering algorithm.
These cannot be predicted by analytic calculations.

%
%
\begin{figure*}[!ht]
\begin{center}
\begin{tabular}{cc}
\psfig{figure=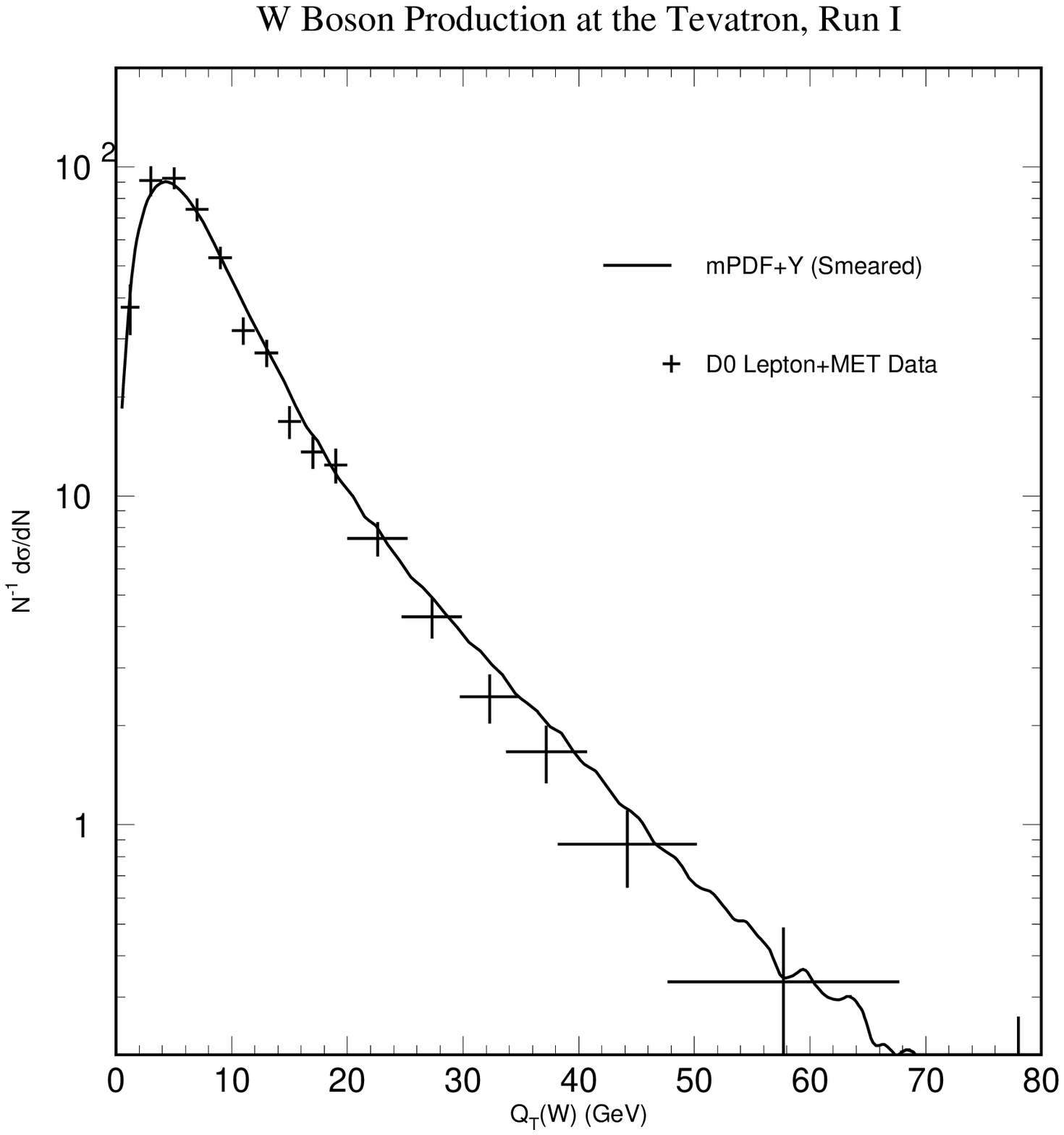,width=8.6cm} & 
\psfig{figure=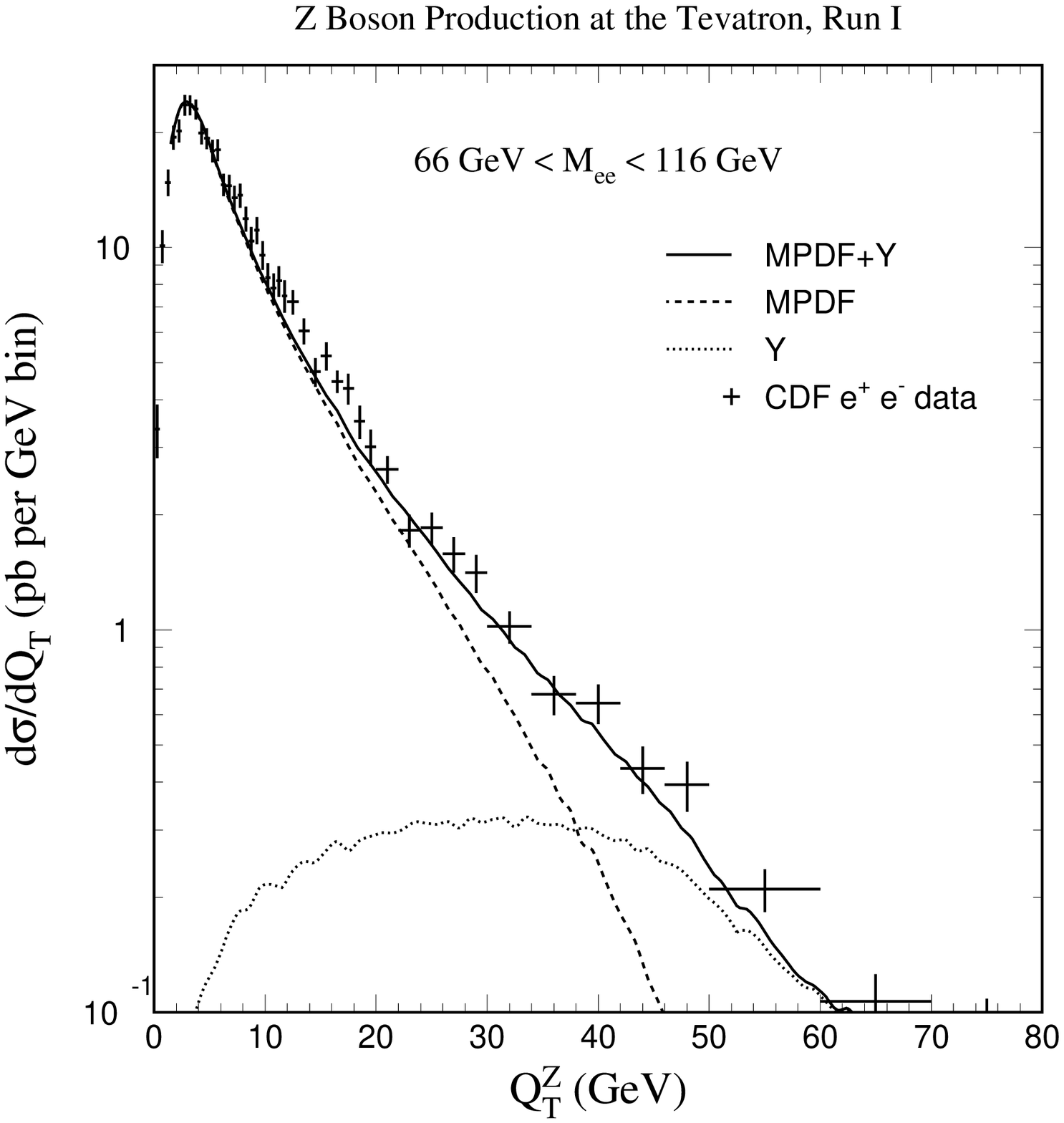,width=8.6cm}
\end{tabular}
\end{center}
\vskip -9.mm
\caption{$(a)$ The prediction of
the $W$ boson transverse momentum distribution in Run I at the
Tevatron (solid line) compared to the D\O~data.
The prediction includes the effects of the modified parton
distribution functions, the correction to the hard scattering
process, and a primordial $k_T$ of 2.0 GeV;
(b) The prediction of
the $Z$ boson transverse momentum distribution in Run I at the
Tevatron (solid line) compared to the CDF~data.
The prediction includes the effects of the modified parton
distribution functions, the correction to the hard scattering
process, and a primordial $k_T$ of 2.0 GeV.}
\label{fig:wpt_runi}
\end{figure*}

In Fig.~\ref{fig:wpt_runi}(a), the transverse momentum of the $W$ boson 
(solid line)
as predicted by the algorithm outlined above
is shown
in comparison to D\O~data\cite{d0data} (crosses).  
The theoretical distribution has been passed through the
{\sf CMS} detector simulation.\footnote{Special thanks go to
Cecilia Gerber, for making the code portable, and to
Michael Seymour for explaining how to properly use it.}
As in analytic calculations,
the position of the peak in the $Q_T$ distribution from parton showering 
depends on non--perturbative physics
\cite{primkt}.
In \pythia, this is implemented through a Gaussian smearing of the
transverse momentum of the incoming partons.  To generate this plot,
we have changed the default Gaussian width from .44 GeV to 2.0 GeV,
which is more in accord with other analyses.  This is the value
used in all subsequent results.  Because of the necessity of
reconstructing the missing $E_T$ in $W$ boson decays to leptons,
the smearing of the $Q_T$ distribution is significant, and the
agreement between the prediction and data is not a rigorous test
of the modified showering algorithm.  Fig.~\ref{fig:wpt_runi}(b)
shows the comparison of the CDF Drell--Yan data \cite{cdfdata}
near the $Z^0$ peak
to the modified showering prediction.  While there is a problem with
the overall normalization, the shape agreement is very good.  We
note that there is also a problem with the overall normalization of
the analytic resummation predictions.\footnote{Csaba Balazs, private
communication.}

Given all the effort necessary to improve the showering, 
it is reasonable to ask if the similar results would have been
obtained by simply renormalizing the usual predictions to the
NLO rate, i.e. using~\pythia~but applying a constant $K$--factor
at the end.  In $W$ boson production, the relative size of the $Q_T$ 
distributions vary by as much as 10\%
in the important regions of small and medium $Q_T$.   Of course, the effect
is much larger for the large $Q_T$ region where there is almost no rate
from the standard parton showering.   If one is worried about
precision measurements or is applying kinematic cuts that bias the 
large $Q_T$ region,
then standard parton showering can yield misleading results.  In most cases,
however,
it appears to be perfectly reasonable to renormalize the parton 
showering results to the total NLO cross section.
We have also checked if our new showering algorithm 
has an impact on jet properties.
For $W$ and $Z$ boson production, 
there are only minor 
differences, which is expected since the Wilson coefficients
for $W$ and $Z$ boson production are nearly unity. 
In general, we do not expect any major changes from using
the modified PDF's,
since the showering depends on the ratio
of the modified PDF's evaluated at two different scales,
which is not as sensitive to the overall normalization
of the PDF.

\subsection{Conclusions}

We have presented a modified, parton showering algorithm that 
produces the total cross section and the event shapes beyond
the leading order.
These modifications are based on the $Q_T$--space
resummation.  The parton showering itself is modified by using
a new PDF (called mPDF) which encodes some information about the hard
scattering process.  Simultaneously, the explicit, hard emission
is included, but only after subtracting out the contribution
already generated by the showering: this correction
is called $Y$.  The presence of $Y$ yields  a smooth
transition from the parton showering to single, hard emission.   
We modified the
\pythia~Monte Carlo to account for these corrections, and
presented comparisons with Run I $W$ and $Z$ boson data.

The scheme works well for the cases considered in
this study, and the correct cross sections, transverse momentum
distributions, and jet properties are generated.  
We have compared our kinematic distributions to the case when
the results of the standard showering are multiplied by a constant
$K$--factor to reproduce the NLO cross section.  We find variations
on the order of 10\% for small and medium transverse momentum.

There are several effects which still need study.
We have not included the exact distributions for the decay
of the leptons \cite{Balazs-Qui-Yuan} for $W$ and $Z$ production,
which are resummed differently.
It is straightforward to include such effects.
In the theoretical discussion and numerical results, we have focussed
on initial state radiation, 
but our results should apply equally well
for final state radiation.  The situation is certainly simpler, since final
state radiation does not require detailed knowledge of the
fragmentation functions.  Also, the case when color flows from
the initial state to the final state requires study.
A resummed calculation already exists for the case of
deep inelastic scattering \cite{Olness}, and
much theoretical progress has been made for heavy quark
production \cite{Sterman}.
We believe that the modified showering scheme outlined in
this study
generalizes beyond NLO, just as the analytic calculations
can be calculated to any given order.
For example, we could include hard $W+2$ jet corrections \cite{Arnold-Reno}
to $Y$.  For consistency, however, higher order
terms ($A$ and $B$)  may
also need to be included in the Sudakov form factor.

The~modified \pythia\ subroutines used in this study and an explanation 
of how to use them are available at the following URL: \\
{\tt moose.ucdavis.edu/mrenna/shower.html}.

{\bf Acknowledgements}
I thank C--P Yuan and T. Sj\"ostrand for many useful discussions and 
encouragement
in completing this work.  This work was supported by United States 
Department of
Energy and the Davis Institute for High Energy Physics.

\section{Recent Progress in \pythia}

\centerline{\it by T. Sj\"ostrand}\vskip 2.mm

\subsection{Introduction}

A general-purpose generator in high-energy physics should address a 
number of physics aspects, such as:
\begin{Itemize}
\item the matrix elements for a multitude of hard subprocesses of interest,
\item the convolution with parton distributions to obtain the
hard-scattering kinematics and cross sections,
\item resonance decays that (more or less) form part of the hard subprocess
(such as $W$, $Z$, $t$ or $h$),
\item initial- and final-state QCD and QED showers (or, as an alternative,
higher-order matrix elements, including a consistent treatment of 
vir\-tual-correction terms), 
\item multiple parton--parton interactions,
\item beam remnants,
\item hadronization,
\item decay chains of unstable particles, and
\item general utility and analysis routines (such as jet finding).
\end{Itemize}
However, even if a Monte Carlo includes all the physics we currently 
know of, there is no guarantee that not some important aspect of the 
physics is missing. 
Certain assumptions and phenomenological models inside the program
are not well tested and will not necessarily hold when extrapolated
to different energy regimes.  For example,
the strong-interaction dynamics in QCD remains unsolved and
thereby unpredictable in an absolute sense.
 
The \pythia~6.1 program was released in March 1997, as a merger of 
{\sc JETSET} 7.4, \pythia~5.7 \cite{pythia} and {\sc SPYTHIA} \cite{spythia}.
It addresses all of the aspects listed above.  The current subversion is
\pythia~6.136, which contains over 50,000 lines of Fortran 77 code.
The code, manuals and sample main programs may be found at\\
\texttt{http://www.thep.lu.se/}$\sim$\texttt{torbjorn/Pythia.html}~.

The two other programs of a similar scope are \herwig \cite{herwig}
\footnote{\verb!http://hepwww.rl.ac.uk/theory/seymour/herwig/!}
and \isajet \cite{isajet}
\footnote{\verb!ftp://penguin.phy.bnl.gov/pub/isajet!}.
For parton-level processes, many more programs have been written.
The availability of several generators provides for useful cross-checks 
and a healthy competition. Since the physics of a complete hadronic 
event is very complex and only partially understood from first principles,
one should not prematurely converge on one single approach.
 
\subsection{\pythia~6.1 Main News}

Relative to previous versions, the main news in \pythia~6.1 includes
\begin{Itemize}
\item a renaming of the old JETSET program elements to begin with 
\texttt{PY}, therefore now standard throughout,
\item new SUSY processes and improved SUSY simulation relative to 
SPYTHIA, and new PDG codes for sparticles,
\item new processes for Higgs (including doubly-char\-ged in 
left--right symmetric models), technicolor, \ldots, 
\item several improved resonance decays, including an alternative 
Higgs mass shape,
\item some newer parton distributions, such as CTEQ5 \cite{cteq5},
\item initial-state showers matched to some matrix elements,
\item new options for final-state gluon splitting to a pair of $c/b$ 
quarks and modified modeling of initial-state flavor excitation,
\item an energy-dependent $p_{\perp\mathrm{min}}$ in multiple interactions, 
\item an improved modeling of the hadronization of small-mass strings,
of importance especially for $c/b$, and
\item a built-in package for one-dimensional histo\-grams (based on GBOOK). 
\end{Itemize}
Some of these topics will be further studied below. Other improvements, 
of less relevance for $\overline{p}p$ colliders, include
\begin{Itemize}
\item improved modeling of gluon emission off $c/b$ quarks in $e^+e^-$,
\item color rearrangement options for $W^+W^-$ events, 
\item a Bose-Einstein algorithm expanded with new options,
\item a new alternative baryon production scheme \cite{patrik},
\item QED radiation off an incoming muon,
\item a new machinery to handle real and virtual photon 
fluxes, cross sections and parton distributions \cite{christer}, and
\item new standard interfaces for the matching to external generators of
two, four and six fermions (and of two quarks plus two gluons) in 
$e^+e^-$.
\end{Itemize}

The current list of over 200 different subprocesses covers topics 
such as hard and soft QCD, heavy flavors, DIS and $\gamma\gamma$, 
electroweak production of $\gamma^*/Z^0$ and $W^{\pm}$ (singly or in 
pairs), production of a light or a heavy Standard Model Higgs, or of 
various Higgs states in supersymmetric (SUSY) or left--right symmetric 
models, SUSY particle production (sfermions, gauginos, etc.), 
technicolor, new gauge bosons, compositeness, and leptoquarks. 

Needless to say, most users will still find that their particular area
of interest is not as well addressed as could be wished. In some areas,
progress will require new ideas, while lack of time and manpower is the 
limiting 
factor in others.

\subsection{Matching To Matrix Elements}

The matrix-element (ME) and parton-shower (PS) approaches to higher-order
QCD corrections both have their advantages and disadvantages. The former
offers a systematic expansion in orders of $\alpha_s$, and a powerful
machinery to handle multiparton configurations on the Born level, 
but loop calculations are tough and lead to messy cancellations at
small resolution scales. Resummed matrix elements may circumvent
the latter problem for specific quantities, but then do not
provide exclusive accompanying events. Parton showers are based
on an improved leading-log (almost next-to-leading-log) approximation, 
and so cannot be accurate for well separated partons, but they offer a 
simple, process-independent machinery that gives a smooth blending of event 
classes (by Sudakov form factors) and a natural match to hadronization.
It is therefore natural to try to combine these descriptions, so
that ME results are recovered for widely separated partons while the
PS sets the subjet structure. 

For final-state showers in $Z^0 \to q\overline{q}$, corrections to
the showering were considered quite a while ago \cite{fsmatch}, 
e.g. by letting the shower 
slightly overpopulate the $q\overline{q}g$ phase space and then using
a Monte Carlo veto technique to reduce down to the ME level. This approach
easily carries over to showers in other color-singlet resonance decays,
although the various relevant ME's have not all been implemented in
PYTHIA so far.

A similar technique is now available for the description of initial-state
radiation in the production of a single color-singlet resonance, such
as $\gamma^*/Z^0/W^{\pm}$ \cite{gabriela}. The basic idea is to map the 
kinematics between the PS and ME descriptions, and to find a correction 
factor that can be applied to hard emissions in the shower so as to bring
agreement with the matrix-element expression. Some simple algebra
shows that, with the PYTHIA shower kinematics definitions,
the two $q\overline{q}' \to gW^{\pm}$ emission rates disagree by a 
factor
\[
R_{q\overline{q}' \to gW}(\hat{s},\hat{t}) = 
\frac{(\mathrm{d}\hat{\sigma}/\mathrm{d}\hat{t})_{\mathrm{ME}} }%
     {(\mathrm{d}\hat{\sigma}/\mathrm{d}\hat{t})_{\mathrm{PS}} } = 
\frac{\hat{t}^2+\hat{u}^2+2 m_W^2\hat{s}}{\hat{s}^2+m_W^4} ~, 
\]
which is always between $1/2$ and 1. 
The shower can therefore be improved in two ways, relative to the 
old description. Firstly, the maximum virtuality of emissions is 
raised from $Q^2_{\mathrm{max}} \approx m_W^2$ to 
$Q^2_{\mathrm{max}} = s$, i.e. the shower is allowed to populate the 
full phase space. Secondly, the emission rate for the final (which 
normally also is the hardest) $q \to qg$ emission on each side is 
corrected by the factor $R(\hat{s},\hat{t})$ above, so as to bring 
agreement with the matrix-element rate in the hard-emission region.
In the backwards evolution shower algorithm \cite{backwards}, this 
is the first branching considered.

The other possible ${\mathcal{O}}(\alpha_s)$ graph is $qg \to q'W^{\pm}$,
where the corresponding correction factor is
\[
R_{qg \to q'W}(\hat{s},\hat{t}) =
\frac{(\mathrm{d}\hat{\sigma}/\mathrm{d}\hat{t})_{\mathrm{ME}} }%
     {(\mathrm{d}\hat{\sigma}/\mathrm{d}\hat{t})_{\mathrm{PS}} } = 
\frac{\hat{s}^2 + \hat{u}^2 + 2 m_W^2 \hat{t}}{(\hat{s}-m_W^2)^2 + m_W^4} ~,
\]
which lies between 1 and 3. A probable reason for the lower shower 
rate here is that the shower does not explicitly simulate the $s$-channel 
graph $qg \to q^* \to q'W$. The $g \to q\overline{q}$ branching 
therefore has to be preweighted by a factor of 3 in the shower, but 
otherwise the method works the same as above. Obviously, the shower 
will mix the two alternative branchings, and the correction factor 
for a final branching is based on the current type.

The reweighting procedure prompts some other chang\-es in the shower. 
In particular, $\hat{u} < 0$ translates into a constraint on the phase
space of allowed branch\-ings. 

Our published comparisons with data on the $W$ $p_{\perp}$ spectrum 
show quite a good agreement with this improved simulation \cite{gabriela}. 
A worry was that an unexpectedly large primordial $k_{\perp}$, around 
4 GeV, was required to match the data in the low-$p_{\perp W}$ region. 
However, at that time we had not realized that the data were not fully 
unsmeared. The required primordial $k_{\perp}$ is therefore likely to
drop by about a factor of two \cite{joey}.

It should be noted that also other approaches to the same problem have
been studied recently. The HERWIG one requires separate treatments in 
the hard- and soft-emission regions \cite{herwigw}. Another, more 
advanced PYTHIA-based one \cite{steve}, also addresses the next-to-leading 
order corrections to the total $W$ cross section, while the one outlined 
above is entirely based on the leading-order total cross section. There 
is also the possibility of an extension to Higgs production \cite{steveh}. 

Summarizing, we now start to believe we can handle initial- and 
final-state showers, with next-to-leading-order accuracy, in cases where 
these can be separated by the production of color singlet resonances
--- even if it should be realized that much work remains to cover the 
various possible cases. That still does not address the big class of 
QCD processes where the initial- and final-state radiation does not 
factorize. Possibly, correction factors to showers could be found
also here. Alternatively, it may become necessary to start showers from
given parton configurations of varying multiplicity and with
virtual-correction weights, as obtained from higher-order ME
calculations. So far, \mbox{PYTHIA} only implements a way to start from a
given four-parton topology in $e^+e^-$ annihilation, picking one
of the possible preceding shower histories as a way to set constraints
for the subsequent shower evolution \cite{johan}. This approach obviously 
needs to be extended in the future, to allow arbitrary parton 
configurations. Even more delicate will be the consistent treatment
of virtual corrections \cite{christerw}, where much work remains.

\subsection{Charm And Bottom Hadronization}

Significant asymmetries are observed between the production of
$D$ and $\overline{D}$ mesons in $\pi^- p$ collisions, with hadrons
that share some of the $\pi^-$ flavor content very much favored at 
large $x_F$ in the $\pi^-$ fragmentation region \cite{casym}. This 
behavior was qualitatively predicted by PYTHIA; in fact, the predictions
were for somewhat larger effects than seen in the data. The new data 
has allowed us to go back and take a critical look at the uncertainties 
that riddle the heavy-flavor description \cite{emanuel}. Many effects 
are involved, and we limit ourselves here to mentioning only one.

A hadronic event can be subdivided into sets of partons
that form separate color singlets. These sets are represented by strings,
that e.g. stretch from a quark end via a number of intermediate gluons
to an antiquark end. The string has a mass, which can be calculated from
the energy-momentum of its partons.
Three different mass regions for the strings may be distinguished in
the process of hadronization.
\begin{Enumerate}
\item {\em Normal string fragmentation}. This is the ideal situation, 
when each 
string has a large invariant mass, and the standard iterative 
fragmentation scheme \cite{lund} works well. In practice, 
this approach can be used for all strings with a mass above a cut-off of a 
few GeV. 
\item {\em Cluster decay}.
If a string is produced with a \mbox{small} invariant mass, then it
is possible that only 
two-body final states are kinematically accessible. The traditional 
iterative Lund scheme is then not applicable. We call such a low-mass 
string a cluster, and treat it separately.  In recent 
program versions, the modeling has been improved to give a smooth 
match onto the standard string scheme in the high-cluster-mass limit.
\item {\em Cluster collapse}.
This is the extreme case of the above situation, where the string 
mass is so small that the cluster cannot decay into even two hadrons.
It is then assumed to collapse directly into a sing\-le hadron, which
inherits the flavor content of the string endpoints. The original 
continuum of string/cluster masses is replaced by a discrete set
of hadron masses. Energy and momentum then cannot be conserved
inside the cluster, but must be exchanged with other objects
within the local neighborhood.
This description has also been improved.  
\end{Enumerate}

Because the mass of the charm and bottom partons are not negligible
in the fragmentation process, the improved treatment of low-mass systems 
will have relatively more impact on charm and bottom hadronization.
In general, flavor asymmetries are predicted to be smaller for bottom 
than for charm, and smaller at higher energies (except possibly at very 
large rapidities). Therefore, we do not expect any spectacular 
effects at the Tevatron. 
However, other nontrivial features of fragmentation 
may persist at higher energies,
like a non-negligible systematic shift between the 
rapidity of a heavy quark parton and that of the hadron produced from it
\cite{emanuel}. The possibility of such effects should be considered 
whenever trying to relate heavy flavor measurements to parton level
calculations.

\subsection{Multiple Interactions}

Because of
the composite nature of hadrons, several parton pairs may interact
in a typical hadron--hadron collision \cite{maria}. Over the years, 
evidence for this mechanism has accumulated, such as the recent di\-rect
observation by CDF \cite{cdfmultint}.  However, the occurrence of two 
hard interactions in one hadronic collision is just the tip of the iceberg.
In the PYTHIA model, most interactions are
at lower $p_{\perp}$, where they are not visible as separate jets but
only contribute to the underlying event structure. As such, they are
at the origin of a number of key features, like the broad 
multiplicity distributions, the significant forward--backward
multiplicity correlations, and the pedestal effect under jets.

Since the perturbative jet cross section is divergent for 
$p_{\perp} \to 0$, it is necessary to regularize it, e.g. by a
cut-off at some $p_{\perp\mathrm{min}}$ scale. That such a 
regularization should occur is clear from the fact that the incoming
hadrons are color singlets --- unlike the colored partons assumed in 
the divergent perturbative cal\-cu\-la\-tions --- and that therefore the 
color charges should screen each other in the $p_{\perp} \to 0$ limit.
Also other damping mechanisms are possible \cite{goga}.
Fits to data typically give $p_{\perp\mathrm{min}} \approx 2$ GeV,
which then should be interpreted as the inverse of some color
screening length in the hadron.   

One key question is the energy-dependence of $p_{\perp\mathrm{min}}$; 
this may be relevant e.g. for comparisons of jet rates at different 
Tevatron energies, and even more for any extrapolation to LHC energies. 
The problem actually is more pressing now than at the time of our 
original study \cite{maria}, since nowadays parton distributions are 
known to be rising more steeply at small $x$ than the flat $xf(x)$ 
behavior normally assumed for small $Q^2$ before HERA. This 
translates into a more dramatic energy dependence of the 
multiple-interactions rate for a fixed $p_{\perp\mathrm{min}}$. 

The larger number of partons also should increase the amount of
screening, as confirmed by toy simulations \cite{johann}.
As a simple first approximation, $p_{\perp\mathrm{min}}$ is assumed
to increase in the same way as the total cross section, i.e. with some 
power $\epsilon \approx 0.08$ \cite{dl} that, via reggeon phenomenology,
should relate to the behavior of parton distributions at small $x$ 
and $Q^2$. Thus the new default in PYTHIA is
\[
p_{\perp\mathrm{min}} = (1.9~{\mathrm{GeV}}) \left(
\frac{s}{1~\mathrm{TeV}^2} \right)^{0.08} ~.
\]

\subsection{Interconnection Effects}

The widths of the $W$, $Z$ and $t$ are all of the order of 
2 GeV. A Standard Model Higgs with a mass above 200 GeV, as well 
as many supersymmetric and other ``Beyond the Standard Model'' particles
would also have widths in the multi-GeV range. Not far from
threshold, the typical decay times 
$\tau = 1/\Gamma  \approx 0.1 \, {\mathrm{fm}} \ll  
\tau_{\mathrm{had}} \approx 1 \, \mathrm{fm}$.
Thus hadronic decay systems overlap, between a resonance and the
underlying event, or between pairs of resonances, so that the final 
state may not contain independent resonance decays.

So far, studies have mainly been performed in the context of
$W$ pair production at LEP2. Pragmatically, one may here distinguish 
three main eras for such interconnection:
\begin{Enumerate}
\item Perturbative: this is suppressed for gluon energies 
$\omega > \Gamma$ by propagator/time\-scale effects; thus only
soft gluons may contribute appreciably.
\item Non-perturbative in the hadroformation process:
normally model-led by a color rearrangement between the partons 
produced in the two resonance decays and in the subsequent parton
showers.
\item Non-perturbative in the purely hadronic phase: best exemplified 
by Bose--Einstein effects.
\end{Enumerate}
The above topics are deeply related to the unsolved problems of 
strong interactions: confinement dyna\-mics, $1/N^2_{\mathrm{C}}$ 
effects, quantum mechanical interferences, etc. Thus they offer 
an opportunity to study the dynamics of unstable particles,
and new ways to probe confinement dynamics in space and 
time \cite{GPZ,ourrec}, {\em but} they also risk 
to limit or even spoil precision measurements.

It is illustrative to consider the impact of 
interconnection effects on the $W$ mass measurements 
at LEP2. Perturbative effects are not likely
to give any significant contribution to the systematic error,
$\langle \delta m_W \rangle \lessim 5$~MeV \cite{ourrec}. 
Color rearrangement is not understood from first principles,
but many models have been proposed to model effects 
\cite{ourrec,otherrec,HR}, and a conservative estimate gives 
$\langle \delta m_W \rangle \lessim 40$~MeV. 
For Bose--Einstein again there is a wide spread in models, and an 
even wider one in results, with about the same potential systematic
error as above \cite{ourBE,otherBE,HR}.
The total QCD interconnection error is thus below $m_{\pi}$ in 
absolute terms and 0.1\% in relative ones, a small number that 
becomes of interest only because we aim for high accuracy. 

A study of $e^+e^- \to t\overline{t} \to b W^+ \overline{b} W^- 
\to b \overline{b} \ell^+ \nu_{\ell} {\ell'}^- \overline{\nu}'_{\ell}$
near threshold gave a realistic interconnection 
uncertainty of the top mass of around 30 MeV, but also showed that 
slight mistreatment of the combined color and showering structure 
could blow up this error by a factor of ten \cite{intertop}. 
For hadronic top decays, errors could be much larger.

The above numbers, when applied to hadronic physics, are maybe not
big enough to cause an immediate alarm. The addition of a colored 
underlying event --- with a poorly-understood multiple-interaction
structure as outlined above --- has not at all been considered so far, 
however, and can only make matters worse in hadronic physics than in 
$e^+e^-$. This is clearly a topic for the future, where we
should be appropriately humble about our current understanding,
at least when it comes to performing precision measurements.

QCD interconnection may also be at the root of a number of 
other, more spectacular effects, such as rapidity gaps and the whole
Pomeron concept \cite{uppsalapom}, and the unexpectedly large rate of
quarkonium production \cite{uppsalaonia}.

\subsection{The Future: On To {C$++$}}

Finally, a word about the future. PYTHIA continues to be developed. 
On the physics side, there is a need to increase the support given 
to different physics scenarios, new and old, and many areas of the 
general QCD machinery for parton showers, underlying events and 
hadronization require further improvements, as we have seen. 

On the technical side, the main challenge is a transition from Fortran 
to C++, the language of choice for Run II (and LHC). To address this,
the PYTHIA 7 project was started in January 1998, with L. L\"onnblad bearing
the main responsibility. A similar project, but more ambitious and better 
funded, is now starting up for HERWIG, with two dedicated postdoc-level 
positions and a three-year time frame. 

For PYTHIA, what exists today is a strategy document \cite{leif}, 
and code for the event record, the particle object, some particle data 
and other data base handling, and the event generation handler structure. 
All of this is completely new relative to the Fortran version, and is 
intended to allow for a much more general and flexible formulation of 
the event generation process. The first piece of physics, the string 
fragmentation scheme, is being implemented by M. Bertini, and is nearing 
completion. The subprocess generation method is being worked on for the 
simple case of $e^+e^- \to Z^0 \to q\overline{q}$. The hope is to have 
a ``proof of concept'' version soon, and some of the current 
PYTHIA functionality up and running by the end of 2000.
It will, however, take much further effort after that to provide a 
program that is both more and better than the current \pythia~6 
version. It is therefore unclear whether \pythia~7 will be of
much use during Run II, except as a valuable exercise for the future. 


\section{A Comparison of the Predictions from Monte Carlo Programs and 
Transverse Momentum Resummation}

\centerline{\it by C. Bal\'azs, J. Huston, I. Puljak, S. Mrenna} \vskip 2.mm

        
\subsection{Introduction}

        Monte Carlo programs including parton showering,
such as \pythia \cite{pythia}, 
\herwig \cite{herwig} and \isajet \cite{isajet},
are commonly used by experimentalists, both as a way of comparing
experimental data to theoretical predictions, and also as a means of 
simulating experimental signatures in kinematic regimes for which there
is no experimental data (such as that appropriate to the LHC). The final 
output of the Monte Carlo programs consists of the 4-vectors of a set of 
stable particles (e.g., $e,\mu,\pi,\gamma$); 
this output can either be compared to reconstructed experimental
quantities or, when coupled with a simulation of a detector response, 
can be directly compared to raw data taken by the experiment, and/or
passed through the same reconstruction procedures as the raw data.
In this way, the parton shower programs can be more useful to 
experimentalists than analytic calculations performed at high orders
in perturbation theory. 
Indeed, almost all of the 
physics plots in the ATLAS physics TDR~\cite{TDR} involve comparisons to 
\pythia (version 5.7). 

        Here, we are concerned with the predictions of parton shower Monte
Carlo programs and those from certain analytic calculations which resum
logarithms associated with the transverse momentum of partons initiating 
the hard scattering.
Most analytic calculations of this kind are either based on 
or originate from the formalism developed by J. Collins, D. Soper, and G. 
Sterman (CSS), which we choose as the analytic ``benchmark'' of this section.
Both the parton showering and analytic calculations
describe the effects of 
multiple soft gluon emission from the incoming partons,
which can have a profound effect on the kinematics of gauge or Higgs
bosons and their decay products produced in hadronic collisions.
This may have an impact on the 
signatures of physics processes at both the trigger and analysis 
levels, and thus
it is important to understand the reliability of such predictions. The best 
method for testing the reliability is a direct comparison of the predictions
to experimental data. If no experimental data is available,
then some understanding of the reliability may by gained by
simply comparing the predictions of different calculational methods.

\subsection{Parton Showering and Resummation}

Parton showering is the {\it backwards} evolution of an initial hard scattering
process, involving only a few partons at a high scale $Q^2_{\rm max}$ 
reflecting
large virtuality, into a complicated, multi-parton
configuration at a much lower scale $Q^2_{\rm min}$ typical of hadronic 
binding energies.
In practice, one does not calculate the probability of arriving at
a specific multi-parton configuration all at once.  Instead, the full shower
is constructed in steps, with evolution down in virtuality $Q^2$ with 
no parton
emission, followed by parton emission, and then a further evolution 
downward 
with no emission, etc., until the scale $Q^2_{\rm min}$ is reached.  
The essential
ingredient for this algorithm is the probability of evolving down in 
scale with no
parton emission or at least no resolvable parton emission.  This can 
be derived
from the DGLAP equation for the evolution of parton distribution 
functions.  One finds
that the probability of no emission $P$ equals $1-\exp(-S)$, where 
$S$ is the Sudakov
form factor, a function of virtuality and the momentum fraction $x$ 
carried by a parton.  

A key ingredient in the parton showering algorithm is the conservation 
of energy-momentum
at every step in the cascade.  The transverse momentum of the final 
system partly
depends on the opening angle between the mother and daughter partons 
in each emission.
Furthermore, after each emission, the entire multi-parton system
is boosted to the center-of-mass frame of the two virtual partons, until
at the end of the shower one is left with two primordial partons which are
on the mass shell and
essentially parallel with the incoming hadrons.  These boosts also 
influence the final transverse momentum.

Parton showering resums primarily the leading logarithms -- those resummed
by the DGLAP equations -- which are
universal, i.e. process independent, and depend only on the given 
initial state. 
In this lies one of the strengths of the parton shower approach, 
since it can be
incorporated into a wide variety of physical processes.
An analytic calculation, in comparison, can resum many other types of
potentially large logarithms, including process dependent ones. 
For example, the CSS formalism in principle sums all of the logarithms 
with $Q^2/p_T^2$ in their arguments,
where, for the example of Higgs boson production, 
$Q$ is the four momentum of the Higgs boson
and $p_T$ is its transverse momentum.
All of the ``dangerous logs'' are included in the 
Sudakov exponent, which can be written in impact parameter ($b$) space as:
\begin{eqnarray*}
{\cal S}(Q,b)=\int_{1/b^2}^{Q^2}\frac{d\overline{\mu }^2}{%
\overline{\mu }^2}\left[A\left(\alpha_s(\overline{\mu })\right) \ln
\left( \frac{Q^2}{\overline{\mu }^2}\right) {+B}\left(\alpha_s(\overline{%
\mu })\right) \right],
\end{eqnarray*}
with the $A$ and $B$ functions being free of 
large logarithms and calculable in fixed--order perturbation theory:
\begin{equation}
\begin{split}
A\left( \alpha_s({\bar{\mu}})\right) & =  \sum_{n=1}^\infty \left( 
\frac{\alpha_s({\bar{\mu}})}\pi \right) ^nA^{(n)} ,\\[1.mm]
B\left( \alpha_s({\bar{\mu}})\right) & =  \sum_{n=1}^\infty
\left( \frac{\alpha_s({\bar{\mu}})}\pi \right) ^nB^{(n)} .
\end{split}
\end{equation}

These functions contain an infinite number of coefficients, 
with the $A^{(n)}$ being universal to a given
initial state, while the $B^{(n)}$ are process dependent. 
In practice, the number of towers of logarithms included in the 
Sudakov exponent
depends on the level to which a fixed order calculation was performed for a 
given process. 
For example, if only a next-to-leading order calculation is available, 
only the coefficients $A^{(1)}$ and $B^{(1)}$ can be included.
If a NNLO calculation is available, then $A^{(2)}$ and $B^{(2)}$ can be 
extracted and incorporated into a resummation calculation, and so on. This is 
the case, for example, for $Z^0$ boson production. So far, only the $A^{(1)}$,
$A^{(2)}$ and $B^{(1)}$ coefficients are known for Higgs production, but the 
calculation of $B^{(2)}$ is in progress~\cite{carlschmidt}.
If we try to interpret parton showering in the same language, 
then we can say that the parton shower Sudakov exponent always 
contains a term analogous to $A^{(1)}$. 
It was shown in Reference~\cite{webber} that a suitable modification of the 
Altarelli-Parisi splitting function, or equivalently the strong 
coupling constant
$\alpha_s$, also effectively approximates the $A^{(2)}$ 
coefficient.\footnote{This is rigorously true only for the high parton $x$ or 
$\sqrt{\tau}$ region.}

In contrast with parton showering, analytic resummation 
calculations integrate over the kinematics of the soft gluon emission, with the
result that they are limited in their predictive power. 
While the parton shower maintains an exact treatment of the 
branching kinematics, the original CSS formalism imposes no kinematic penalty
for the emission of the soft gluons, although an approximate treatment of 
this can be incorporated into a numerical implementation, like 
ResBos~\cite{csabaref}.
Neither parton showering nor analytic resummation
reproduces kinematic configurations
where one hard parton is emitted at large $p_T$. In the parton shower, matrix 
element corrections can be imposed~\cite{gabriela,herwigw}, 
while, in the analytic resummation calculation, 
matching is necessary.

With the appropriate input from higher order cross sections,
a resummation calculation has the corresponding higher order normalization and
scale dependence. 
The normalization and
scale dependence for the Monte Carlo, though, remains that of a leading 
order calculation  -- though see Ref.~\cite{steve} and the related contribution
to these proceedings for an idea of how to include these at NLO.  
The parton showering occurs with unit probability after the 
hard scattering, so it does not change 
the total cross section.\footnote{Technically, 
one could add the branching for 
$q\to q$+Higgs in the shower, which would have the 
capability of increasing somewhat the Higgs cross section; however, the main 
contribution to the higher order $K$-factor comes from the virtual 
corrections and the `Higgs Bremsstrahlung' contribution is negligible.}

Given the above discussion, one quantity which should be well-described 
by both 
calculations is the shape of the transverse momentum ($p_T$) distribution
of the final state electroweak boson in a subprocess such
as $q\overline{q} \to WX$, $ZX$ or $gg \to H X$, where most of the 
$p_T$ is provided by initial state parton showering. The parton showering 
supplies the same sort of transverse kick as the soft gluon radiation
in a resummation calculation. Indeed, very similar 
Sudakov form factors appear in both approaches, with the caveats about
the $A^{(n)}$ and $B^{(n)}$ terms mentioned previously.

At a point in its evolution corresponding to a virtuality on the order of a 
few GeV, 
the parton shower is stopped and the effects of gluon emission at
softer scales must be parameterized and inserted by hand.  
Typically, a Gaussian probability distribution function is used to assign
an extra ``primordial'' $k_T$ to the primordial partons of the shower
(the ones which are put on the mass shell at the end of the backwards 
showering).  In {\tt PYTHIA}, the default is a constant value of $k_T$.
Similarly, there is a
somewhat arbitrary division between perturbative and non-perturbative 
regions in a resummation calculation.  
Sometimes the non-perturbative
effects are also parametrized by Gaussian distributions in $b$ or $Q_T$ space.
In general, the value for the non-perturbative
$\langle k_T \rangle$ needed in a Monte Carlo program will depend on 
the particular kinematics being investigated. 
In the case of the resummation calculation 
the non-perturbative physics is determined from fits to fixed 
target data and then automatically evolved to the kinematic regime of interest.

        A value  for the  average non-perturbative 
$k_T$ of greater than 1 GeV does not imply that there
is an anomalous intrinsic $k_T$ associated with the parton size; rather this
amount of $\langle k_T \rangle$ needs to be supplied to provide what 
is missing in the 
truncated parton shower. If the shower is cut off at a higher virtuality, more
of the ``non-perturbative'' $k_T$ will be needed. 

\subsection{$Z^0$ Boson Production at the Tevatron}

\begin{figure}[!th]
\centerline{\psfig{file=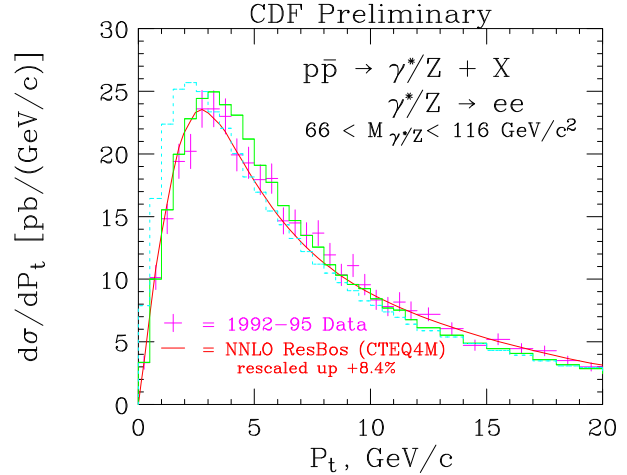,width=8cm}}
\caption{
The $Z^0$ $p_T$ distribution (at low $p_T$) from CDF for Run 1 compared to 
predictions from ResBos and from {\tt PYTHIA}. The two {\tt PYTHIA} 
predictions use the default (rms) value for the non-perturbative $k_T$ 
(0.44 GeV) and the value that gives the best agreement with the shape of 
the data (2.15 GeV). 
}
\label{fig:run1_ee_pt}
\end{figure}

The 4-vector of a $Z^0$ boson, and thus its transverse momentum, can be 
measured with great precision in the $e^+e^-$ decay mode. Resolution 
effects are relatively minor and are easily corrected for. Thus, the $Z^0$ 
$p_T$ distribution is a great testing ground for both the resummation and 
Monte Carlo formalisms for soft gluon emission. The corrected
$p_T$ distribution for $Z^0$ bosons in the low $p_T$ region for the CDF 
experiment\footnote{We thank Willis Sakumoto for providing the figures for 
$Z^0$ production as measured by CDF} is shown in 
Figure~\ref{fig:run1_ee_pt}, compared 
to both the resummed prediction from ResBos, and to two predictions from 
{\tt PYTHIA} (version 6.125). One {\tt PYTHIA} prediction uses the default 
(rms)\footnote{For a Gaussian distribution, $k_T^{rms}=1.13\langle k_T 
\rangle$.} value of  intrinsic $k_T$ of 0.44 GeV and the second a value 
of 2.15 GeV per incoming parton.\footnote{A previous 
publication~\cite{gabriela} indicated the need for a substantially larger 
non-perturbative $\langle k_T \rangle$, of the order of 4 GeV for the 
case of $W$ production at the Tevatron. The data used in the comparison, 
however, were not corrected for resolution smearing, a fairly large effect 
for the case of $W \to e{\nu}$ production and decay.} The latter 
value was found to give the best agreement between {\tt PYTHIA} and the
data.\footnote{A similar conclusion has been reached for comparisons of 
the CDF $Z^0$ $p_T$ data with \herwig.~\cite{corcella}} 
All of the predictions use the CTEQ4M parton distributions~\cite{cteq4}.
The shift between 
the two \pythia~predictions at low $p_T$ is clearly evident. As might have 
been expected, the high $p_T$ region (above 10 GeV) is unaffected by the 
value of the non-perturbative $k_T$. 
Note the $k_T$ imparted 
to the incoming partons at their lowest virtuality, $Q_0$, is greatly reduced
in its effect on the $Z^0$ $p_T$ distribution.
This dilution arises because the center-of-mass energy of the 
``primordial'' partons
is typically much larger than that of the original hard scattering.  
Therefore, the
transverse $\beta$ of the boost applied to the $Z^0$ boson to transform it 
to the frame where the ``primordial'' partons have transverse momentum $k_T$
is small.

As an exercise, one can transform the resummation formula in order to 
bring it to a form where the non-perturbative function acts as a Gaussian 
type smearing term. Using the Ladinsky-Yuan parameterization~\cite{LY} of 
the non-perturbative function in ResBos leads to an rms value for the 
effective $k_T$ smearing parameter, for $Z^0$ production at the Tevatron, 
of 2.5 GeV. This is similar to that needed for {\tt PYTHIA} and {\tt 
HERWIG} to describe the $Z^0$ production data at the Tevatron. 

In Figure~\ref{fig:run1_ee_pt}, the normalization of the resummed 
prediction has been rescaled upwards by 8.4\%. The {\tt PYTHIA} prediction 
was rescaled by a factor of 1.3-1.4 (remember that this is only a leading 
order comparison) for the shape comparison. 

As stated previously, the resummed prediction correctly describes the 
shape of the $Z^0$ $p_T$ distribution at low $p_T$, 
although there is still a noticeable difference in 
shape between the Monte Carlo and the resummed prediction. It is 
interesting to note that if the process dependent coefficients ($B^{(1)}$ and 
$B^{(2)}$) were not incorporated into the resummation prediction, the result 
would be an increase in the height of the peak and a decrease in the rate 
between 10 and 20 GeV, leading to a better agreement with the {\tt PYTHIA} 
prediction~\cite{csaba}.

The {\tt PYTHIA} and ResBos predictions both 
describe the data well over a wider $p_T$ range than shown in the figure.
Note especially the agreement of {\tt PYTHIA} with 
the data at high $p_T$, made possible by explicit matrix element 
corrections (from the subprocesses $q\overline{q} \to Z^0g$ and 
$gq \to Z^0q$) to the $Z^0$ production process.\footnote{Slightly 
different techniques are used for the matrix element corrections by {\tt 
PYTHIA}~\cite{gabriela} and by {\tt HERWIG}~\cite{herwigw}.  In {\tt 
PYTHIA}, the parton shower probability distribution is applied over the 
whole phase space and the exact matrix element corrections are applied 
only to the branching closest to the hard scatter.  In {\tt HERWIG}, the 
corrections are generated separately for the regions of phase space 
unpopulated by {\tt HERWIG} (the `dead zone') and the populated region. In 
the dead zone, the radiation is generated according to a distribution 
using the first order matrix element calculation, while the algorithm for 
the already populated region applies matrix element corrections whenever a 
branching is capable of being `the hardest so far'.}

\subsection{Diphoton Production}

\begin{figure}[!th]
\centerline{\psfig{file=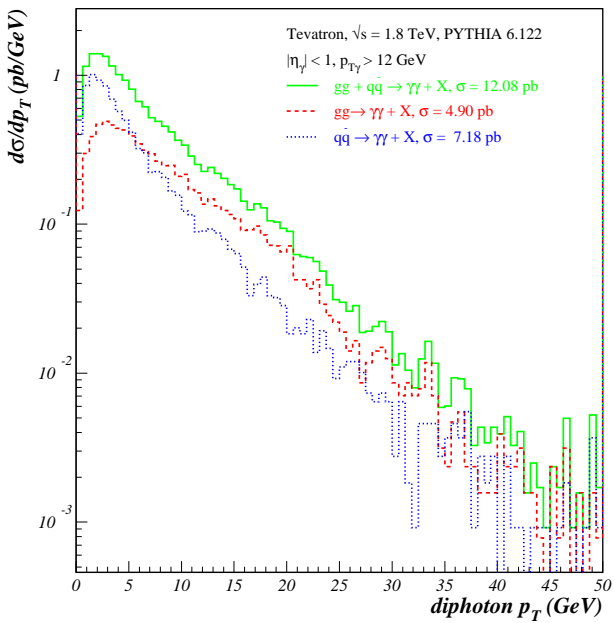,width=10cm}}
\caption{
A comparison of the {\tt PYTHIA} predictions for di-photon production 
at the 
Tevatron for the two different subprocesses, $q\overline{q}$ and $gg$. 
The same cuts are applied to {\tt PYTHIA} as in the CDF di-photon analysis.
} 
\label{fig:pythiakt}
\end{figure}

Most of the comparisons between resummation 
calculations/Monte Carlos and data have been performed for Drell-Yan 
production, i.e. 
$q\overline{q}$ initial states. It is also interesting to examine diphoton 
production at the Tevatron, where a large fraction of the contribution at 
low diphoton mass is due to $gg$ scattering. The prediction for the 
di-photon $k_T$ 
distribution at the Tevatron, from {\tt PYTHIA} (version 6.122), is shown 
in Figure~\ref{fig:pythiakt}, using the experimental cuts applied in the 
CDF analysis~\cite{cdfdiphot}. 

It is interesting to note that about half 
of the di-photon cross section at the Tevatron is due to the $gg$ 
subprocess, and that the di-photon $p_T$ distribution is noticeably broader 
for the $gg$ subprocess than the $q\overline{q}$ subprocess.
The $gg$ subprocess predictions in ResBos agree well 
with those from {\tt PYTHIA} 
while the $q\overline{q}$ $p_T$ distribution is noticeably broader in ResBos. 
The latter behavior is due to the presence of the $Y$ piece 
(fixed-order corrections)
in ResBos at moderate $p_T$, and the matching of the  $q\overline{q}$ cross 
section to the fixed order $q\overline{q} \to {\gamma}{\gamma}g$ at high 
$p_T$. The corresponding matrix element correction is not in {\tt PYTHIA}. 
It is interesting to note that the {\tt PYTHIA} and ResBos predictions for $gg 
\to {\gamma}{\gamma}$ agree in the moderate $p_T$ region, even though the 
ResBos prediction has the $Y$ piece present and is matched to the matrix 
element piece $gg \to {\gamma}{\gamma}g$ at high $p_T$, while there is no 
such matrix element correction for {\tt PYTHIA}. This shows that 
the $Y$ piece correction is not important
for the $gg$ subprocess, which is the same conclusion that was 
reached in Ref.~\cite{BalazsNadolskySchmidtYuan}.
This is probably a result of steep decline in the $gg$ parton-parton 
with increasing partonic center of mass energy, 
$\sqrt{\hat{s}}$. This falloff 
tends to suppress the size of the $Y$ piece since the production of the 
di-photon pair at higher $p_T$ requires larger $x_1$, $x_2$ values. In the 
default CSS formalism, there is no such kinematic penalty in the resummed 
piece since the soft gluon radiation comes for ``free.'' (Larger $x_1$  
and $x_2$ values are not required.)

A comparison of the CDF di-photon data to NLO \cite{owens} and resummed 
(ResBos) QCD predictions has been performed, but the analysis is still in
progress, so the results are not presented here.
The transverse 
momentum distribution, in particular, is sensitive to the effects of the 
soft gluon radiation and better agreement can be observed with the ResBos 
prediction than with the NLO one.  A much more precise comparison with 
the effects of soft gluon radiation will be possible with the 2 fb$^{-1}$ 
or greater data sample that is expected for both CDF and D\O~in Run 2.

\subsection{Higgs Boson Production}

\begin{figure}[!th]
\centerline{\psfig{file=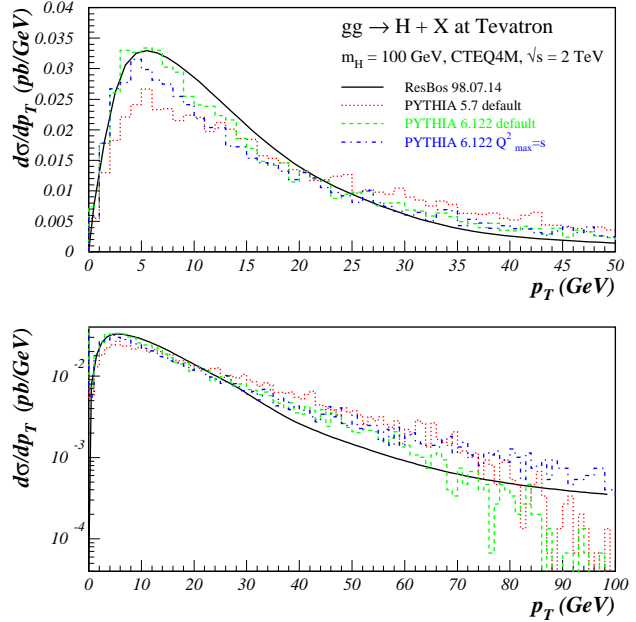,width=10cm}}
\caption{
A comparison of predictions for the Higgs $p_T$ distribution at the 
Tevatron from ResBos and from two recent versions of {\tt PYTHIA}. The 
ResBos and {\tt PYTHIA} predictions have been normalized to the same area. 
} 
\label{fig:resbos_pythia_higgs_tev}
\end{figure}

A comparison of the two versions of {\tt PYTHIA} and of ResBos is 
shown in Figure~\ref{fig:resbos_pythia_higgs_tev} for the case of the
production of a Higgs boson with mass 100 GeV
at the Tevatron with center-of-mass 
energy of 2.0 TeV. The same qualitative  features are observed at 
the LHC: the newer version of {\tt PYTHIA} agrees better with ResBos in 
describing the low $p_T$ shape, and there is a falloff at high $p_T$ 
unless the hard scale for showering is increased.  The
default (rms) value of the non-perturbative $k_T$ (0.44 GeV)  was used for 
the {\tt PYTHIA} predictions.
Note that the peak of the resummed distribution has moved to $p_T \approx$ 
7 GeV (compared to about 3 GeV for $Z^0$ production at the Tevatron). 
This is due primarily to the larger color factors associated with 
initial state gluons ($C_A = 3$) rather than quarks ($C_F = 4/3$).

The newer version of {\tt PYTHIA} agrees well with ResBos at low to 
moderate $p_T$, but falls below the resummed prediction at high $p_T$. 
This is easily understood: ResBos switches to the NLO Higgs + jet matrix 
element at high $p_T$ while the default {\tt PYTHIA} can generate the 
Higgs $p_T$ distribution only by initial state gluon radiation, using as 
default a maximum scale equal to the Higgs boson mass.
High $p_T$ Higgs boson production is 
another example where a $2 \to 1$ Monte Carlo calculation with parton 
showering can not completely reproduce the exact matrix element 
calculation without the use of matrix element corrections. The high $p_T$ 
region is better reproduced if the maximum virtuality $Q_{max}^2$ is set 
equal to the collider center-of-mass energy, $s$, rather than 
subprocess $\hat{s}$. 
This is equivalent to applying the parton shower to all of phase 
space. However, the consequence is that the low $p_T$ region is
now depleted of events, since the parton showering does not change
the total production cross section.
The 
appropriate scale to use in {\tt PYTHIA} (or any Monte Carlo) depends on 
the $p_T$ range to be probed.  If matrix element information is used to 
constrain the behavior, the correct high $p_T$ cross section can be 
obtained while still using the lower scale for showering. The 
incorporation of matrix element corrections to Higgs production (involving 
the processes $gq \to qH$,$q{\overline{q}} \to gH$, $gg \to gH$) is the 
next logical project for the Monte Carlo experts, in order to accurately 
describe the high $p_T$ region.

The older version of {\tt PYTHIA} produces too many Higgs events at 
moderate $p_T$ (in comparison to ResBos) at both the Tevatron and the LHC. 
Two changes have been implemented in the newer version. 
The first change 
is that a cut is placed on the combination of $z$ and $Q^2$ values in a 
branching: $\hat{u} = Q^2/z-\hat{s}(1-z) < 0$, where $\hat{s}$ refers to the 
subsystem of the hard scattering plus  the shower partons considered to 
that point.  The association with $\hat{u}$ is relevant if the branching 
is interpreted in terms of a $2 \to 2$ hard scattering. 
This requirement is not fulfilled when the $Q^2$ value 
of the space-like emitting parton is little changed and the $z$ value of 
the branching is close to unity. This affects mainly the hardest 
emission (largest $Q^2$). The net result of this requirement is a 
substantial reduction in the total amount of gluon radiation 
\cite{pythiaman}.  Such branchings are kinematically allowed, but 
since matrix element corrections would assume initial state partons to 
have $Q^2=0$, a non-physical $\hat{u}$ results (and thus  no possibility 
to impose matrix element corrections). The correct behavior is beyond the 
predictive power of leading log Monte Carlos. 

In the second change, the 
parameter for the minimum gluon energy emitted in space-like showers is 
modified by an extra factor roughly corresponding to the $1/\gamma$ factor 
for the boost to the hard subprocess frame~\cite{pythiaman}. The effect of 
this change is to increase the amount of gluon radiation. Thus, the two 
effects are in opposite directions but with the first effect being 
dominant. 

This difference in the $p_T$ distribution  between the two versions of 
{\tt PYTHIA} could have an impact on the analysis strategies for Higgs 
searches at the LHC. For example, for the CMS detector, the higher $p_T$ 
activity associated with Higgs production in version 5.7 would have 
allowed for a more precise determination of the event vertex from which 
the Higgs (decaying into two photons) originated. Vertex pointing with the 
photons is not possible in CMS, and the large number of interactions 
occurring with high intensity running will mean a substantial probability 
that  at least one of the interactions will produce jets at low to 
moderate $E_T$.~\cite{denegri}
In principle, this problem could affect the $p_T$ distribution for all 
{\tt PYTHIA} processes. In practice, the effect has manifested itself
only in $gg$ initial states, 
due to the enhanced branching probability.

As an exercise, an 80 GeV $W$ and an 80 GeV Higgs were generated at the 
Tevatron using {\tt PYTHIA}5.7~\cite{mrennarun2}. A comparison of the 
distribution of values of $\hat{u}$ and the virtuality $Q$ for the two
processes indicates a greater tendency for the Higgs virtuality to be near
the maximum  value and for there to be a larger number of Higgs events
with positive $\hat{u}$ (than W events).

\subsection{Comparison with {\tt HERWIG}}

The variation between versions 5.7 and 6.1 of {\tt PYTHIA} gives an 
indication of the uncertainties due to the types of choices that can be 
made in Monte Carlos. The requirement that $\hat{u}$ be negative for all 
branchings is a choice rather than an absolute requirement.  Perhaps the 
better agreement of version 6.1 with ResBos is an indication that the 
adoption of the $\hat{u}$ restrictions was correct. Of course, there may 
be other changes to {\tt PYTHIA} which would also lead to better agreement 
with ResBos for this variable. 

\begin{figure}[!th]
\centerline{\psfig{file=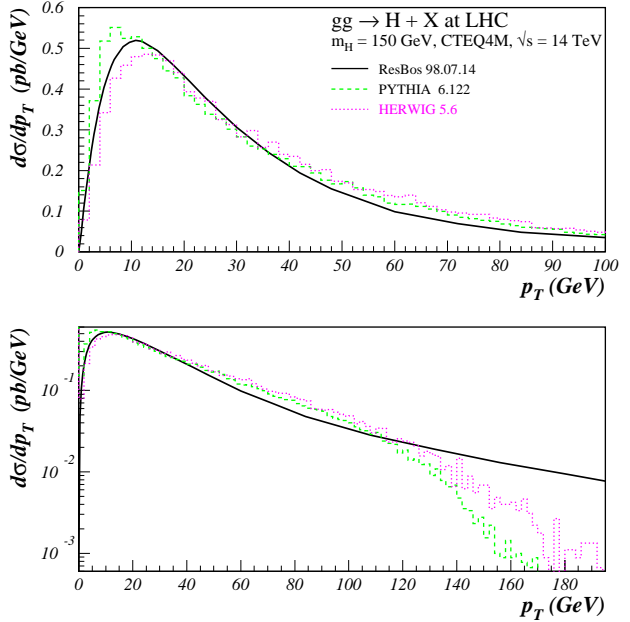,width=10cm}}
\caption{
A comparison of predictions for the Higgs $p_T$ distribution at 
the LHC from 
ResBos, two recent versions of {\tt PYTHIA} and {\tt HERWIG}. The 
ResBos, {\tt PYTHIA} and {\tt HERWIG}
 predictions have been normalized to the same area.
} 
\label{fig:comparison_lhc}
\end{figure}
Since there are a variety of choices that can be made in Monte Carlo 
implementations, it is instructive to compare the predictions for the 
$p_T$ distribution for Higgs boson production from ResBos and 
{\tt PYTHIA} with 
that from {\tt HERWIG} (version 5.6, also using the CTEQ4M parton 
distribution functions). The {\tt HERWIG} prediction is shown in 
Figure~\ref{fig:comparison_lhc} along with the {\tt PYTHIA} and ResBos 
predictions, all normalized to the ResBos prediction.~%
\footnote{The normalization factors (ResBos/Monte Carlo) are {\tt 
   PYTHIA} (both versions)(1.61) and {\tt HERWIG} (1.76).} 
In all cases, the CTEQ4M parton distribution was used. The predictions 
from {\tt HERWIG} and {\tt PYTHIA} 6.1 are very similar, with the {\tt 
HERWIG} prediction matching the ResBos shape somewhat better at low $p_T$.

\subsection{Non-perturbative $k_T$}

\begin{figure}[!th]
\centerline{\psfig{file=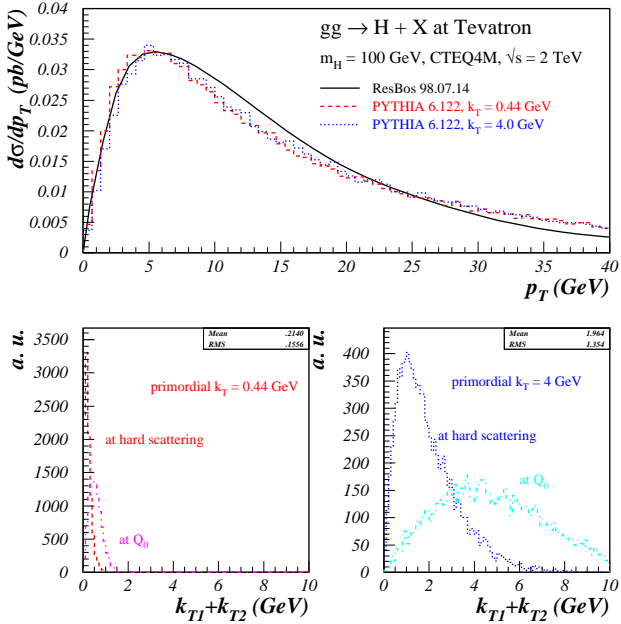,width=10cm}}
\caption{
(top) A comparison of the {\tt PYTHIA} predictions for the $p_T$ 
distribution of a 100 GeV Higgs at the Tevatron using the default (rms) 
non-perturbative $k_T$ (0.44 GeV) and a larger value (4 GeV), at the 
initial scale $Q_0$ and at the hard scatter scale. 
Also shown is the ResBos prediction 
(bottom) The vector sum of the intrinsic $k_T$ ($k_{T1}$+$k_{T2}$) for the 
two initial state partons at the initial scale $Q_0$ and at the hard 
scattering scale for the two values of intrinsic $k_T$.
}
\label{fig:kt_higgs_tev}
\end{figure}

A question still remains as to the appropriate value of non-perturbative 
$k_T$ to input in the Monte Carlos to achieve a better agreement in shape, 
both at the Tevatron and at the LHC. Figure~\ref{fig:kt_higgs_tev} 
compares the ResBos and {\tt 
PYTHIA} predictions for the Higgs boson $p_T$ distribution at the Tevatron.
The {\tt PYTHIA} prediction (now version 6.1 alone) is shown with 
several values of non-perturbative $k_T$. Surprisingly, no difference is 
observed between the predictions with the  different values of $k_T$, with 
the peak in {\tt PYTHIA} always being somewhat below that of ResBos. This 
insensitivity can be understood from the plots at the bottom of the two 
figures which show the sum of the non-perturbative initial state $k_T$ 
($k_{T1}$+$k_{T2}$) at $Q_0$ and at the hard scatter scale $Q$. Most of 
the $k_T$ is radiated away, with this effect being larger (as expected) 
at the LHC. The large gluon radiation probability from a gluon-gluon 
initial state (and the greater phase space available at the LHC) lead to a 
stronger degradation of the non-perturbative $k_T$ than was observed with 
$Z^0$ production at the Tevatron.

\subsection{Conclusions}

An understanding of the signature for Higgs boson production at either the 
Tevatron or LHC depends upon the understanding of the details of soft 
gluon emission from the initial state partons.  This soft gluon emission 
can be modeled either in a Monte Carlo or in a $k_T$ resummation program, 
with various choices possible in both implementations.  A comparison of the 
two approaches is useful to understand the strengths and weaknesses of 
each. The data from the Tevatron that either exists now, or will exist in 
Run 2, will be extremely useful to test both approaches. 

{\bf Acknowledgements}

We would like to thank Claude Charlot, 
Gennaro Corcella, Willis Sakumoto, Torbjorn Sj\"ostrand and  
Valeria Tano for useful conversations and for providing some
of the plots.


\section{MCFM: a parton-level Monte Carlo at NLO Accuracy}

\def\TeV{\mbox{TeV}}
\def\GeV{\mbox{GeV}}

\centerline{\it by John Campbell and R.K. Ellis }\vskip 2.mm

\subsection{Introduction}
\label{sec:mcfm}

In Run II, experiments at the Tevatron will be sensitive to processes
occurring at the femtobarn level. Of particular interest are processes
which involve heavy quarks, leptons and missing energy, since so many 
of the signatures for physics beyond the standard model produce 
events containing these features. We have therefore written the program 
MCFM \cite{EV,16} which calculates the rates for a number of standard 
model processes.
These processes are included beyond the leading order 
in the strong coupling constant where possible; in QCD 
this is the first order in which the normalization of the cross sections 
is determined. Because the 
program produces weighted Monte Carlo events, we can implement experimental
cuts allowing realistic estimates of event numbers for an ideal 
detector configuration. MCFM is expected to give more reliable results 
than parton shower Monte Carlo programs, especially in phase space regions 
with well separated jets. On the other hand it gives little information 
about the phase space regions which are dominated by multiple parton emission.
In addition, because the final state contains partons rather than hadrons, 
a full detector simulation cannot be performed directly using the output 
of MCFM.

The processes already included in MCFM at NLO are as follows 
($H_1,H_2=p$ or $ \bar{p}$),
\begin{itemize}
\item   $H_1 + H_2 \rightarrow W^\pm$
\item   $H_1 + H_2 \rightarrow Z$
\item   $H_1 + H_2 \rightarrow W^\pm + \mbox{1 jet}$
\item   $H_1 + H_2 \rightarrow Z + \mbox{1 jet}$
\item   $H_1 + H_2 \rightarrow W^\pm + H$
\item   $H_1 + H_2 \rightarrow Z + H$
\item   $H_1 + H_2 \rightarrow W^+ W^-$
\item   $H_1 + H_2 \rightarrow W^\pm Z$
\item   $H_1 + H_2 \rightarrow Z Z$
\item   $H_1 + H_2 \rightarrow W^+ + g^*(\rightarrow b \bar{b})
,~\mbox{massless}~$b$\mbox{-quarks}$
\item   $H_1 + H_2 \rightarrow Z + g^*(\rightarrow b \bar{b})
,~\mbox{massless}~$b$\mbox{-quarks}$
\item   $H_1 + H_2 \rightarrow H \rightarrow W^+ W^-, ZZ~\mbox{or}~t \bar{t} $
\item   $H_1 + H_2 \rightarrow \tau^+ + \tau^- \; . $
\end{itemize} 
The decays of vector bosons and/or Higgs bosons are included.
We have also included the leptonic decays of the $\tau$-lepton.
As described below the implementation of NLO corrections requires 
the calculation of both the amplitude for real radiation and 
the virtual corrections to the Born level process. We have extensively used 
the one loop results of Bern, Dixon, Kosower \etal \cite{BDKW}, 
\cite{dks} to obtain the virtual corrections to above processes.

A future development path for the program would be to include the following 
processes at NLO:
\begin{itemize}
\item   $H_1 + H_2 \rightarrow W^\pm + \mbox{2 jets}$
\item   $H_1 + H_2 \rightarrow Z + \mbox{2 jets} \; . $
\end{itemize}

In addition there are an number of processes which we have 
included only at leading order. This restriction to leading order 
is both a matter of expediency and because the theoretical framework 
for including radiative corrections to processes involving massive 
particles is not yet complete.
\begin{itemize}
\item   $H_1 + H_2 \rightarrow t + \bar{t}$
\item   $H_1 + H_2 \rightarrow t + \bar{t} + \mbox{1 jet}$
\item   $H_1 + H_2 \rightarrow t + \bar{b}$
\item   $H_1 + H_2 \rightarrow t + \bar{b} + \mbox{1 jet}$
\item   $H_1 + H_2 \rightarrow t + \bar{t} + H$
\item   $H_1 + H_2 \rightarrow t + \bar{t} + Z$
\end{itemize} 
$H,Z$ and top quark decays are included.

\subsection{General structure}
In order to evaluate the strong radiative corrections to a given
process, we have to consider Feynman diagrams describing real
radiation, as well as the diagrams involving virtual corrections to the tree
level graphs.  The corrections due to real radiation are dealt with
using a subtraction algorithm\cite{ERT} as formulated by Catani and
Seymour \cite{CS}.  This algorithm is based on the fact that the
singular parts of the QCD matrix elements for real emission can be
singled out in a process-independent manner.  By exploiting this
observation, one can construct a set of counter-terms that cancel all
non-integrable singularities appearing in real matrix elements. The
NLO phase space integration can then be performed numerically in four
dimensions.

The counter-terms that were subtracted from the real matrix elements
have to be added back and integrated analytically over the phase space
of the extra emitted parton in $n$ dimensions,
leading to poles in $\epsilon=(n-4)/2$.
After combining those poles with the ones coming from the virtual
graphs, all divergences cancel, so that one can safely perform the limit
$\epsilon \rightarrow 0$ and carry out the remaining phase space
integration numerically.

\begin{figure}[!ht]
\centerline{\psfig{file=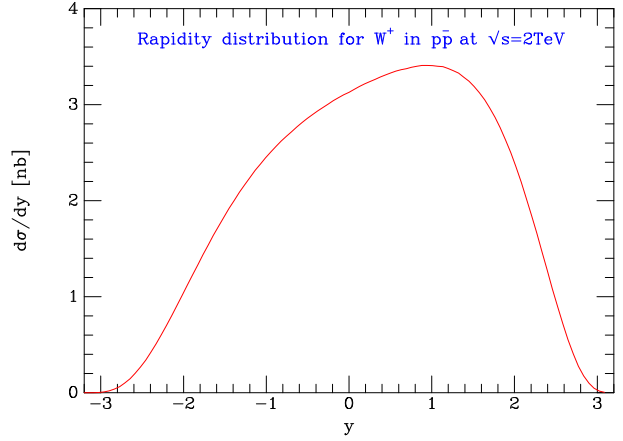,width=8cm,angle=-90}}
    \caption{The rapidity distribution for $W^+$ production in 
$p\bar{p}$ collisions at $\sqrt{s}=2\TeV$.}
    \label{fig:ellis1}
\end{figure}

As an example of this procedure we consider the production of an 
on-shell $W$ boson decaying to a lepton-antilepton pair.
\begin{multline} \label{DYLO}
q(p_1)+\bar{q}(p_2) \rightarrow W^+(\nu(p_3)+e^+(p_4)),\\[1.mm]
p_1+p_2=p_3+p_4, \;\; (p_3+p_4)^2=M_W^2.
\end{multline}
In this  case, the $W$ boson rapidity distribution 
is calculable analytically in $O(\alpha_s)$ \cite{AEM,kubar}. 
Fig.~\ref{fig:ellis1} shows the result 
calculated in the $\overline{MS}$ scheme.
The virtual corrections to (\ref{DYLO}) are of the Drell-Yan type and
are well known \cite{AEM}.
They are expressible as an overall factor multiplying the lowest order matrix
element squared,
\begin{multline}
\sigma^{\rm V} = \sigma^{\rm LO} \times \frac{\alpha_s C_F}{2\pi}
\left(\frac{4\pi\mu^2}{Q^2}\right)^\epsilon \frac{1}{\Gamma(1-\epsilon)}
\\[1.mm]
\left[ -\frac{2}{\epsilon^2}-\frac{3}{\epsilon}-6+\pi^2 \right]
\label{eq:dyvirt}
\end{multline}
and must be combined with the real radiation contribution.
For example, gluon radiation from the $q \bar{q}$ initial state
yields the subprocess
\begin{multline}
q(p_1)+\bar{q}(p_2) \rightarrow W(\nu(p_3)+e^+(p_4))+g(p_5), \\[1.mm]
p_1+p_2=p_3+p_4+p_5.
\end{multline}
To eliminate the singular part of this subprocess, 
we generate a counter event with the kinematics of the
$2 \to 2$ process as follows
\begin{multline}
q(x_a p_1)+\bar{q}(p_2) \rightarrow W(\nu(\tilde{p}_3)+e^+ (\tilde{p}_4)),
\\[1.mm]
x_a p_1+p_2=\tilde{p}_3+\tilde{p}_4
\end{multline}
where a Lorentz transformation has been performed on all $j$
final state momenta
\begin{equation}
\tilde{p}^\mu_j = \Lambda^\mu_\nu p_j^\nu,\;\; j=3,4
\end{equation}
such that $\tilde{p}_j^\mu \rightarrow p_j$ for $p_5$ collinear or soft.
Thus the energy of the emitted gluon $p_5$ is absorbed by $p_1$, and 
the momentum components are absorbed by the transformation of the final 
state vectors.  The phase space has a convolution structure,
\begin{multline}
d \Phi^{(3)}(p_5,p_4,p_3;p_2,p_1) = \\[1.mm]
\int_0^1 dx \,
d \Phi^{(2)}(\tilde{p}_4,\tilde{p}_3;p_2,p_1) \times 
[dp_5(p_1,p_2,x)]
\end{multline}
where
\begin{multline}
[dp_5(p_1,p_2,x)] = \\[1.mm]
\frac{d^dp_5}{(2 \pi)^{d-1}}\,\delta^+(p_5^2) 
\Theta(x) \Theta(1-x) \delta(x-x_a)
\end{multline}
This phase space may be used to integrate out the dipole term $D^{15,2}$,
which is chosen to reproduce the singularities in the real matrix
elements as the gluon ($5$) becomes soft or collinear to the quark ($1$),
\begin{equation}
D^{15,2} = \frac{4\pi\alpha_s C_F \mu^{2\epsilon}}{p_1 \cdot p_5}
 \left(\frac{2}{1-x_a}-1-x_a \right)
\end{equation}
Performing the integration yields,
\begin{eqnarray}
&& \int_0^1 dx \, D^{15,2} \, [dp_5(p_1,p_2,x)] = \nonumber \\
&&\frac{\alpha_s C_F}{2\pi} \left(\frac{4\pi\mu^2}{2p_1 
\cdot p_2}\right)^\epsilon
\frac{1}{\Gamma(1-\epsilon)} \times \nonumber \\
&&\left[
 -\frac{1}{\epsilon} p_{qq}(x) + \delta(1-x) \left(
 \frac{1}{\epsilon^2}+\frac{3}{2\epsilon}-\frac{\pi^2}{6} \right)\right . 
\nonumber \\
&& \left . +2(1+x^2)\left[ \frac{\log(1-x)}{1-x} \right]_+
 \right]
\end{eqnarray}
with the Altarelli-Parisi function $p_{qq}(x)$ given by
\begin{equation}
p_{qq}(x)= \frac{2}{(1-x)_+}-1-x+\frac{3}{2} \delta(1-x)
\end{equation}
In order to obtain the complete counter-term, one must add the
(identical) contribution from the dipole
configuration $D^{25,1}$ that accounts for the gluon becoming collinear
with the anti-quark. In a more complicated process, we would sum over
a larger number of distinct dipole terms involving partons both in the
initial and final states. In this simple case, we find the total
counter-term contribution to the $q {\bar q}$ cross-section to be
\begin{eqnarray*}
&&\sigma^{\rm CT} = \frac{\alpha_s C_F}{2\pi}
\left(\frac{4\pi\mu^2}{Q^2}\right)^\epsilon
\frac{1}{\Gamma(1-\epsilon)} \times \Biggl[ \Biggr. \\
&& \Biggl.  -\frac{2}{\epsilon} p_{qq}(x) + \delta(1-x) \left(
 \frac{2}{\epsilon^2}+\frac{3}{\epsilon}-\frac{\pi^2}{3} \right) \\
&& -2p_{qq}(x)\log x + 4(1+x^2)\left[ \frac{\log(1-x)}{1-x} \right]_+
 \Biggr]
\end{eqnarray*}
where each of these terms leads to a different type of contribution
in MCFM. The first term, proportional to $p_{qq}(x)$,
is canceled by mass factorization, up to some additional
finite (${\cal O}(\epsilon^0)$) pieces. The terms multiplying the
delta-function $\delta(1-x)$ manifestly cancel the poles generated by
the virtual graphs, given in equation~(\ref{eq:dyvirt}),
leaving an additional $\pi^2$ contribution. The
remaining terms, which don't have the structure of the virtual
contribution, are collected together and added separately in MCFM.

\begin{figure}[!ht]
\centerline{\psfig{file=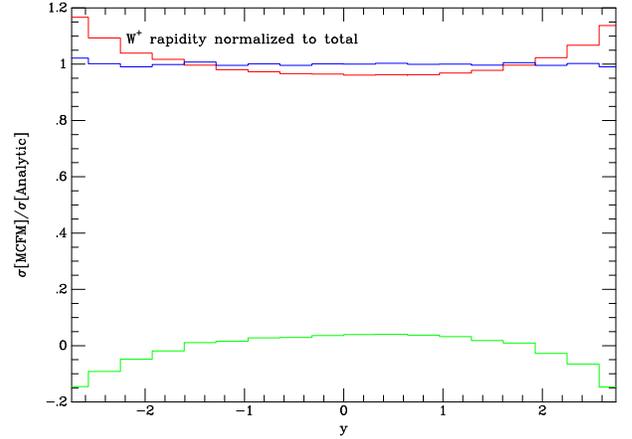,width=8cm,angle=-90}}
\caption{The ratio of the contributions to the 
rapidity distribution of $W^+$ production}
    \label{fig:ellis2}
\end{figure}

In Fig.~\ref{fig:ellis2} we have plotted the three contributions 
to the $W$ rapidity calculated using MCFM. The three contributions 
are $a$) the contribution of (real-counterterm)
[the lower curve], $b$) the contribution 
of leading order + virtual + integrated counter-term 
[the upper-most curve] and  $c$) the total 
contribution. All three terms have been normalized to the $O(\alpha_s)$
rapidity distribution shown in Fig.~\ref{fig:ellis1}.
We see that ($b$), the leading order term, combined with 
the virtual correction and the results from the counterterm provides
the largest contribution to the cross section. The total contribution is 
a horizontal line at unity, showing the agreement between MCFM and the 
analytically calculated result. 
Only at the boundaries of the phase 
space at large $y$ can the contribution of the real emission minus the 
counterterm become sizeable. 

\subsection{Examples of MCFM results}
We first detail the input parameters used in our phenomenological estimates.
The electroweak theory is specified by four numbers, $M_W,M_Z,\alpha(M_Z)$,
and $G_F$, the values of which are given in Table~\ref{param},
together with other necessary constants. Other derived parameters are
$e,g_W$ and $\sin^2 \theta_W$ which, when defined
as below, are effective parameters including the leading effects of 
top quark loops\cite{Georgi}.  We use the the first of the MRS99 parton 
distributions\cite{mrs99} which has $\alpha_S(M_Z)=0.1175$.
\begin{eqnarray}
e^2 &=& 4 \pi \alpha(M_Z) \nonumber \\ \nonumber
g^2 _W &=& 4 \sqrt{2} G_F M_W^2 \\
\sin^2 \theta_W &=&\frac{e^2}{g^2_W}
\end{eqnarray}   
\begin{table}[t]
\caption{Input parameters}
\label{param}
\begin{center}
\begin{tabular}{|c|c|}
\hline
$M_Z,\Gamma_Z$   & $91.187,2.49~\GeV$\\
$M_W,\Gamma_W$ & $80.41,2.06~\GeV$ \\
$m_t,\Gamma_t$ & $175,1.4~\GeV$ \\
$ \alpha(M_Z)  $ &  1/128.89 \\
$ G_F  $& $  1.16639\times 10^{-5}$\\
$ \sin^2 \theta_W $  & 0.228534483\\
$V_{ud}$&0.97500 \\
$V_{us}$&0.22220 \\
$V_{cd}$&0.22220 \\
$V_{cs}$&0.97500\\
\hline
Higgs mass (GeV) & BR($H\to b\bar b$) \\ \hline
$100$ & $0.8119$\\
$110$ & $0.7697$\\
$120$ & $0.6778$\\
$130$ & $0.5254$\\
\hline
\end{tabular}
\end{center}
\end{table}
\begin{table}[th]
\caption{Diboson cross sections (in pb) at the Tevatron and the LHC}
\label{ellis:diboson}
\begin{center}
{\footnotesize
\begin{tabular}{|c|c|c|c|c|}
\hline
$\sqrt{s}$   & $ \sigma(W^+ W^-)  $ & $ \sigma(W^+ Z) $ 
 & $  \sigma(W^- Z) $ & $ \sigma(ZZ) $  \\
\hline
\hline
$2~\TeV~(p\bar{p}) $   & $ 12.2 $     & \multicolumn{2}{c|}{$ 2.02 $} & $
1.75 $ \\
\hline
$ 14~\TeV~(pp) $         & $ 103.6 $     & $ 27.2 $   & $ 17.7$ &$16.7$ \\
\hline
\end{tabular}
}
\end{center}
\end{table}
Table \ref{ellis:diboson} shows the production cross sections for di-boson
production calculated using MCFM for $p \bar{p}$ collisions at 
$\sqrt{s}=2$~TeV and for $p p$ collisions at 
$\sqrt{s}=14$~TeV. The next-to-leading order corrections vary between
approximately $30\%$ and $50\%$ of leading order and are almost
entirely due to the virtual graphs.
The numbers here are slightly different than the results
in \cite{16}, because  of the different 
choices made both for the input EW parameters and parton distributions 
as detailed above. 

Much effort has been devoted to the study of Higgs production at the
Tevatron at $\sqrt{s}=2$~TeV. These studies indicate that, given
enough luminosity, a light Higgs boson can be discovered at the
Tevatron using the associated production channels $WH$ and $ZH$.  In
this report we present results of an analysis that incorporates as many
of the backgrounds as possible at next-to-leading order for the $WH$
channel. Whilst we use no detector simulation and do not attempt to
include non-physics backgrounds, the results presented here can
provide a normalization for more detailed studies. This is of
importance since more detailed studies are often performed using
shower Monte Carlo programs which can give misleading results for well
separated jets.

\begin{figure}[!ht]
\centerline{\psfig{file=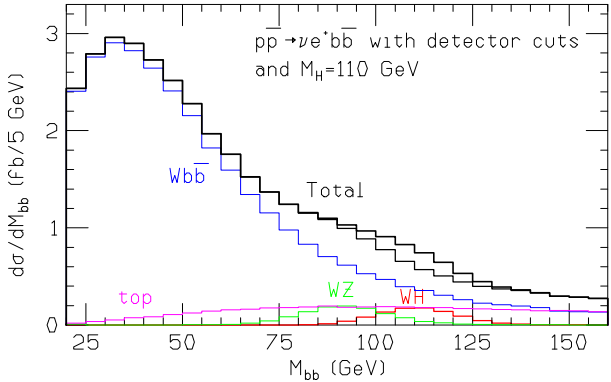,width=8cm,angle=90}}
    \caption{Signal and backgrounds for WH. `top' represents the
    sum of all the backgrounds including a top quark.}
    \label{fig:mbb}
\end{figure}

In particular, we will consider the light Higgs case ($M_H < 130$~GeV)
in the channel $p \bar{p} \rightarrow b \bar{b} \nu e^+$.
In addition to the usual cuts on rapidity and transverse momentum,
\begin{equation}
\begin{array}{rcl}
|y_b|, |y_{\bar{b}}| &<& 2\ , \\
|y_e| &<& 2.5\ , \\
|p^T_b|, |p^T_{\bar{b}}| &>& 15~\GeV \ ,\\
|p^T_e|, |p^T_{\nu}| &>& 20~\GeV \ ,
\end{array}
\label{cuts1}
\end{equation}
we also impose isolation cuts,
\begin{equation}
R_{b\bar{b}},R_{eb},R_{e\bar{b}}>0.7\ ,
\label{cuts2}
\end{equation}
as well as a cut on the scattering angle of the $b \bar{b}$
system \cite{KKY} (the Higgs scattering angle)
in the Collins-Soper frame \cite{CSFrame},
\begin{equation}
|\cos{\theta_{b\bar{b}}}|< 0.8\ .
\label{cuts3}
\end{equation}
Note that imposing the cut on $\cos{\theta_{b\bar{b}}}$
requires knowledge of the
longitudinal component of a neutrino momentum.
Our results for the signal, backgrounds and significance are shown
in Table~\ref{table_w_2}, where we use $\epsilon_{b {\bar b}}=0.45$ and
integrate the cross-sections over a $b{\bar b}$ mass range appropriate
for the Higgs mass under consideration,
\begin{equation}
|M_{H}-M_{b \bar{b}} | < \sqrt{2} \sigma_M, \sigma_M = .1 M_H.
\end{equation}
\begin{table*}
\caption{Signal, backgrounds (in fb) and significance for the
$W$-channel at $\sqrt{s}=2~\TeV$}
\label{table_w_2}
\begin{center}
\begin{tabular}{|l|c|c|c|c|c|}
\hline
$M_H$~[GeV] & Scale       & 100  &110  & 120 & 130     \\
\hline
$W^{\pm} H(\rightarrow b \bar{b})$ & $m_H$ & $8.8$ & $6.4$ &
$4.2$ & $2.5$   \\
\hline
\hline
$W^{\pm} g^*(\rightarrow b \bar{b})$
 &$(m_W+100 ~\GeV)/2$ & $25.7$ & $22.7$ & $18.5$ & $15.5$   \\
$W^{\pm} Z(\rightarrow b \bar{b})$
 &$(m_W+100 ~\GeV)/2$ & $6.7$ & $4.3$ & $2.0$ & $1.0$   \\
$t(\rightarrow bW^+) \bar{t}(\rightarrow \bar{b}W^-_{\rm lept})$ & 
$100 ~\GeV$ & 
 $3.3$ & $3.7$ & $3.9$ & $3.9$   \\
$t(\rightarrow bW^+) \bar{t}(\rightarrow \bar{b}W^-_{\rm hadr})$ & 
$100 ~\GeV$ & 
 $0.3$ & $0.4$ & $0.5$ & $0.6$   \\
$W^{\pm *} (t(\rightarrow bW^+) \bar{b})$
 & $100 ~\GeV$ & $5.1$ & $5.8$ & $6.0$ &$6.0$   \\
$q^{\prime} t(\rightarrow bW^+)$ & $100 ~\GeV$ & $0.3$ & $0.4$ & $0.5$ 
&$0.6$   \\
Total $B$ & - & $41.4$ & $37.3$ & $31.4$ & $27.6$ \\
\hline
$S/B $ & - & $0.21$ & $0.17$ & $0.13$ & $0.09$ \\
\hline
$S/\sqrt{B}$ & - & $1.37$ & $1.05$ & $0.75$ & $0.48$ \\
\hline
\end{tabular}
\end{center}
\end{table*}
From this table, one can see that, even with a fairly restrictive set
of cuts, the $Wg^*$ process in particular provides a challenging
background. This is further emphasized in Figure~\ref{fig:mbb}, where
the cross-sections for $M_H=110$~GeV are presented in $5$~GeV bins
across the entire $m_{b {\bar b}}$ spectrum. The signal, the two largest
backgrounds and the sum of all the backgrounds including top quark 
production are
plotted separately, as well as the totals with and without the Higgs
signal.
The sharp peak of the Higgs signal becomes only a small shoulder in the
total distribution. 

\subsection{Conclusions}

We have introduced the program MCFM, which calculates the rates for a
number of standard model processes that are particularly relevant in
Run II. These calculations are performed in fixed-order perturbation
theory, mainly at next-to-leading order in the strong coupling, and as
such differ from other approaches such as parton shower Monte Carlos.
As illustrations of the use of MCFM, we have presented total di-boson
cross-sections and a primitive study (lacking detector effects and
non-physics backgrounds) of $WH$ production as a search for a light
Higgs. Such calculations can be used to provide normalizations for
more detailed studies in the future.


\section{Experimental handles on the backgrounds to new physics searches}

\centerline{\it by Regina Demina}\vskip 2.mm
       
\subsection{Introduction}

Significant work has been done in the course of the SUSY/Higgs \cite{1} and 
Strong Dynamics \cite{2} Workshops to understand the Tevatron discovery 
potential for new physics. 
Several promising signatures have been identified and 
the discovery reach has been estimated. 
In these studies, it was assumed that the systematic error on the
signal and background normalization is similar in size to
the statistical error, which is about 10\%. 
Thus, the systematic error of each individual background process must be 
kept under 5\%.  
Though it is probably a reasonable assumption, this will not happen 
automatically and dedicated 
studies are needed to achieve this goal.
In this paper, we review the most important backgrounds to new physics 
and ways to estimate them in signal-depleted control samples.

\subsection{New physics signatures}
Associated vector boson and heavy flavor jets production is probably the 
most promising signature for new physics searches at the Tevatron. 
Standard Model (SM) Higgs boson \cite{3},
Supersymmetry \cite{4}, technicolor and topcolor \cite{5} 
and even extra-dimension \cite{6} signatures
may appear in these channels.

\begin{table*}[!ht]
\begin{center}
\begin{tabular}{||c|l|l|l||} \hline\hline
\# & Process & Model & Special features \\ \hline
1  & $WH, H\to b\bar b$ & SM Higgs & Resonance in $M_{b\bar b}$ \\ 
2  & $\rho_T^\pm\to W^\pm\pi_T^0, \pi_T^0\to b\bar b$ & Technicolor & 
Resonance in $M_{Wb\bar b}$ and $M_{b\bar b}$ \\ 
3  & $\rho_T^0\to W^\pm\pi_T^\mp, \pi_T^\pm\to c\bar b$ & Technicolor & 
Resonance in $M_{Wc\bar b}$ and $M_{c\bar b}$ \\ 
4  & $\widetilde\chi^+_1\widetilde\chi^0_2, 
\widetilde\chi^+_1\to\ell\nu\widetilde\chi^0_1,\widetilde\chi^0_2\to 
b\bar b\widetilde\chi^0_1$ 
   & SUSY & $M_T(\ell\slashchar{E}_T)$ inconsistent with $W$ \\
5  & $t\bar t, t\to bW, \bar t\to\widetilde{\bar t} 
\widetilde\chi^0_1,\widetilde{\bar t}\to c\widetilde\chi^0_1$ & SUSY & 
$M_T(\ell\slashchar{E}_T)$ inconsistent with $W$ \\ \hline\hline
\end{tabular}
\end{center}
\caption{Potential new physics signatures in the $W+2$ jet channel.  
From the experimental point of view,
a ``$W$'' is a high $p_T$ lepton accompanied by significant missing 
energy (e.g., CDF cuts are
$p_T(e,\mu)>20$ GeV/c, $\slashchar{E}_T>20$ GeV).  In that sense, 
$\widetilde\chi^+_1$ looks like a ``$W$.''}
\label{table1}
\end{table*}

\begin{table*}[!ht]
\begin{center}
\begin{tabular}{||c|l|l|l||} \hline\hline
\# & Process & Model & Special features \\ \hline
1  & $t\bar t, t\to bW, \bar t\to\widetilde{\bar 
t}\widetilde\chi^0_1,\widetilde{\bar t}\to \bar b\widetilde\chi^-_1$ 
   & SUSY & $M_T(\ell\slashchar{E}_T)$ inconsistent with $W$ \\
2  & $\tilde t\widetilde{\bar t}, \tilde t\to b\widetilde\chi^+_1$ & 
SUSY & $M_T(\ell\slashchar{E}_T)$ inconsistent with $W$ \\
3  & $\tilde t\widetilde{\bar t}, \tilde t\to b\ell\tilde\nu$ & SUSY & 
$M_T(\ell\slashchar{E}_T)$ inconsistent with $W$ \\
4  & $\widetilde{g}\widetilde{g}, \widetilde{g}\to t\tilde{\bar t}$ 
   & SUSY & $M_T(\ell\slashchar{E}_T)$ inconsistent with $W$ \\
5  & $Z'(V_8,\eta_t)\to t\bar t$ & Topcolor & Resonance in $M_{t\bar t}$ 
\\ \hline\hline
\end{tabular}
\end{center}
\caption{Potential new physics signatures in the $W+3$ or more jet channel.}
\label{table2}
\end{table*}

Tables~\ref{table1} and \ref{table2}, 
show examples of new physics processes that can produce $W+2$ jet and 
$W+3$ or more jet signatures.
From the experimental point of view, a ``$W$'' is 
usually a high $p_T$ 
lepton accompanied by a significant missing
energy (e.g. CDF Run I cuts are $P_T(e,\mu)>20$ GeV/c, 
$\slashchar{E}_T>20$ GeV \cite{7}).  In that sense, 
the supersymmetric partner of 
$W$ -- $\widetilde\chi^+_1$ -- looks like a $W$, 
except its transverse mass will be inconsistent with the $W$ hypothesis, 
but this
will become obvious only when significant statistics is accumulated.  
Some models predict special features, like
resonance behavior in the $b\bar b$ invariant mass, while others do not.

\begin{table*}[!ht]
\begin{center}
\begin{tabular}{||c|l|l|l||} \hline\hline
\# & Process & Model & Special features \\ \hline
1  & $ZH, H\to b\bar b$ & SM Higgs & Resonance in $M_{b\bar b}$ \\ 
2  & $\rho_T^+\to Z\pi_T^+, \pi_T^+\to c\bar b$ & Technicolor & 
Resonance in $M_{Zc\bar b}$ and $M_{c\bar b}$ \\ 
3  & $\widetilde\chi^+_1\widetilde\chi^0_2, \widetilde\chi^+_1\to 
c\bar s\widetilde\chi^0_1,
   \widetilde\chi^0_2\to \ell^+\ell^-\widetilde\chi^0_1$ 
   & SUSY & $M_{\ell\ell}$ inconsistent with $Z$, extra $\slashchar{E}_T$ \\
4  & $\tilde b\widetilde{\bar b}, \tilde b\to b\widetilde\chi^0_1, 
\widetilde{\bar b}\to \bar b\widetilde\chi^0_2,
\widetilde\chi^0_2\to\ell\ell\widetilde\chi^0_1$ & SUSY & extra 
$\slashchar{E}_T$ \\ \hline\hline
\end{tabular}
\end{center}
\caption{Potential new physics signatures in the $Z+2$ jet channel.  
From the experimental point of view,
a ``$Z$'' is two high $p_T$ leptons, usually with a $Z$ mass window cut.
In that sense, $\widetilde\chi^0_2$ looks like a ``$Z$'' only in some 
regions of SUSY parameter space.}
\label{table3}
\end{table*}

\begin{table*}[!ht]
\begin{center}
\begin{tabular}{||c|l|l|l||} \hline\hline
\# & Process & Model & Special features \\ \hline
1  & $\tilde t\widetilde{\bar t}, \tilde t\to c\widetilde\chi^0_1$ 
   & SUSY & 2 charm jets and $\slashchar{E}_T$ \\
1  & $\tilde b\widetilde{\bar b}, \tilde b\to b\widetilde\chi^0_1$ 
   & SUSY & 2 bottom jets and $\slashchar{E}_T$ \\
3  & $LQ_2 LQ_2, LQ_2\to c\nu$ 
   & Leptoquarks & 2 charm jets and $\slashchar{E}_T$ \\
3  & $LQ_3 LQ_3, LQ_3\to b\nu$ 
   & Leptoquarks & 2 bottom jets and $\slashchar{E}_T$ \\ \hline\hline
\end{tabular}
\end{center}
\caption{Potential new physics signatures in the $\slashchar{E}_T+2$ 
jet channel. 
This does not include processes complementary to those in 
Table~\ref{table3}, where a $Z$ decays
to a pair of neutrinos $Z\to\nu\bar\nu$, thus producing missing energy.}
\label{table4}
\end{table*}

Table~\ref{table3} presents new physics processes that can produce 
$Z+2$ jet signatures.  Here we assume that
the $Z$ decays to a pair of leptons.  Usually, a $Z$ mass window cut 
is applied.  In that sense, the
supersymmetric partner of the $Z$ -- $\widetilde\chi^0_2$ -- looks 
like a $Z$ in only some regions
of SUSY parameter space.  If the $Z$ 
decays to a pair of neutrinos,
it produces missing energy.  In this case, all the processes 
presented in Table~\ref{table3} produce
a $\slashchar{E}_T+2$ jet signature.  Table~\ref{table4} shows 
additional new physics processes that result
in a $\slashchar{E}_T$ signature.  As we see, these channels are very 
important for new physics searches, and
the Standard Model backgrounds to these signatures must 
be thoroughly understood before
any claims of discovery are made.  

\subsection{Backgrounds to new physics}

The $W(Z)b\bar b$ signature was studied in the course of the SUSY/Higgs 
Workshop for the
Higgs discovery
potential estimate \cite{3}.  The $\slashchar{E}_T$+heavy flavor $(c/b)$ 
signature was studied in 
the CDF stop/sbottom search \cite{8}.  We use these analyses as examples 
in our discussion.

\paragraph{$W(Z)b\bar b$ signature\label{section3.1}}
\subparagraph{Selection and sample composition}

\begin{figure}
  \centerline{\psfig{file=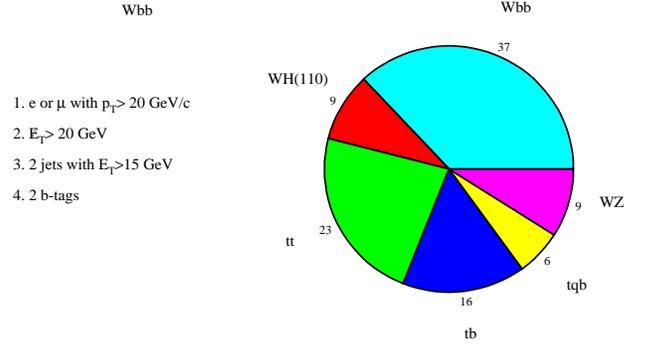,width=6cm,bbllx=120pt,bblly=150pt,bburx=450pt,bbury=400pt}}
\caption{\label{figure1} Selection cuts and composition of the 
$Wb\bar b$ sample.}
\end{figure}

\begin{figure}
  \centerline{\psfig{file=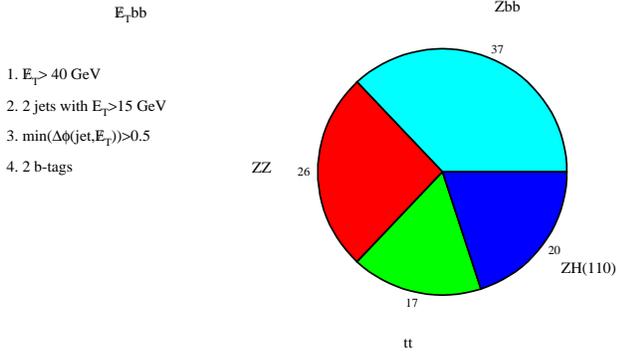,width=6cm,bbllx=120pt,bblly=150pt,bburx=450pt,bbury=400pt}}
\caption{\label{figure2} Selection cuts and composition of the 
$\slashchar{E}_T b\bar b$ sample.}
\end{figure}

The $Wb\bar b$ ($Z(\to\nu\bar\nu)b\bar b$) selection criteria and 
the resultant sample composition are
summarized in Fig.~\ref{figure1} (Fig.~\ref{figure2}).
The dominant contribution to both samples is QCD production of a vector 
boson accompanied by two $b$-jets.

\subparagraph{QCD $W(Z)b\bar b$ production.  Experimental studies 
of gluon splitting to heavy flavor.}

\begin{figure*}
  \centerline{\psfig{file=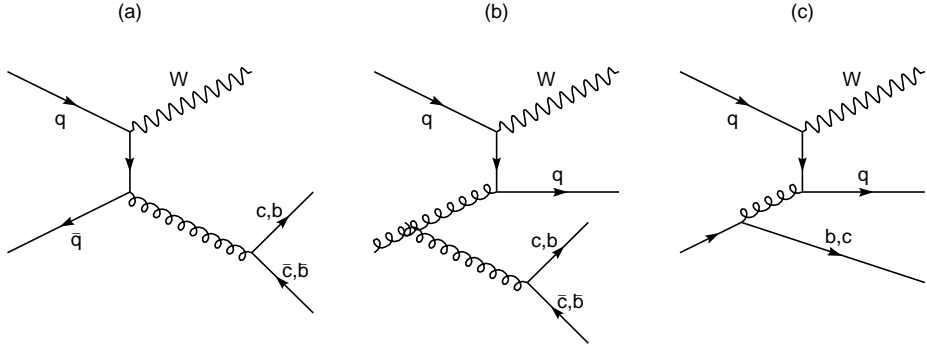,width=10cm,bbllx=100pt,bblly=50pt,bburx=450pt,bbury=220pt}}
\caption{\label{figure3} Diagrams of QCD associated production of 
$W(Z)$ and heavy flavor jets.}
\end{figure*}

Diagrams of QCD associated production of $W(Z)$ and heavy flavor 
jets are presented in Figure~\ref{figure3}.
The leading contribution is $W(Z)$+gluon production with subsequent 
gluon splitting to
a $b\bar b$ or $c\bar c$ pair, shown in Figure~\ref{figure3}(a).

Though a next-to-leading-order calculation of the QCD $Wb\bar b$ 
production exists \cite{9}, even the
authors themselves recommend that it should be tested experimentally.  
This is a particularly hard
task in the presence of a potential signal contribution. In the case 
of the Higgs search,
an invariant mass of two $b$-jets could be used as an additional 
handle, since gluon splitting
contributes mainly to the low part of the $M_{b\bar b}$ spectrum, 
while the Higgs is a resonance at
110-130 GeV/c$^2$.  This is not the case for some other potential 
signal process, e.g. process 5 in Table~\ref{table1}.

The probability for a gluon to split to two heavy flavor jets can be 
studied experimentally in
different samples. The signal contamination becomes negligible, if
the presence of a vector boson is not required.

\begin{figure*}[!ht]
    \centerline{\psfig{file=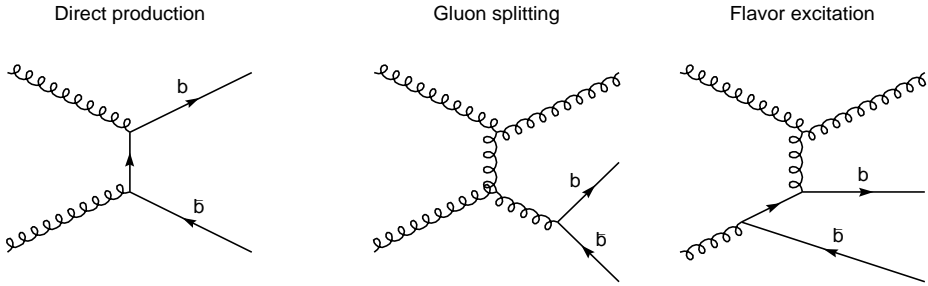,width=10cm,bbllx=100pt,bblly=100pt,bburx=450pt,bbury=200pt}}
  \caption{Diagrams of QCD heavy flavor production.}
  \label{figure4}
\end{figure*}

Three heavy flavor production mechanisms can be isolated -- direct 
production, 
final state gluon splitting and initial state gluon splitting, 
also called flavor excitation. 
Diagrams of these processes are presented in Figure~\ref{figure4}.

\begin{figure}[!ht]
  \centerline{\psfig{file=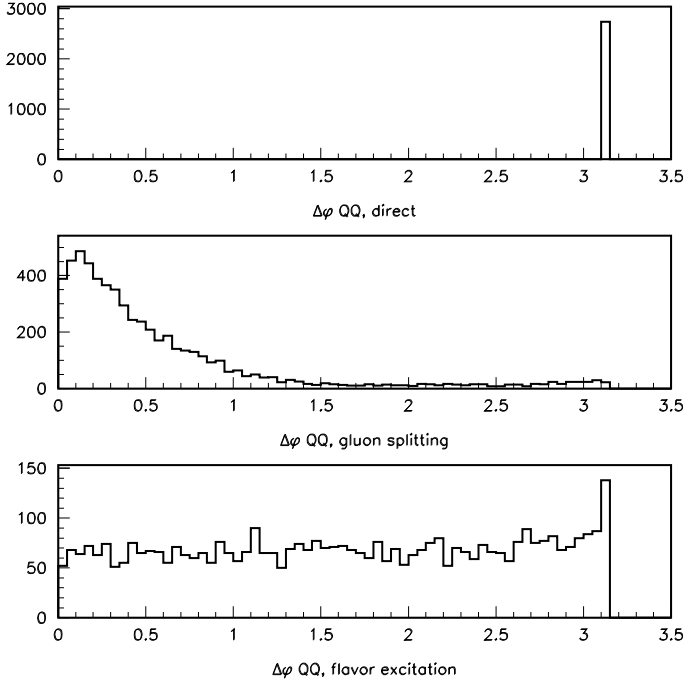,width=9cm}}
  \caption{Distribution $\Delta\phi$ between two $b$-quarks from Monte 
Carlo events.}
  \label{figure5}
\end{figure}

Though direct production is the lowest order process, 
it is responsible for the production of only $\sim~20$\% of heavy flavor 
jets with energy above 20 GeV; 
about 35\% are produced by flavor excitation and 45\% by gluon splitting. 
The relative contribution of different processes changes after 
$b$-tagging is applied. 
Tagging is usually more efficient on directly produced jets, 
which tend to be back-to-back in the azimuthal plane.
Heavy flavor quarks produced from gluon splitting are not well separated,
and are often assigned to the same jet. 
Thus the relative contribution of gluon splitting to the double-tagged 
jet sample is quite low. 
Flavor excitation involves an initial state gluon splitting to two 
heavy flavor quarks, one of which undergoes hard scattering. 
The other quark, being a part of the proton remnant, is often outside 
the detector acceptance. 
Thus, the contribution of flavor excitation to the double-tagged 
sample is significantly depleted. 
An analysis of the angular correlation between two heavy flavor tagged jets
can be used to 
isolate the gluon-splitting component in heavy flavor production, 
as depicted in Fig.~\ref{figure5}.

Different methods can be used to tag heavy flavor jets.
\begin{enumerate}
\item Impact parameter or secondary vertex tagging (JETPROB or SECVTX 
in CDF jargon) \cite{7} 
  are the ones most commonly used. 
  These samples have relatively high statistics. 
  Using the same tagging method for the background and the signal 
sample eliminates the 
  systematic uncertainty.
  The main disadvantage of these methods is the relatively low purity 
of these samples -- 
  contamination from $c$-jets and mistags is non-negligible. 
  Usually, to get a stable fit, the $b/c$ ratio has to be fixed to the 
one predicted by Monte Carlo, 
  which is not without its own uncertainty.

\item One of the heavy flavor jets is tagged by the presence of a high 
$p_T>8$ GeV/c lepton -- electron or muon -- and JETPROB or SECVTX tags 
another jet \cite{10}. 
These samples have high statistics as well, but again suffer from 
charm and mistag contamination. Nonetheless,
it is an interesting independent study.

\item Both heavy flavor jets can be tagged by leptons. 
In this case it is possible to go lower in lepton momentum, usually 
$p_T>3$ GeV/c \cite{11}.
Compared to the first two cases, these studies probe a lower energy 
region, where the direct production 
mechanism dominates.
Thus not much information about gluon splitting probability can be gained.

\item Study \#2 can be modified to increase the purity by 
reconstructing exclusive or semi-exclusive final states 
  in one of the jets:
  \begin{enumerate}
  \item Reconstructing a decay chain $D^*\to D^0\pi,D^0\to Ke(\mu)\nu$ 
can isolate the charm contribution \cite{12}.
The presence of a high $p_T$ lepton guarantees that the contribution 
from $b\to c$ decay is at the order of 
10\% or lower.  Studying the angular correlation between $D^*$ jet and 
an impact parameter tagged jet isolates 
the gluon splitting to charm contribution. 
This probability can then be applied to study \#2 to extract the 
probability of gluon splitting to $b$-quarks.

  \item A similar trick can be used to isolate the $b$-contribution 
in the lepton sample. 
Promising decay chains are \cite{13}:
     $B\to\ell\nu D^*$, $D^*\to D^0\pi$, $D^0\to K\pi$ or $K 3\pi$;
     $B\to\ell\nu D^+$, $D^+\to K\pi\pi$; $B\to\ell\nu D^0$, $D^0\to 
K\pi$; and $B\to J/\Psi K$.
  \end{enumerate}
\end{enumerate}

\begin{table*}[!ht]
\begin{center}
  \begin{tabular}{||c|l|r|r|r|r||} \hline\hline
\# & Sample & N(Run I) & $\sigma(g\to Q\bar Q)$ Run I & N(Run II) & 
$\sigma(g\to Q\bar Q)$ Run II \\ \hline
1  & Double tagged jets & 700             & 20\%      & 28000     & 3.2\% \\
2  & Muon+JETPROB tag   & 2620            & 16\%      & 104800    & 2.5\% \\
3  & $c\to D^*\to D^0(\to K\ell\nu)\pi$ & 18000 & 15\% & 720000    & 2.4\% \\
4  & $B\to D^*\ell\nu\to D^0(\to K\pi/K3\pi)\ell\nu\pi$ & 1700 & n.a. & 
68000    & n.a. \\
5  & $B\to D^+(\to K\pi\pi)\ell\nu$  & 1900 & n.a.    &  76000 & n.a. \\
6  & $B\to D^0(\to K\pi)\ell\nu$     & 2700 & \#4-\#7 & 108000 & \#4-\#7 \\
7  & $B\to J/\Psi K^{(*)}$           & 1300 & 23\%    &  52000 & 3.6\% 
\\ \hline\hline
  \end{tabular}
\end{center}
  \caption{Data samples for heavy flavor production study.  Numbers in 
samples \#1 and \#2 are double tags, while in samples \#3-\#7 numbers 
of exclusively reconstructed events are shown, without requiring a tag 
on the opposite side.}
  \label{table5}
\end{table*}

In Table~\ref{table5}, we present the number of events in each of the 
discussed samples collected in 
Run I and expected in Run II and associated statistical uncertainty on 
the gluon splitting probability. 

Run I numbers are based on CDF results. 
In Run II, both CDF and D\O~will have similar tracking and vertexing 
capabilities, 
thus these numbers are applicable to both detectors.
Statistics in Run II is increased by a factor of 40, where 20 is gained 
from the luminosity increase and 2 
from increased acceptance of the silicon microvertex detectors. 
The tagging efficiency increase is not taken into account. 
With these dedicated studies, the statistical uncertainty on the 
probability of gluon splitting 
to heavy flavor quarks can be significantly reduced in Run II, 
and will become adequate to the needs of new physics searches.

\subparagraph{Top, single top and diboson production}
Other backgrounds to new physics searches are top pair \cite{14}, 
single top \cite{15} and diboson \cite{16} production. 
The theoretical predictions for these backgrounds are more reliable, 
because they do not involve gluon radiation and 
splitting, yet they still have to be tested experimentally. 
This is more or less a straightforward task for top pair and diboson 
production, 
where final states can be exclusively identified. 
It is less so for single top production, where the final state is exactly 
the same -- $Wb\bar b$ -- as in new physics channels in Table~\ref{table1}. 
Additional mass constraints, e.g. on the $Wb$
mass can be used to isolate this process, but it is not at all obvious 
that adequate uncertainty can be reached for this channel.

\paragraph{Missing energy and heavy flavor signatures}
Here, we summarize the selection criteria and 
composition of the missing energy and 
heavy flavor sample used for top squark searches \cite{8}:

\begin{figure}[!ht]
  \centerline{\psfig{file=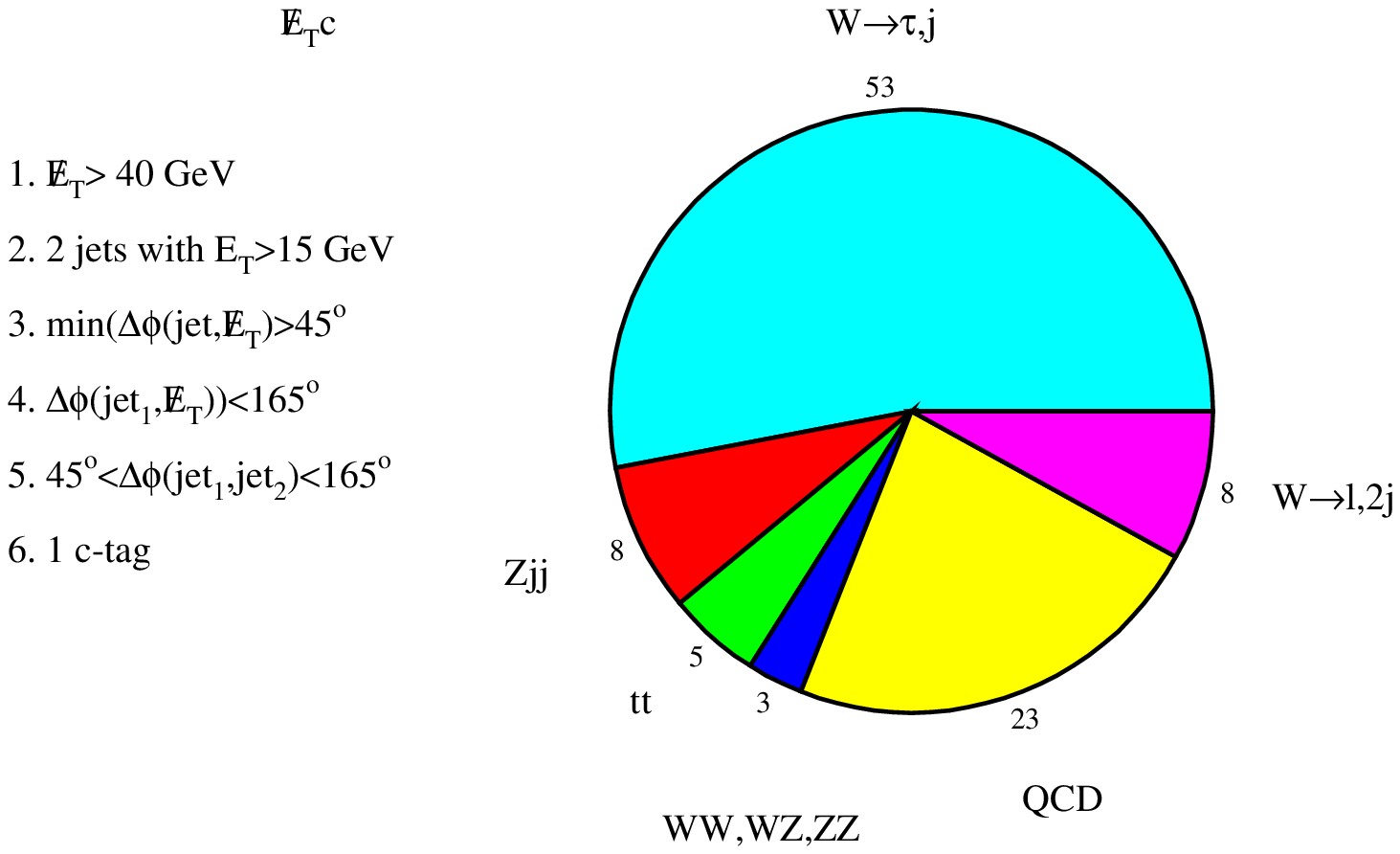,width=6.5cm,bbllx=120pt,bblly=150pt,bburx=450pt,bbury=400pt}}
\caption{Selection cuts and composition of $\slashchar{E}_T c$ sample.
\label{figure6}}
\end{figure}

More than 50\% of the background is composed of
$W(\to\tau\nu)+1$ jet events.

\subparagraph{$W+c$ production}

The leading order production process for $W(\to\tau\nu)+1$ jet, 
where this jet is identified as charm, is $sg\to Wc$.
The main uncertainty of the production rate for this process comes 
from the PDF of sea $s$ quarks $f_s(x)$, 
which is measured by NuTeV \cite{17} in the neutrino scattering process 
$\nu_\mu s\to\mu c$. 

\begin{figure}[!ht]
  \centerline{\psfig{file=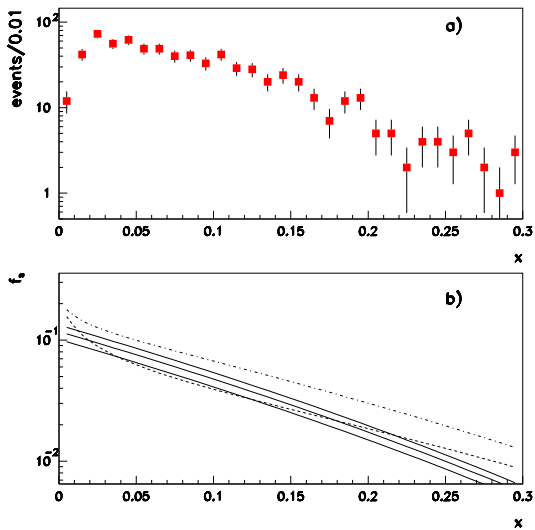,width=7cm}}
\caption{(a) Distribution in $x$ of sea $s$-quarks that contribute to 
$sg\to Wc$ production at the Tevatron, 
generated with \pythia~6.1+CTEQ4LO. Selection cuts have been applied.  
(b) Solid lines represent $f_s(x)$ and its uncertainty, as measured by NuTeV. 
It is compared to GRV94LO (dashed line) and CTEQ4LO (dot-dashed line) 
strange sea distributions.
\label{figure7}}
\end{figure}

Figure~\ref{figure7} (a) shows the distribution in $x$ of sea $s$-quarks 
that contribute to $sg\to Wc$ 
production at the Tevatron after the selection cuts from 
Figure~\ref{figure6} have been applied. 
Figure~\ref{figure7} (b) shows the $f_s(x)$ and its uncertainty 
measured by NuTeV. 
As we see, the region of NuTeV sensitivity is relevant for Tevatron studies.
The overall uncertainty on $f_s(x)$ is 13.5\%. 
$f_s(x)$ measured by NuTeV is in a good agreement with the results of 
CCFR \cite{18}, which has an uncertainty of 10.5\%.

Since these uncertainties are dominated by the experimental statistics, 
we can hope that the combined result will have an uncertainty near 8.5\%. 
The strange sea parton density function was also measured by the 
CHARM~II \cite{19} and CDHS \cite{20} 
experiments. Combination of results of all four experiments is 
certainly desirable, but non--trivial, since somewhat 
different techniques where used in each analysis.

In Figure~\ref{figure7}(b), $f_s(x)$ measured by NuTeV is compared to 
the one provided by the GRV94LO \cite{21} and CTEQ4LO \cite{22} PDF's, 
which are shown by dashed and dot-dashed lines, respectively.
None of the PDF's provide an adequate description of the strange sea data. 
In the Tevatron search experiments, the systematic uncertainty on the 
background due to PDF's was typically 
estimated by switching from one PDF to another. In this case, the 
systematic uncertainty on the number of $Wc$ 
events that pass our cuts is 36\%. If instead of CTEQ4LO, $f_s(x)$ 
measured by NuTeV were used, 
the number of expected $Wc$ events would go down by 30\%. This is 
within the estimated uncertainty, 
but clearly the uncertainty has been overestimated. 
The correct uncertainty to use is 13.5\%, or 8.5\%, 
when the results of NuTeV and CCFR will be combined. 
This is a significant reduction compared to 36\%, yet it is still not 
down to the desirable level of 5\%. 
We can probably do better by studying $Wc$ production when the $W$ is 
identified by its decay to a muon or an electron. 
The expected number of events in the $Wc, W\to\ell\nu (\ell=e,\mu)$ 
channel is about 2800, after applying the cuts listed in 
Figure~\ref{figure6}, which corresponds to the statistical uncertainty 
for this background of about 2\%. 
The systematic uncertainty on missing energy and charm identification 
are common to the two channels,
and the difference is in lepton vs. tau identification uncertainty, 
which can be expected to be below 5\% with Run II statistics. 

\subparagraph{QCD background} 

The next dominant background in the $\slashchar{E}_T c$ channel is QCD 
multijet production, where missing energy comes from 
jet energy mis-measurement. This background is the toughest one to 
estimate, because it involves multiple gluon radiation and splitting. 
Not only the overall rate, but also the angular correlation 
between jets may not be predicted reliably.  
To isolate this component,
the usual trick is to apply all the signal selection cuts except 
tagging, subtract other known backgrounds and call the rest ``QCD.''
The tagging probability derived from an independent jet sample 
is then applied
to estimate the QCD contribution to the tagged sample \cite{23}.
One obvious drawback is that the heavy flavor fraction can change after 
the cuts are applied. 
Another is that the signal contribution is not always 
negligible even before tagging, and
to some extent, it may be normalized away.  

\subparagraph{Other backgrounds in the $\slashchar{E}_T c$ channel.}

Other backgrounds in the $\slashchar{E}_T c$
channel are top pair production, di-bosons, $Z+$ jets and $W+$ jets, 
where leptons were not identified. 
The discussion of these processes in Section~\ref{section3.1} applies 
to the $\slashchar{E}_T c$ channel as well.   

\section{Variable flavor number schemes for heavy quark electroproduction}

\centerline {\it by J. Smith } \vskip 2.mm

Heavy quark production has been a major topic of investigation
at hadron-hadron, electron-proton and electron-positron
colliders. Here a review is given of some topics which are of interest
primarily for electron-proton colliders. We concentrate on this
reaction because a theoretical treatment can be based on the 
operator product expansion, and also because data are available 
for deep-inelastic charm production at HERA. How all this relates to 
Fermilab experiments will be discussed at the end. 

In QCD perturbation theory, one needs to introduce a renormalization 
scale and a mass factorization scale to perform calculations. 
We choose both equal to $\mu^2$, which will be a function 
of $Q^2$ and the square of the mass of the charm quark $m^2$.
At small $\mu^2$, where kinematic effects due to quark masses are important, 
the best way to describe charm quark production is via heavy quark 
pair production from light quark $u,d,s$ and gluon initial states.
The mass $m$ only appears in 
the heavy quark coefficient functions (or partonic cross sections) like
$H_{i,g}^{\rm S,(2)}(z,Q^2,m^2,\mu^2)$, etc., \cite{lrsn}. Here the 
superscripts refer to their flavor decomposition and the order in 
perturbation theory, while the subscripts refer to the projection
$i=2,L$ and the partonic initial state. The arguments refer 
to the partonic Bjorken variable $z = Q^2/(s+Q^2)$ 
and to the fact that these functions depend upon invariants and scales. 
The renormalization necessary to calculate
these NLO expressions follows the CWZ method \cite{cwz}. 
The symbol $H$ refers to those coefficient functions which are 
derived from Feynman diagrams where the virtual
photon couples to a heavy quark line. 
Analytic expressions for these functions are not known, but
numerical fits are available in \cite{rsn}. Asymptotic 
expressions in the limit $Q^2 \gg m^2$ are available in \cite{bmsmn}.
These contain 
terms like $\ln^2(Q^2/m^2)$ and $\ln(Q^2/m^2)\ln(Q^2/\mu^2)$ 
multiplied by functions of $z$; they are manifestly
singular in the limit that $m\rightarrow 0$.

There are other heavy quark coefficient functions such as
$L_{i,q}^{\rm NS,(2)}(z,Q^2,m^2)$, which arise from tree diagrams where the 
virtual photon attaches to the an initial state light quark line, so the
heavy-quark is pair produced via virtual gluons. Analytic expressions
for these functions are known for all $z$, $Q^2$ and $m^2$, which, in the limit
$Q^2 \gg m^2$ contain powers of $\ln(Q^2/m^2)$ multiplied by functions
of $z$. 
The three-flavor light mass $\overline{\rm MS}$ parton densities 
can be defined in terms of matrix elements of operators and are now 
available in parton density sets. This is a fixed order perturbation 
theory (FOPT) description of heavy quark production with three-flavor 
parton densities.  Due to the work in \cite{lrsn}, the 
perturbation series is now known up to second order. In regions of 
moderate scales and invariants, this NLO description is well defined 
and can be combined with a fragmentation function to predict exclusive 
distributions \cite{hs} for the outgoing charm meson, 
the anti-charm meson and the additional parton.
This NLO massive charm approach agrees well with the recent
$D$-meson inclusive data in \cite{ZEUS} and \cite{H1}. 
The charm quark structure functions in this NLO description
will be denoted
{\boldmath $F^{\rm EXACT}_{i,c}(x,Q^2,m^2,n_f=3)$}.

A different description, which should be more 
appropriate for large scales where terms in $m^2$ are 
negligible, is to represent charm production by a parton density 
$f_c(x,\mu^2)$, with a boundary condition that the density vanishes
at small values of $\mu^2$. 
Although at first sight these approaches appear to
be completely different, they are in fact intimately related. 
It was shown in \cite{bmsn1} that the large terms in $\ln(Q^2/m^2)$ 
which arise when $Q^2 \gg m^2$, can be resummed to all orders in perturbation
theory. In this reference, all the two-loop corrections to the
matrix elements of massive quark and massless gluon operators
in the operator product expansion were calculated. These contain
the same type of logarithms mentioned above multiplied by functions
of $z$ (which is the last Feynman integration parameter).
After operator renormalization and suitable reorganization of convolutions
of the operator matrix elements (OME's) and the coefficient functions,
the expressions for the infrared-safe charm quark 
structure functions $F_{i,c}(x,Q^2,m^2,\Delta)$
take on a simple form. 
After resummation, they are convolutions of light-mass,
four-flavor parton coefficient functions, commonly denoted by expressions 
like ${\cal C}^{\rm S,(2)}_{i,g}(Q^2/\mu^2)$
(available in \cite{zn}, \cite{rijk}), with four-flavor light-parton densities,
which also include a charm quark density $f_c(x,\mu^2)$. 
Since the corrections to the OME's contain terms in $\ln(Q^2/m^2)$ 
and $\ln(m^2/\mu^2)$ as well as non-logarithmic terms, it is simplest 
to work in the $\overline{\rm MS}$ scheme with the scale 
$\mu^2 = m^2$ for $Q^2 \le m^2$
and $\mu^2 = m^2 + Q^2(1-m^2/Q^2)^2/2$ for $Q^2 > m^2$ and discontinuous
matching conditions on the flavor densities at $\mu^2 = m^2$. 
Then all the logarithmic terms vanish at $Q^2=\mu^2=m^2$ 
and the non-logarithmic terms 
in the OME's are absorbed into the boundary conditions on the charm density,
the new four-flavor gluon density and the new light-flavor
u,d,s densities. The latter are convolutions of the previous 
three-flavor densities with the OME's given 
in the Appendix of \cite{bmsn1}. 

The above considerations lead to a precise description 
through order $\alpha_s^2$ of how, in the limit $m\rightarrow 0$, 
to re-express the $F_{i,c}^{\rm EXACT}(x,Q^2,m^2)$
written in terms of convolutions of heavy quark coefficient functions 
with three-flavor light parton densities into a description in terms of 
four-flavor light-mass parton coefficient functions
convoluted with four-flavor parton densities. 
This procedure leads to the so-called zero-mass variable-flavor-number 
scheme (ZM-VFNS) for $F_{i,c}(x,Q^2,\Delta)$ where the $m$ dependent
logarithms are absorbed into the new four-flavor densities. 
To implement this scheme, one has to be careful to use inclusive quantities 
which are collinearly finite in the limit $m \rightarrow 0$ and $\Delta$
is an appropriate parameter which enables us to do this. In the 
expression for $F_{i,c}$ there is a cancellation of terms 
in $\ln^3(Q^2/m^2)$ between the two-loop corrections to the light 
quark vertex function (the Sudakov form factor) and the convolution of 
the densities with the soft part of the $L_{2,q}$-coefficient function. 
This is the reason for the split of $L_{i,q}$ into soft and hard 
parts, via the introduction of a constant $\Delta$. Details
and analytic results for $L_{i,c}^{\rm SOFT}$ and $L_{i,c}^{\rm HARD}$ 
are available in \cite{csn1}.  All this analysis yielded 
and used the two-loop matching
conditions on variable-flavor parton densities across flavor thresholds,
which are special scales where one makes transitions from say a three-flavor
massless parton scheme to a four-flavor massless parton scheme. 
The threshold is a choice of $\mu$
which has nothing to do with the actual kinematical heavy flavor
pair production threshold at $Q^2(x^{-1} -1) = 4 m^2$. 
In \cite{bmsn1},\cite{bmsn2} it was shown that the
$F^{\rm EXACT}_{i,c}(x,Q^2,m^2, n_f=3)$ tend numerically to the 
known asymptotic results in $F^{\rm ASYMP}_{i,c}(x,Q^2,m^2,n_f=3)$,
when $Q^2 \gg m^2 $, which also equal the ZM-VFNS results.
The last description is good for large (asymptotic) scales and 
contains a charm density $f_c(x,\mu^2)$ which satisfies
a specific boundary condition at $\mu^2 = m^2$. 
We denote the charm quark structure functions
in this description by {\boldmath $F^{\rm PDF}_{i,c}(x,Q^2,n_f=4)$}.

For moderate values of $Q^2$,
a third approach has been 
introduced to describe the charm components of $F_{i}(x,Q^2)$.
This is called a variable flavor number scheme (VFNS). A first 
discussion was given in \cite{acot}, where a VFNS prescription 
called ACOT was given in lowest order only. A proof of factorization 
to all orders was recently given in \cite{col} for the total structure 
functions $F_{i}(x,Q^2)$, but the NLO expressions for $F_{i,c}(x,Q^2,m^2)$ 
in this scheme were not provided. An NLO version of a VFNS scheme has 
been introduced in \cite{csn1} and will be called the {\bf CSN} scheme.
A different approach, also generalized to all orders, 
was given in \cite{bmsn1},\cite{bmsn2}, which is called the {\bf BMSN} scheme. 
Finally another version of a VFNS was presented 
in \cite{thro}, which is called the TR scheme. 
The differences between the various schemes can be attributed to
two ingredients entering the construction of a VFNS. 
The first one is the mass factorization procedure carried out before the 
large logarithms can be resummed. The second one is the matching 
condition imposed on the charm quark density, which has to vanish 
in the threshold region of the production process. 
All VFNS approaches require two sets of parton densities. One set 
contains three-flavor number densities whereas the second set
contains four-flavor number densities. The sets have to satisfy the 
$\overline{\rm MS}$ matching relations derived in \cite{bmsn1}. 
Appropriate four-flavor densities have been constructed in \cite{csn1} 
starting from the three-flavor LO and NLO sets of parton densities recently 
published in \cite{grv98}.

\begin{figure}[th]
\phantom{trick}
\vskip -8.mm
\centerline{\psfig{file=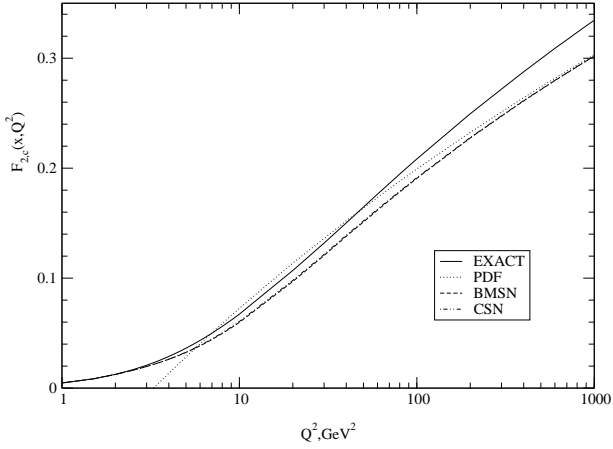,width=9.5cm,angle=-90}}
\vskip -8.mm
\caption{
The charm quark structure functions
$F_{2,c}^{\rm EXACT}(n_f=3)$ (solid line)
$F_{2,c}^{\rm CSN}(n_f=4)$, (dot-dashed line)
$F_{2,c}^{\rm BMSN}(n_f=4)$, (dashed line) and
$F_{2,c}^{\rm PDF}(n_f=4)$, (dotted line)
in NNLO for $x=0.005$
plotted as functions of $Q^2$.}
\label{fig:smith1}
\end{figure}
Since the formulae for the heavy quark structure functions
are available in \cite{csn1}, we only mention a few points here. 
The BMSN scheme avoids the introduction of any new coefficient
functions other than those above. 
Since the asymptotic limits for $Q^2 \gg m^2$ of all the operator matrix 
elements and coefficient functions are known, we define (here $Q$ refers to the
heavy charm quark)
\begin{multline}
F_{i,Q}^{\rm BMSN}(x,Q^2,m^2,\Delta,n_f=4)= \\[1.mm]
F_{i,Q}^{\rm EXACT}(x,Q^2,m^2,\Delta,n_f=3) \\[2ex]
- F_{i,Q}^{\rm ASYMP}(x,Q^2,m^2,\Delta,n_f=3)\\[1.mm]
+ F_{i,Q}^{\rm PDF}(x,Q^2,m^2,\Delta,n_f=4)\,.
\end{multline}
The scheme for $F_{i,Q}^{\rm CSN}$ introduces a new heavy quark OME 
$A_{QQ}^{\rm NS,(1)}(z,\mu^2/m^2)$ \cite{neve} and coefficient functions
$H_{i,Q}^{\rm NS,(1)}(z,Q^2/m^2)$ \cite{bnb}
because it requires an incoming heavy
quark $Q$, which did not appear in the NLO corrections in \cite{lrsn}.
The CSN coefficient functions are defined via the following equations. 
Up to second order we have 
\begin{multline}
{\cal C}_{i,q,Q}^{\rm CSN,SOFT,NS,(2)}
\Big (\Delta,\frac{Q^2}{m^2},\frac{Q^2}{\mu^2}\Big) = 
A_{qq,Q}^{\rm NS,(2)}\Big(\frac{\mu^2}{m^2}\Big) 
{\cal C}_{i,q}^{\rm NS,(0)} \\[1.mm]
-\beta_{0,Q} \ln \left ( \frac{\mu^2}{m^2} \right ) 
\times {\cal C}_{i,q}^{\rm NS,(1)} \Big(\frac{Q^2}{\mu^2}\Big)
-{\cal C}_{i,q}^{\rm VIRT,NS,(2)}(\frac{Q^2}{m^2})\\[1.mm]
- L_{i,q}^{\rm SOFT,NS,(2)}\Big 
(\Delta,\frac{Q^2}{m^2},\frac{Q^2}{\mu^2}\Big) \,, 
\end{multline}
with the virtual term the second order Sudakov form factor.
\begin{figure}[th]
\phantom{trick}
\vskip -8.mm
\centerline{\psfig{file=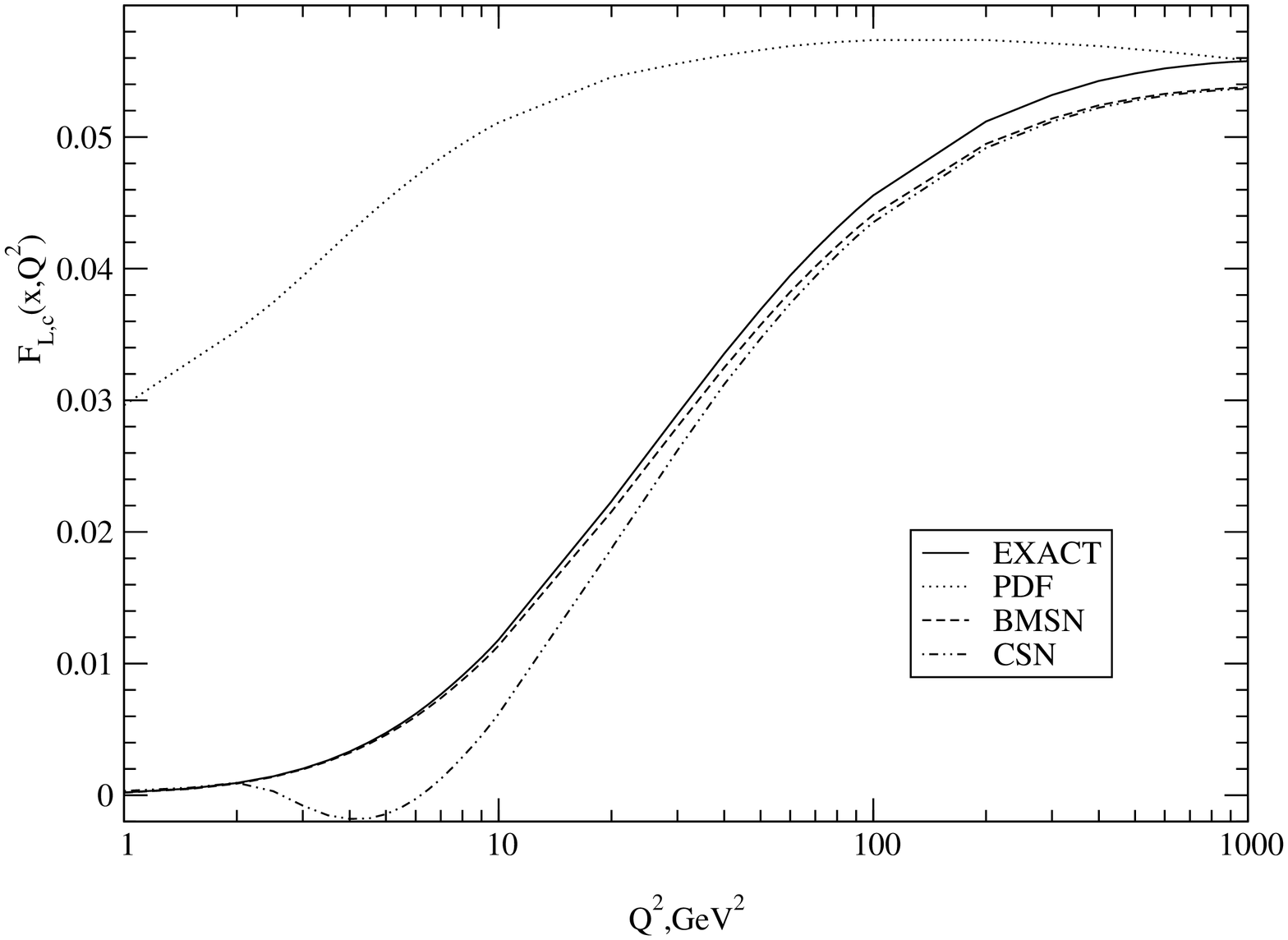,width=9.5cm,angle=-90}}
\vskip -8.mm
    \caption{The charm quark structure functions
$F_{L,c}^{\rm EXACT}(n_f=3)$ (solid line)
$F_{L,c}^{\rm CSN}(n_f=4)$, (dot-dashed line)
$F_{L,c}^{\rm BMSN}(n_f=4)$, (dashed line) and
$F_{L,c}^{\rm PDF}(n_f=4)$, (dotted line)
in NNLO for $x=0.005$ plotted as functions of $Q^2$.
}
    \label{fig:smith2}
\end{figure}
The other CSN coefficient functions are defined by 
equations like (we only give one of the longitudinal terms
for illustration)
\begin{multline}
{\cal C}_{L,g}^{\rm CSN,S,(1)}\Big (\frac {Q^2}{m^2},
\frac {Q^2}{\mu^2}\Big)= \\[1.mm]
H_{L,g}^{\rm S,(1)}\Big (\frac {Q^2}{m^2}\Big)-
A_{Qg}^{\rm S,(1)}(\frac{\mu^2}{m^2})\,
{\cal C}_{L,Q}^{\rm CSN,NS,(0)}\Big (\frac {Q^2}{m^2}\Big) \,,
\label{eq:subtraction}
\end{multline}
with 
${\cal C}_{L,Q}^{\rm CSN,NS,(0)}={4m^2}/{Q^2}$.
The CSN and BMSN schemes are designed to have the following two properties. 
First of all, suppressing unimportant labels, 
\begin{multline}
F_{i,Q}^{\rm CSN}(n_f=4) = F_{i,Q}^{\rm BMSN}(n_f=4) \\[1.mm]
= F_{i,Q}^{\rm EXACT}(n_f=3) \quad \mbox{for}
\quad Q^2 \le m^2 \,.
\label{eq:csn}
\end{multline}
Since $f_Q(m^2)^{\rm NNLO} \not = 0$ (see \cite{bmsn1}) this condition can be
only satisfied when we truncate the perturbation series at the same order.
The second requirement is that 
\begin{multline}
 \mathop{\mbox{lim}}\limits_{\vphantom{
\frac{A}{A}} Q^2 \gg m^2}
F_{i,Q}^{\rm BMSN}(n_f=4)
= \mathop{\mbox{lim}}\limits_{\vphantom{
\frac{A}{A}} Q^2 \gg m^2} F_{i,Q}^{\rm CSN}(n_f=4) \\[1.mm]
= \mathop{\mbox{lim}}\limits_{\vphantom{\frac{A}{A}} Q^2 \gg m^2}
F_{i,Q}^{\rm PDF}(n_f=4) \,.
\label{eq:bmsn}
\end{multline}
The only differences between the two schemes arises from 
terms in $m^2$ so they may not be equal 
just above $Q^2 = m^2$. This turns out to be the case for the
longitudinal structure function, which is more sensitive to mass effects.

Figure~\ref{fig:smith1} shows NNLO results for the $Q^2$
dependence of $F_{2,c}^{\rm EXACT}(n_f=3)$,
$F_{2,c}^{\rm CSN}(n_f=4)$, $F_{2,c}^{\rm BMSN}(n_f=4)$, and 
$F_{2,c}^{\rm PDF}(n_f=4)$ at $x=0.005$. Note that the results
satisfy the requirements in Eqs.~(\ref{eq:csn}) and (\ref{eq:bmsn}). 
The ZM-VFNS description is poor at small $Q^2$.
Figure~\ref{fig:smith2} shows the results for $F_{L,c}^{\rm EXACT}(n_f=3)$,
$F_{L,c}^{\rm CSN}(n_f=4)$, $F_{L,c}^{\rm BMSN}(n_f=4)$, and 
$F_{L,c}^{\rm PDF}(n_f=4)$ at $x=0.005$. We see that the CSN result
is negative and therefore unphysical 
for $2.5 < Q^2 < 6$ $({\rm GeV}/c)^2$ which is due
to the term in $4m^2/Q^2$ and the subtraction in Eq.~(\ref{eq:subtraction}).

One way this research work is of relevance to Fermilab experiments 
is that it produces more precise ZM-VFNS parton densities.
Such densities are used extensively to predict cross sections 
at high energies, for example for 
single top quarks. Therefore the previous work on four-flavor parton 
densities has been extended in \cite{cs1}
to incorporate the two-loop discontinuous 
matching conditions across the bottom flavor threshold 
at $\mu = m_b$ and provided a set of five-flavor densities,
which contains a bottom quark density $f_b(x,\mu^2)$.
The differences between the five-flavor densities and those
in \cite{mrst98} and \cite{cteq5} are also discussed. 
Results for deep-inelastic electroproduction of bottom quarks 
will be presented in \cite{csn2}.

\section{The Underlying Event in Hard Scattering Processes}

\centerline{\it by Rick Field and David Stuart}\vskip 2.mm
       
\subsection{Introduction}

\begin{figure*}[htbp]
    \centerline{\psfig{file=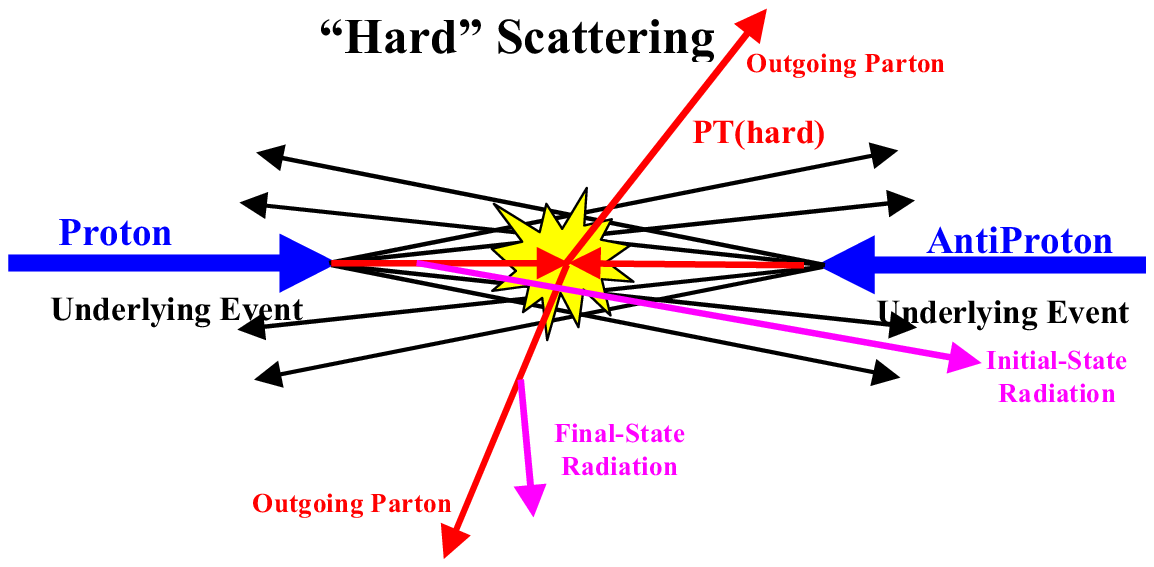,width=7.5cm}\hskip 10.mm
\psfig{file=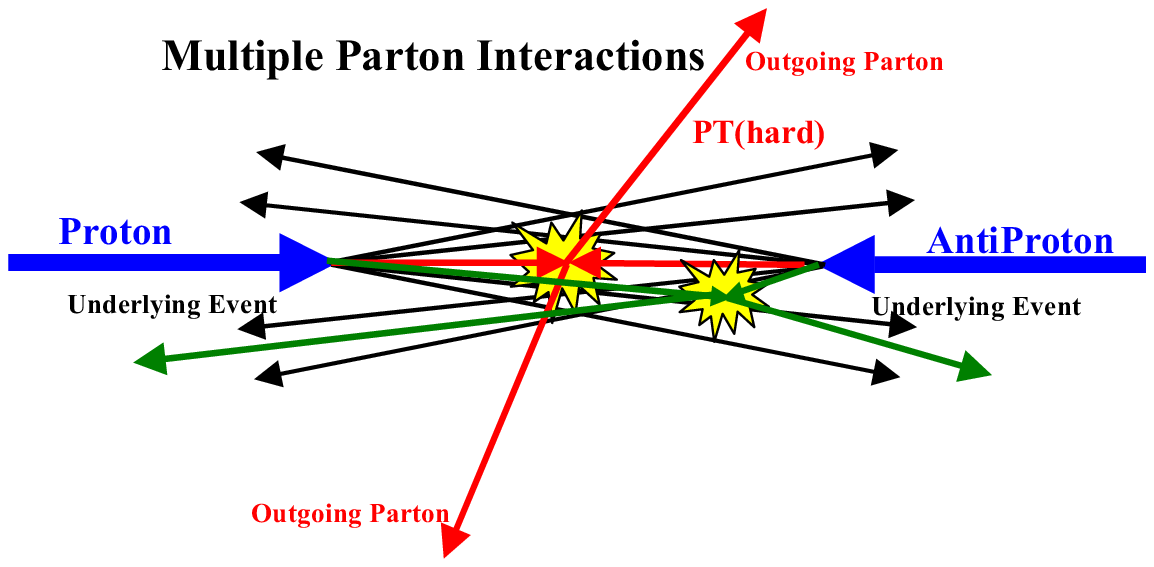,width=7.5cm}
}
\caption{
(a) Illustration of a $p \bar p$ collision in which a ``hard'' 
2-to-2 parton scattering  with transverse momentum, $p_T({\rm hard})$, has 
occurred.  The resulting event contains particles that originate from the 
two outgoing partons (plus final-state radiation) and particles that come 
from the breakup of the $p$ and $\bar p$ (\ie ``beam-beam 
remnants'').  The ``underlying event'' consists of the beam-beam remnants 
plus initial-state radiation;
(b) Illustration of a $p \bar p$ collision in which a multiple parton 
interaction has occurred. In addition to the 
``hard'' 2-to-2 parton scattering  with transverse momentum, 
$p_T({\rm hard})$, there is an additional ``semi-hard''
parton-parton scattering that contributes particles to the 
``underlying event.''  For \pythia, we include the 
contributions from multiple parton scattering in the beam-beam remnant 
component.}
\label{rdf_fig1}
\label{rdf_fig2}
\end{figure*}

The total proton-antiproton cross section is the sum of the elastic and 
inelastic cross sections.  The inelastic cross section consists of a 
single-diffractive, double-diffractive, and a ``hard core'' component, where
the ``hard core'' is everything else.
``Hard core'' does not necessarily imply ``hard scattering.''
A ``hard scattering'' collision, such as that illustrated in 
Fig.~\ref{rdf_fig1}(a), is one in which a ``hard'' (i.e. large 
transverse momentum) 2-to-2 parton-parton subprocess has occurred.  
``Soft'' hard core collisions correspond to events in which no ``hard'' 
interaction has occurred.  When there is no large $p_T$ 
subprocess in the collision, one is not probing short distances 
and it probably does not make any sense to talk about partons. 
The QCD ``hard scattering'' cross section grows with 
increasing collider energy and
becomes a larger and larger fraction of the total inelastic 
cross section.  In this analysis, we used the CDF Min-Bias 
trigger data sample in conjunction with the JET20 trigger data sample to 
study the growth and development of  
``charged particle jets'' from $p_T({\rm jet}) = 0.5$ to $50\gev$.  
We compared several ``local'' jet observables with 
the QCD ``hard scattering'' Monte-Carlo models of \herwig \cite{herwig}, 
\isajet \cite{isajet}, and \pythia \cite{pythia}.

A ``hard scattering'' event, like that illustrated in Fig.~\ref{rdf_fig1}(a) 
consists of large-$p_T$ outgoing hadrons that 
originate from the large-$p_T$ partons (\ie outgoing hard 
scattering ``jets'') and also hadrons that 
originate from the break-up of the proton and antiproton 
(\ie the ``beam-beam remnants'').  The ``underlying event''
is an interesting object that is not very well understood.  In addition to 
beam-beam remnants, it  may contain hadrons 
resulting from initial-state radiation.  Also, it is possible that multiple 
parton scattering occurs in hadron-hadron 
collisions as illustrated in Fig.~\ref{rdf_fig2}(b).  This is a 
controversial issue, but the underlying event might also contains hadrons 
that originate from multiple parton interactions.   \pythia, for example, 
uses multiple parton interactions as a way 
to enhance the activity of the underlying event \cite{pythia}.  

In this analysis, we studied a variety of  ``global'' observables
to probe the growth and structure of the 
underlying event.  We find that the underlying ``hard scattering'' 
event is not the same as a ``soft'' $p \bar p$ 
collision.  For the same available energy, the underlying event in a 
hard scattering is considerably more active (\ie 
higher charged particle density and more $p_T$) than a ``soft'' 
collision.  This is not surprising since a 
violent hard scattering has occurred!  We find that none of the QCD 
Monte-Carlo models ({\it with their default 
parameters}) describe correctly all the properties of the underlying event.

\subsection{Data Selection and Monte-Carlo Models}

\paragraph{\bf (1) Data Selection}

The CDF detector, described in detail in Ref. \cite{rdf_CDF}, 
measures the trajectories and transverse momenta, 
$p_T$, of charged particles in the pseudorapidity region $|\eta|< 1.1$ 
with the central tracking chamber (CTC), 
silicon vertex detector (SVX), and vertex time projection chamber 
(VTX), which are immersed in a $1.4$ T 
solenoidal magnetic field. In this analysis we consider only charged 
particles measured in the central tracking 
chamber (CTC) and use the two trigger sets of data listed in 
Table~\ref{rdf_table1}. The minimum bias (min-bias) 
data were selected by requiring that at least one particle interacted 
with the forward beam-beam counter BBC ($3.4 <\eta < 
5.9$) {\it and/or} the backward BBC ($-5.9 <\eta < -3.4$). The min-bias trigger 
selects predominately the ``hard core'' component of the inelastic 
cross section. 

Charged particle tracks are found with high efficiency as long as the 
density of particles is not high. To remain in a 
region of high efficiency, we consider only charged particles with 
$p_T > 0.5\gev$ and $|\eta| < 1$. The observed 
tracks include some fake tracks that result from secondary interactions 
between primary particles, including neutral 
particles, and the detector material. There are also particles 
originating from other $p \bar p$ collisions. To 
reduce the contribution from these sources, we consider only tracks 
which point to the primary interaction vertex 
within $2$ cm along the beam direction and $1$ cm transverse to the 
beam direction. Detector simulations indicate 
that this impact parameter cut is very efficient and that the number 
of fake tracks is about $3.5\%$ when a $1$ cm 
impact parameter cut is applied in conjunction with a $2$ cm vertex 
cut.  Without the impact parameter cut the 
number of fake tracks is approximately $9\%$.  

This dependence of the number of fake tracks on the CTC impact 
parameter cut provides a method of estimating 
systematic uncertainties due to fakes.  Every data point $P$ on every 
plot in this analysis was determined three times 
by using a $2$ cm vertex cut in conjunction with three different CTC 
$d_0$ cuts; a $1$ cm CTC $d_0$  cut ($P$), a 
$0.5$ cm CTC $d_0$  cut ($P_1$),  and no CTC $d_0$  cut ($P_2$).  The 
$1$ cm cut determined the value of the 
data point, $P$, and the difference between the $0.5$ cm cut value and 
no cut value of the data point determined the 
systematic error of the data point as follows: sys-error 
$= P|P_2-P_1|/P_1$  This systematic error was then added in 
quadrature with the statistical error.  We do not correct the data for 
the CTC track-finding efficiency. Instead, the 
theoretical Monte-Carlo model  predictions are corrected.
\begin{table*}[htbp]
\begin{center}
\caption{\label{rdf_table1} Data sets and selection 
criterion used in this analysis.}
\begin{tabular}{||c|c|c|c||} \hline \hline
 {\bf CDF Data Set}  & {\bf Trigger}  & {\bf Events}  & {\bf Selection} 
\\ \hline\hline
 Min-Bias & Min-Bias Trigger &{\small  626,966}  
 & zero or one vertex in $|z|<100$ cm \\
 & & & $|z_c-z_v|<2$ cm, $|{\rm CTC}\ d_0|<1$ cm \\
 & & & $p_T^{\rm track}>0.5\gev$, $|\eta|<1$  \\ \hline
JET20 & Calorimeter Tower cluster &{\small  78,682}  
 & zero or one vertex in $|z|<100$ cm \\
 & with $E_T>20\gev$ & & $|z_c-z_v|<2$ cm, $|{\rm CTC}\ d_0|<1$ cm \\
 & & & $p_T^{\rm track}>0.5\gev$, $|\eta|<1$  \\ \hline\hline
\end{tabular}
\end{center}
\end{table*}

\vskip 0.1in
\paragraph{\bf (2) QCD ``Hard Scattering'' Monte-Carlo Models}

The ``hard'' scattering QCD Monte-Carlo models used in this analysis 
are listed in Table~\ref{rdf_table2}.  The 
QCD perturbative 2-to-2 parton-parton differential cross section 
diverges as the $p_T$ of the 
scattering, $p_T^{\rm hard}$, goes to zero (see Fig.~\ref{rdf_fig1}).  
One must set a minimum $p_T^{\rm hard}$ large enough 
so that the resulting cross section is not larger that the total 
``hard core'' inelastic cross section, and also large 
enough to ensure that QCD perturbation theory is applicable.  
In this analysis we take $p_T^{\rm hard}> 3\gev$.
\begin{table*}[tbph]
\begin{center}
\caption{\label{rdf_table2} Theoretical QCD ``hard'' scattering 
Monte-Carlo models studied in this analysis.  In all cases 
we take $p_T({\rm hard}) > 3\gev$.}
\begin{tabular}{||c|c|c||} \hline \hline
 {\bf Monte-Carlo Model}  & {\bf Subprocess}  & {\bf Comments}  \\ \hline\hline
 \herwig~5.9 & { QCD 2-to-2 parton scattering} & 
{ Default values for all parameters} \\
 & { IPROC = 1500} & \\ \hline
\isajet~7.32 & { QCD 2-to-2 parton scattering} & 
{ Default values for all parameters} \\
 & { TWOJET} & \\ \hline
\pythia~6.115 & { QCD 2-to-2 parton scattering} & 
{ Default values for all parameters:} \\ 
 & { MSEL = 1} &  { PARP(81) = $1.4$} \\ \hline
\pythia~6.125 & { QCD 2-to-2 parton scattering} & 
{ Default values for all parameters:} \\
 & { MSEL = 1} &  { PARP(81) = $1.9$} \\ \hline
\pythia~No MS & { QCD 2-to-2 parton scattering} & 
{\footnotesize Default values for all parameters:} \\ 
 & { MSEL = 1} &  { MSTP(81) = $0$} \\ \hline\hline
\end{tabular}
\end{center}
\end{table*}

Each of the QCD Monte-Carlo models handle the ``beam-beam remnants''  
in a similar fashion. A hard 
scattering event is basically the superposition of a hard parton-parton 
interaction on top of a ``soft'' collision.  \herwig \cite{herwig} 
assumes that the underlying event is a soft collision between the two 
``beam clusters.''
\isajet \cite{isajet} uses a model similar to the one it uses for 
soft ``min-bias'' events (\ie ``cut Pomeron''), but 
with different parameters, to describe the underlying beam-beam 
remnants. \pythia \cite{pythia} assumes that 
each incoming beam hadron leaves behind a ``beam remnant,'' which do 
not radiate initial state radiation, and simply 
sail through unaffected by the hard process.  However, unlike 
\herwig~and \isajet, \pythia~also uses multiple 
parton interactions to enhance the activity of the underlying event as 
illustrated in Fig.~\ref{rdf_fig2}.

In this analysis we examine two versions of \pythia, \pythia~6.115 and 
\pythia~6.125 both with the default 
values for all the parameters.  The default values of the parameters 
are different in version 6.115 and 6.125.  In 
particular, the effective minimum $p_T$ for multiple parton 
interactions, PARP(81), changed from 
$1.4\gev$ in version 6.115 to $1.9\gev$ in version 6.125.  Increasing 
this cut-off decreases the multiple parton 
interaction cross section which reduces the amount of multiple parton 
scattering. For completeness, we also consider 
\pythia~with no multiple parton scattering (MSTP(81)=0).

Since \isajet~employs ``independent fragmentation'' it is possible to 
trace particles back to their origin and divide 
them into three categories: particles that arise from the break-up of 
the beam and target ({\it beam-beam remnants}), 
particles that arise from initial-state radiation, and particles that 
result from the outgoing hard scattering jets plus 
final-state radiation.  The ``hard scattering component'' consists of  
the particles that arise from the outgoing hard 
scattering jets plus initial and final-state radiation ({\it sum of the 
last two categories}).  Particles from the first two 
categories ({\it beam-beam remnants plus initial-state radiation}) are
normally what is referred to as the underlying 
event (see Fig.~\ref{rdf_fig1}). 
Of course, these categories are not directly observable experimentally.
Nevertheless, it is 
instructive to examine how particles from various origins affect the 
experimental observables.

Since \herwig~and \pythia~do not use independent fragmentation, it is not 
possible to distinguish particles that 
arise from initial-state radiation from those that arise from final-state 
radiation, but we can identify the beam-beam 
remnants.  When, for example,  a color string breaks into hadrons it 
is not possible to say which of the two partons 
producing the string was the parent.  For \herwig~and \pythia, we divide 
particles into two categories: particles 
that arise from the break-up of the beam and target ({\it beam-beam 
remnants}), and particles that result from the 
outgoing hard scattering jets plus initial and final-state radiation 
({\it hard scattering component}).  For \pythia, we 
include particles that arise from multiple parton interactions in the 
beam-beam remnant component.

\vskip 0.1in
\paragraph{\bf (3) Method of Comparing Theory with Data}

Our philosophy in comparing the theory with data in this analysis is to 
select a region where the data is very ``clean.'' 
The CTC efficiency can vary substantially for very low $p_T$ tracks and 
in dense high $p_T$ jets.  To avoid this 
we have considered only the region $p_T > 0.5\gev$ and $|\eta| < 1$ 
where the CTC efficiency is high and stable 
(estimated to be $92\%$ efficient) and we restrict ourselves to jets 
less than $50\gev$.  The data presented here are 
uncorrected.  Instead the theoretical Monte-Carlo predictions are 
corrected for the track finding efficiency and have 
an error ({\it statistical plus systematic}) of about $5\%$. The errors
on the ({\it uncorrected}) data include both 
statistical and correlated systematic uncertainties.   

In comparing the QCD ``hard scattering'' Monte-Carlo models with the 
data, we require that the Monte-Carlo events 
satisfy the CDF min-bias trigger and we apply an $8\%$ correction for 
the CTC track finding efficiency.  The 
corrections are small.  On the average, $8$ out of every $100$ charged 
particles predicted by the theory are removed 
from consideration. Requiring the theory to satisfy the min-bias 
trigger is important when comparing with the Min-
Bias data, but does not matter when comparing with the JET20 data 
since essentially all high $p_T$ jet events satisfy 
the min-bias trigger.

\subsection{The Evolution of Charge Particle ``Jets'' from $0.5$ to $50$ GeV}

\begin{figure}[htbp]
    \centerline{\psfig{file=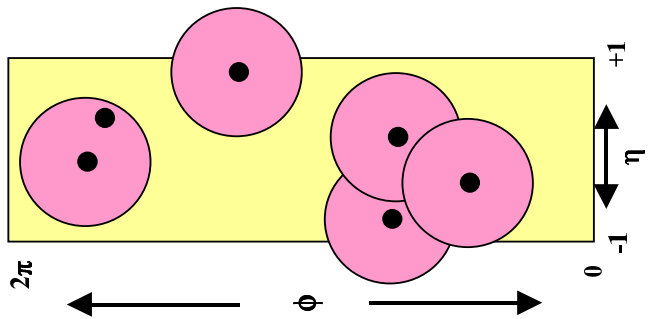,width=8.5cm,angle=90}}
\caption{\footnotesize 
Illustration of an event with six charged particles ($p_T > 0.5\gev$ 
and $|\eta| < 1$) and five charged ``jets'' (circular 
regions in $\eta$-$\phi$ space with $R = 0.7$).
}
\label{rdf_fig3}
\end{figure}

We define charged particle ``jets'' and examine the evolution of these 
``jets'' from $p_T^{\rm jet} = 0.5$ to 
$50\gev$. As illustrated in Fig.~\ref{rdf_fig3}, 
``jets'' are defined as ``circular regions'' ($R = 0.7$) in 
$\eta$-$\phi$ space and 
contain charged particles from the underlying event  as well as particles 
which originate from the fragmentation of 
high $p_T$ outgoing partons (see Fig.~\ref{rdf_fig1}).  
Also, every charged particle in the event is assigned to a ``jet,'' with the 
possibility that some ``jets'' might consist of just one charged 
particle.  We adapt a very simple jet definition since 
we will be dealing with ``jets'' that consist of only a few low $p_T$ 
charged particles.  The standard jet algorithm 
based on calorimeter clustering is not applicable at low $p_T$.

\vskip 0.1in
\paragraph{\bf (1) Jet Definition (charged particles)}

We define jets as circular regions in $\eta$-$\phi$ space with ``distance''
$R=\sqrt{(\Delta\eta)^2+(\Delta\phi)^2}$. Our jet algorithm is as follows:

\begin{itemize}
\item{} Order all charged particles according to their $p_T$.
\item{} Start with the highest $p_T$ particle and include in  the 
``jet'' all particles within $R = 0.7$.
\item{} Go to the next highest $p_T$ particle ({\it not already included 
in a ``jet''}) and add to the ``jet'' all particles 
({\it not already included in a ``jet''}) within $R =0.7$.
\item{} Continue until all particles are in a ``jet.''
\end{itemize}

We consider all charged particles ($p_T > 0.5\gev$ and $|\eta| < 1$) 
and allow the jet radius to extend outside $|\eta| 
< 1$.  Fig.~\ref{rdf_fig3} illustrates an event with six charged 
particles and five jets.  We define the $p_T$ of the 
jet to be the {\it scalar} $p_T$ sum of all the particles within the 
jet (\ie it is simply the scalar $p_T$ sum within the 
circular region).

We know that the simple charged particle jet definition used here is 
not theoretically favored since if applied at the 
parton level it is not infrared safe.  Of course, all jet definitions 
({\it and in fact all observables}) are infrared safe at 
the hadron level.  We have done a detailed study comparing the naïve 
jet definition used here with a variety of more 
sophisticated charge particle jet definitions.  This analysis will be 
presented in a future publication. Some of the 
observables presented here do, of course, depend on one's definition 
of a jet and it is important to apply the same 
definition to both the theory and data.

\vskip 0.1in
\paragraph{\bf (2) Charged Jet Multiplicity versus $p_T({\rm jet\# 1})$}

\begin{figure}[htbp]
    \centerline{\psfig{file=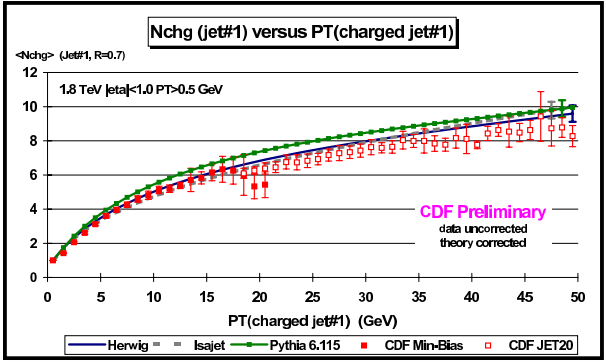,width=8cm}}
\caption{The average number of charged particles ($p_T > 0.5\gev$ and 
$|\eta| < 1$) within the leading charged jet 
($R = 0.7$) as a function of the $p_T$ of the leading charged jet.  
The solid (open) points are Min-Bias (JET20) 
data. The errors on the ({\it uncorrected}) data include both
statistical and correlated systematic uncertainties.  The 
QCD ``hard scattering'' theory curves (\herwig~5.9, \isajet~7.32, 
\pythia~6.115) are corrected for the track 
finding efficiency and have an error ({\it statistical plus systematic}) 
of around $5\%$.
}
\label{rdf_fig4}
\end{figure}

Fig.~\ref{rdf_fig4} shows the average number of charged particles 
($p_T > 0.5\gev$ and $|\eta| < 1$) within jet\#1  ({\it leading 
charged jet}) as a function of $p_T({\rm jet\# 1})$.  The solid points 
are Min-Bias data and the open points are the 
JET20 data. The JET20 data connect smoothly to the Min-Bias data and 
allow us to study observables over the 
range $ 0.5 < p_T({\rm jet\# 1}) < 50\gev$.  There is a small overlap 
region where the Min-Bias and JET20 
data coincide. The errors on the data include both statistical and 
correlated systematic uncertainties, however, the data have 
not been corrected for efficiency.  Fig.~\ref{rdf_fig4} shows a sharp 
rise in the leading 
charged jet multiplicity at low $p_T({\rm 
jet\# 1})$ and then a flattening out and a gradual rise at high 
$p_T({\rm jet\# 1})$.  The data are compared with the 
QCD ``hard scattering'' Monte-Carlo predictions of \herwig~5.9, 
\isajet~7.32, and \pythia~6.115.  The theory 
curves are corrected for the track finding efficiency and have an 
error ({\it statistical plus systematic}) of around $5\%$.

\vskip 0.1in
\paragraph{\bf  (3) Charged Jet ``Size'' versus $p_T({\rm jet\# 1})$}

\begin{figure}[htbp]
    \centerline{\psfig{file=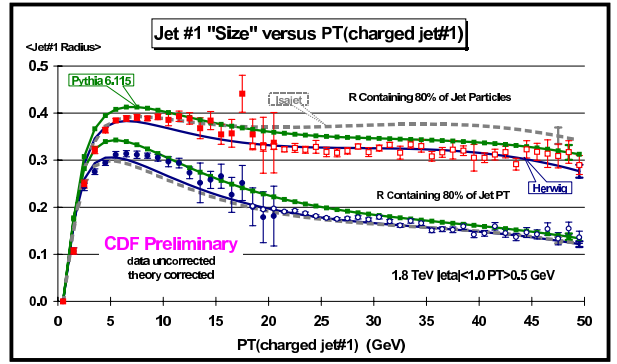,width=8cm}}
\caption{The average radius in $\eta$-$\phi$ space containing $80\%$ of the 
charged particles (and $80\%$ of the 
charged $p_T$) as a function of the $p_T$ of the leading charged jet. 
The errors on the ({\it 
uncorrected}) data include both statistical and correlated systematic 
uncertainties. The QCD ``hard scattering'' theory 
curves (\herwig~5.9, \isajet~7.32, \pythia~6.115)  are corrected for 
the track finding efficiency and have an 
error ({\it statistical plus systematic}) of around $5\%$.}
\label{rdf_fig5}
\end{figure}

Although the charged particle jets are defined as circular regions in 
$\eta$-$\phi$ space with $R = 0.7$, 
this is not  the ``size'' of the jet.  The ``size'' of a jet can be 
defined in two ways: size according to particle number and size 
according to $p_T$.  The first corresponds to the radius in 
$\eta$-$\phi$ space that contains $80\%$ of the charged particles in the jet, 
and the second corresponds to the radius in $\eta$-$\phi$ space that contains 
$80\%$ of the jet $p_T$. The data on the average ``jet size'' of the 
leading charge particle jet are 
compared with the QCD ``hard scattering'' Monte-Carlo predictions of 
\herwig~5.9, \isajet~7.32, and 
\pythia~6.115 in Fig.~\ref{rdf_fig5}.  
A leading $20\gev$ charged jet has $80\%$ of its charged particles 
contained, on the average, 
within a radius in $\eta$-$\phi$ space of about $0.33$, and $80\%$ 
of its $p_T$ contained, on the 
average, within a radius of about $0.20$.  Fig.~\ref{rdf_fig5} 
clearly illustrates the ``hot core'' of jets.  The radius containing 
$80\%$ of the $p_T$ is smaller than the radius that contains $80\%$ 
of the particles.  Furthermore, 
the radius containing $80\%$ of the $p_T$ decreases as the overall 
$p_T$ of the jet 
increases due to limited momentum perpendicular to the jet direction.

\subsection{The Overall Event Structure as a Function of $p_T({\rm jet\# 1})$}

In the previous section, we studied ``local'' leading jets observables.  
The QCD Monte-Carlo models did not have to 
describe correctly the entire event in order to fit the observable.  
They only had to describe correctly the properties 
of the leading charge particle jet, and all the models fit the data 
fairly well 
({\it although not perfectly}).  Now we will study  ``global''
observables, where to fit the observable 
the QCD Monte-Carlo models will have to describe 
correctly the entire event structure.

\vskip 0.1in
\paragraph{\bf (1) Overall Charged Multiplicity versus $p_T({\rm jet\# 1})$}

\begin{figure}[htbp]
    \centerline{\psfig{file=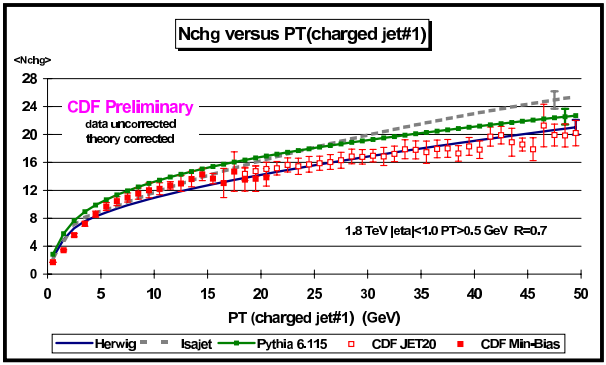,width=8cm}}
\caption{
The average total number of charged particles in the event ($p_T > 
0.5\gev$ and $|\eta| < 1$ including 
jet\#1)  as a function of the $p_T$ of the leading charged jet. The 
solid (open) points are the Min-
Bias (JET20) data. The errors on the ({\it uncorrected}) data include 
both statistical and 
correlated systematic uncertainties.  The QCD ``hard scattering'' 
theory curves (\herwig~5.9, \isajet~7.32, \pythia~6.115) are 
corrected for the track finding efficiency and have an error ({\it 
statistical plus systematic}) of around $5\%$.}
\label{rdf_fig6}
\end{figure}

Figure~\ref{rdf_fig6} shows the average number of charged particles in 
the event with 
$p_T > 0.5\gev$ and $|\eta| < 1$ ({\it including jet\#1}) as a function 
of $p_T({\rm jet\# 1})$ ({\it leading charged jet}) for the Min-Bias and 
JET20 data. Again the JET20 data connect smoothly to the Min-Bias data, 
and there is a small overlap region where the Min-Bias 
and JET20 data coincide. Figure~\ref{rdf_fig6} 
shows a sharp rise in the overall charged multiplicity at low 
$p_T({\rm jet\# 1})$ and 
then a flattening out and a gradual rise at high $p_T({\rm jet\# 1})$ 
similar to Fig.~\ref{rdf_fig4}. We would like to investigate 
where these charged particles are located relative to the direction of 
the leading charged particle jet.

\vskip 0.1in
\paragraph{\bf (2) Correlations in $\Delta\phi$ relative to 
$p_T({\rm jet\# 1})$}

\begin{figure}[htbp]
    \centerline{\psfig{file=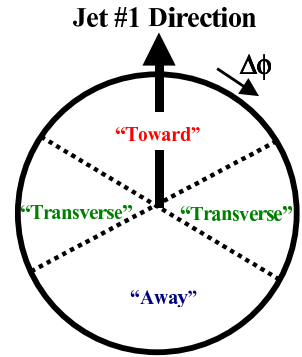,width=1.5in}}
\caption{
Illustration of correlations in azimuthal angle $\Delta\phi$ relative 
to the direction of the leading charged jet in the 
event, jet\#1.  The angle $\Delta\phi=\phi-\phi_{\rm jet\#1}$ is the 
relative azimuthal angle between charged 
particles and the direction of jet\#1.  The region $|\Delta\phi| < 
60^\circ$ is referred to as ''toward'' jet\#1 ({\it 
includes particles in jet\#1}) and the region $|\Delta\phi| > 120^\circ$ 
is called ``away'' from jet\#1.  The 
``transverse'' to jet\#1 region is defined by  $60^\circ < |\Delta\phi| 
< 120^\circ$.  Each region, ``toward,'' 
``transverse,'' and ``away'' covers the same range $|\Delta\eta|\times 
|\Delta\phi| = 2\times 120^\circ$.
}
\label{rdf_fig7}
\end{figure}

As illustrated in Fig.~\ref{rdf_fig7}, 
the angle $\Delta\phi$ is defined to be the relative azimuthal angle 
between charged particles 
and the direction of the leading charged particle jet.  We label the 
region 
$|\phi-\phi_{\rm jet\#1}| < 60^\circ$ as ``toward'' jet\#1 and the region 
$|\phi-\phi_{\rm jet\#1}| > 120^\circ$ is as ``away'' from jet\#1.  
The ``transverse'' to 
jet\#1 region is defined by  $60^\circ < |\phi-\phi_{\rm jet\#1}| 
< 120^\circ$.  Each region, ``toward,'' ``transverse,'' 
and ``away'' covers the same range $|\Delta\eta|\times |\Delta\phi| 
= 2\times 120^\circ$.   The ``toward'' region 
includes the particles from jet\#1 as well as a few particles from 
the underlying event.  As we will see, the 
``transverse'' region is very sensitive to the underlying event.  
The ``away'' region is a mixture of the 
underlying event and the ``away-side'' hard scattering jet. 

\begin{figure}[htbp]
    \centerline{\psfig{file=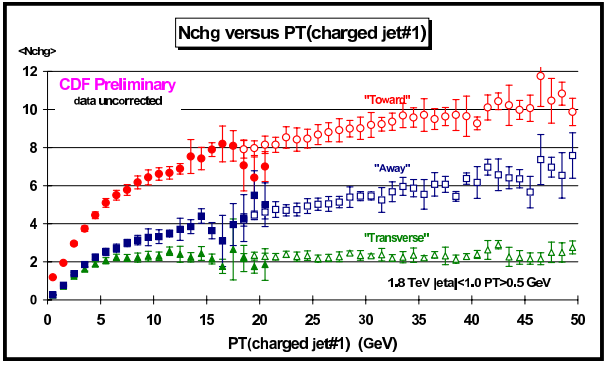,width=8cm}}
\caption{
The average number of  ``toward'' ($|\Delta\phi| < 60^\circ$), 
``transverse'' ($60^\circ < |\Delta\phi| < 120^\circ$), 
and ``away'' ($|\Delta\phi| > 120^\circ$) charged particles 
($p_T > 0.5\gev$ and $|\eta| < 1$ including jet\#1) as a 
function of the $p_T$ of the leading charged jet. Each point 
corresponds to the $\langle N_{chg} \rangle$  in a 
$1\gev$ bin. The solid (open) points are the Min-Bias (JET20) data. 
The errors on the ({\it uncorrected}) data 
include both statistical and correlated systematic uncertainties. 
The ``toward,'' ``transverse,'' and ``away'' regions are 
defined in Fig.~\ref{rdf_fig7}.}
\label{rdf_fig8}
\end{figure}

Figure~\ref{rdf_fig8} shows the 
data on the average number of  charged particles ($p_T > 0.5\gev$ 
and $|\eta|< 1$) as a function of 
$p_T({\rm jet\# 1})$ for the three regions.  Each point corresponds 
to the ``toward,'' ``transverse,'' or ``away'' 
$\langle N_{chg} \rangle$  in a $1\gev$ bin.   The solid points are 
Min-Bias data and the open points are JET20 data.  The 
data in Fig.~\ref{rdf_fig8} define the average event ``shape.''  
For example, for an ``average'' $p \bar p$ collider event at 
$1.8\tev$ with $p_T({\rm jet\# 1}) = 20\gev$  there are, on the 
average, $8.7$ charged particles ``toward'' jet\#1 
({\it including the particles in jet\#1}), $2.5$ ``transverse'' to 
jet\#1, and $4.9$ ``away'' from jet\#1.

\begin{figure}[htbp]
    \centerline{\psfig{file=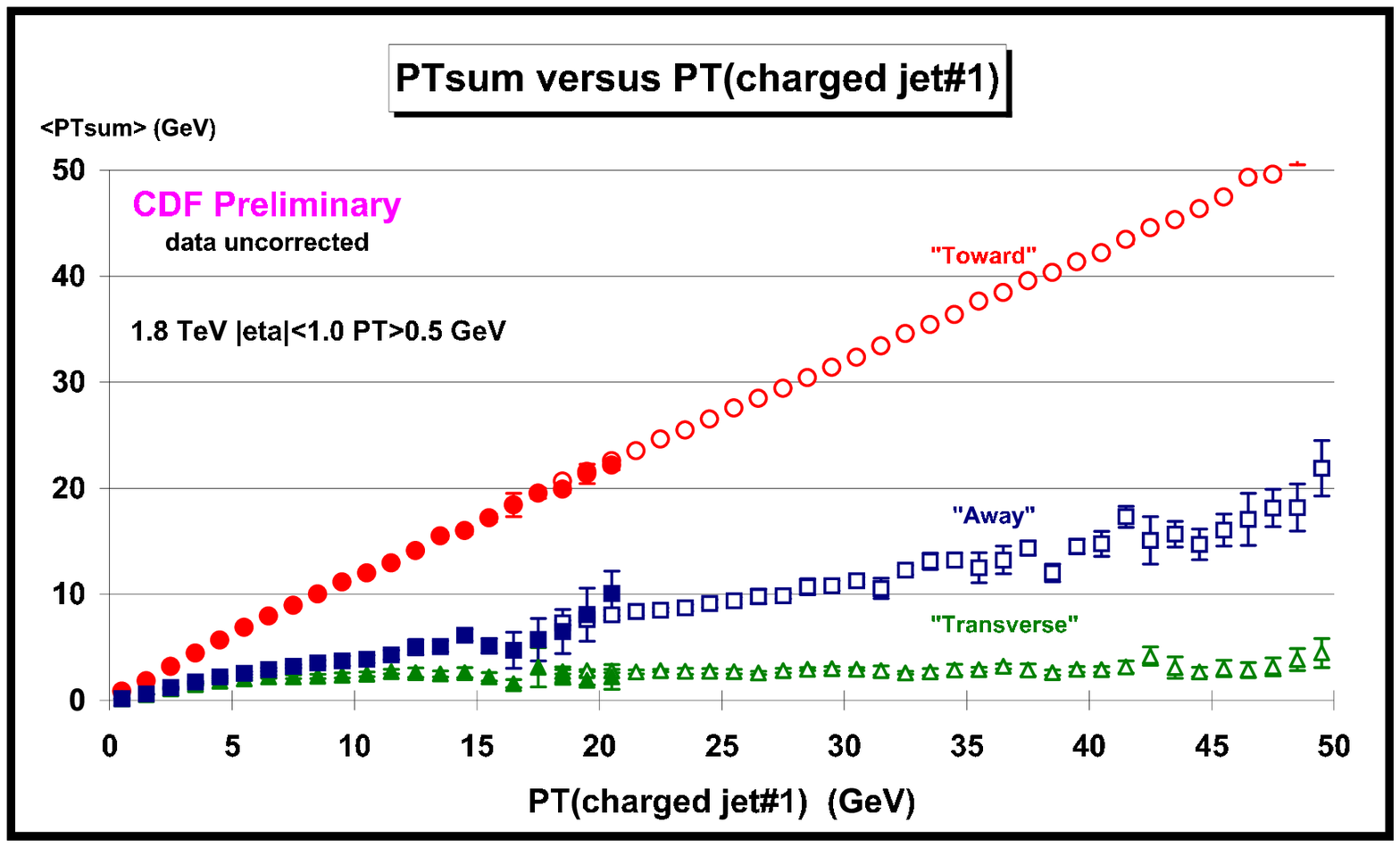,width=8cm}}
\caption{
The average {\it scalar} $p_T$ sum of ``toward'' ($|\Delta\phi| < 
60^\circ$), ``transverse'' ($60^\circ < |\Delta\phi| < 
120^\circ$), and ``away'' ($|\Delta\phi| > 120^\circ$) charged 
particles ($p_T > 0.5\gev$ and $|\eta| < 1$ including 
jet\#1)  as a function of the $p_T$ of the leading charged jet. Each 
point corresponds to the 
$\langle PT_{sum} \rangle$  in a $1\gev$ bin. The solid (open) points 
are the Min-Bias (JET20) data. The errors on the ({\it 
uncorrected}) data include both statistical and correlated systematic 
uncertainties.  The ``toward,'' ``transverse,'' and 
``away'' regions are defined in Fig.~\ref{rdf_fig7}.
}
\label{rdf_fig9}
\end{figure}

\begin{figure}[htbp]
    \centerline{\psfig{file=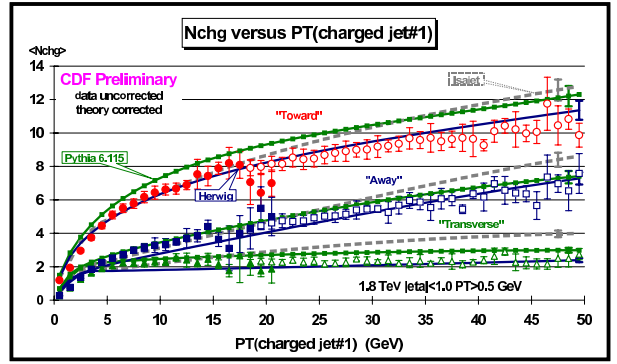,width=8cm}}
\caption{
Data from Fig.~\ref{rdf_fig8} on the average number of  
``toward'' ($|\Delta\phi| < 60^\circ$), ``transverse'' ($60^\circ < 
|\Delta\phi| < 120^\circ$), and ``away'' ($|\Delta\phi| > 120^\circ$) 
charged particles ($p_T > 0.5\gev$ and $|\eta| < 
1$ including jet\#1) as a function of the $p_T$ of the leading charged 
jet compared to QCD ``hard 
scattering'' Monte-Carlo predictions of \herwig~5.9, \isajet~7.32, and 
\pythia~6.115. The errors on the ({\it 
uncorrected}) data include both statistical and correlated systematic 
uncertainties. The theory curves are corrected 
for the track finding efficiency and have an error ({\it statistical
plus systematic}) of around $5\%$.
}
\label{rdf_fig10}
\end{figure}

Figure~\ref{rdf_fig9} shows the data on the average {\it scalar} $p_T$ sum of  
charged particles ($p_T > 0.5\gev$ and $|\eta|< 1$) 
as a function of $p_T({\rm jet\# 1})$ for the three regions.  Here each 
point corresponds to the ``toward,'' 
``transverse,'' or ``away'' $\langle {p_T}_{sum} \rangle$  in a 
$1\gev$ bin.  In Fig.~\ref{rdf_fig10},
data on $\langle N_{chg}\rangle$ as a function 
of $p_T({\rm jet\# 1})$ for the three regions are compared with the 
QCD ``hard scattering'' Monte-Carlo predictions 
of \herwig~5.9, \isajet~7.32, and \pythia~6.115.  
The QCD Monte-Carlo models agree qualitatively ({\it but not 
precisely}) with the data. We will now examine more closely these 
three regions.

\vskip 0.1in
\paragraph{\bf (3) The ``Toward'' and ``Away'' Region versus 
$p_T({\rm jet\# 1})$}

\begin{figure}[htbp]
    \centerline{\psfig{file=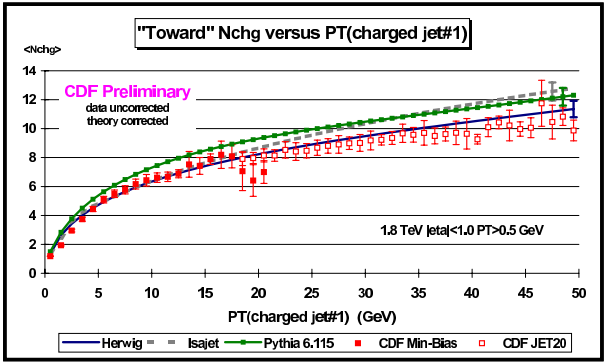,width=8cm}}
\caption{Data from Fig.~\ref{rdf_fig8} 
on the average number of  charged particles ($p_T > 0.5\gev$ and 
$|\eta| < 1$) as a function of 
$p_T({\rm jet\#1})$ ({\it leading charged jet}) for the ``toward'' 
region '' defined in Fig.~\ref{rdf_fig7} 
compared with the QCD 
``hard scattering'' Monte-Carlo predictions of \herwig~5.9, 
\isajet~7.32, and \pythia~6.115.  Each point 
corresponds to the ``toward''  $\langle N_{chg} \rangle$  in a $1\gev$ bin.
The errors on the ({\it uncorrected}) data include 
both statistical and correlated systematic uncertainties. The theory 
curves are corrected for the track 
finding efficiency and have an error ({\it statistical plus systematic}) 
of around $5\%$.}
\label{rdf_fig11}
\end{figure}

\begin{figure}[htbp]
    \centerline{\psfig{file=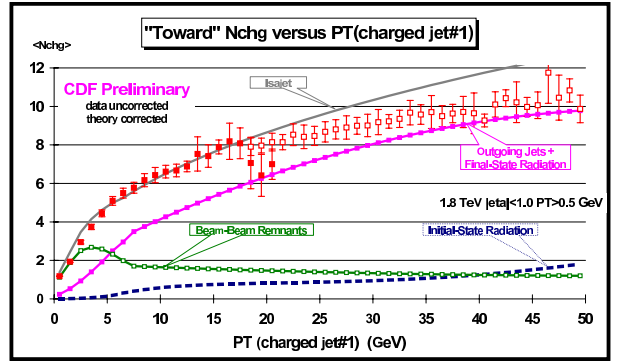,width=8cm}}
\caption{Data from Fig.~\ref{rdf_fig8} 
on the average number of  charged particles ($p_T > 0.5\gev$ and 
$|\eta| < 1$) as a function of 
$p_T({\rm jet\#1})$ ({\it leading charged jet}) for the ``toward''
region defined in Fig.~\ref{rdf_fig7} 
compared with the QCD 
''hard scattering'' Monte-Carlo predictions of \isajet~7.32. The 
predictions of \isajet are divided into 
three categories: charged particles that arise from the break-up of 
the beam and target ({\it beam-beam remnants}), 
charged particles that arise from initial-state radiation, and charged 
particles that result from the outgoing jets plus 
final-state radiation (see Fig.~\ref{rdf_fig1}). 
The errors on the ({\it uncorrected}) data include both statistical and 
correlated 
systematic uncertainties. The theory curves are corrected for the 
track finding efficiency and have an 
error ({\it statistical plus systematic}) of around $5\%$.
}
\label{rdf_fig12}
\end{figure}

Figure~\ref{rdf_fig11} shows the data from Fig.~\ref{rdf_fig8} 
on the average number of  ``toward'' region charged particles compared 
with the 
QCD ``hard scattering'' Monte-Carlo predictions of \herwig~5.9, 
\isajet~7.32, and \pythia~6.115. This plot is 
very similar to the average number of charged particles within the 
leading jet shown in Fig.~\ref{rdf_fig4}. 
At $p_T({\rm jet\# 1}) = 20\gev$,  the ``toward'' region contains, 
on the average, about $8.7$ charged particles with about $6.9$ of these 
charged particles belonging to jet\#1. As expected, the toward region 
is dominated by the leading jet.  This is seen 
clearly in Fig.~\ref{rdf_fig12} 
where the predictions of \isajet~for the ``toward'' region are divided 
into three categories: charged 
particles that arise from the break-up of the beam and target ({\it 
beam-beam remnants}), charged particles that arise 
from initial-state radiation, and charged particles that result from 
the outgoing jets plus final-state 
radiation. For $p_T({\rm jet\# 1})$ values below $5\gev$ the 
``toward'' region charged multiplicity arises 
mostly from the beam-beam remnants, but as $p_T({\rm jet\# 1})$ 
increases the contribution from the 
outgoing jets plus final-state radiation quickly begins to dominate.  
The bump in the beam-beam remnant 
contribution at low $p_T({\rm jet\# 1})$ is caused by leading jets 
composed almost entirely from the remnants.  
 
\begin{figure}[htbp]
    \centerline{\psfig{file=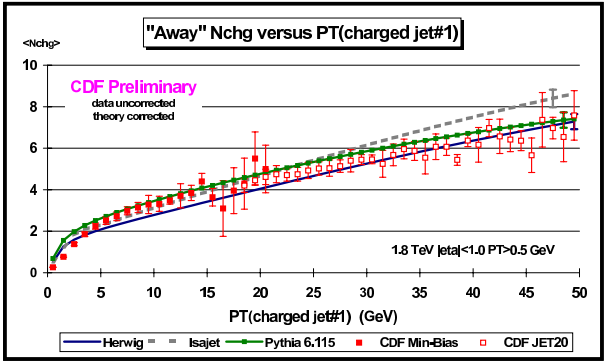,width=8cm}}
\caption{Data from Fig.~\ref{rdf_fig8} on the average number of  
charged particles ($p_T > 0.5\gev$ and $|\eta| < 1$) as a function of 
$p_T({\rm jet\#1})$ ({\it leading charged jet}) for the ``away'' region 
defined in Fig.~\ref{rdf_fig7} compared with the QCD 
``hard scattering'' Monte-Carlo predictions of \herwig~5.9, 
\isajet~7.32, and \pythia~6.115. The errors on the 
({\it uncorrected}) data include both statistical and correlated 
systematic uncertainties. The theory curves are 
corrected for the track finding efficiency and have an error 
({\it statistical plus systematic}) of 
around $5\%$.
}
\label{rdf_fig13}
\end{figure}

Fig.~\ref{rdf_fig13} shows the data from Fig.~\ref{rdf_fig8} on the 
average number of  ``away'' region 
charged particles compared with the QCD ``hard scattering'' Monte-Carlo 
predictions of \herwig~5.9, \isajet~7.32, and \pythia~6.115.  In 
Fig.~\ref{rdf_fig4} the 
data from Fig.~\ref{rdf_fig9} on the average {\it scalar} $p_T$ sum in 
the ``away'' region 
is compared to the QCD Monte-Carlo predictions.  The ``away'' region is a 
mixture of the underlying event and the ``away-side'' outgoing ``hard 
scattering'' jet. This can be seen in Fig.~\ref{rdf_fig15} where the 
predictions of \isajet~for the ``away'' region are divided into 
three categories: beam-beam remnants, initial-state radiation, and 
outgoing jets plus final-state radiation.  Here the 
underlying event plays a more important role since the ``away-side'' outgoing 
``hard scattering'' jet is sometimes 
outside the regions $|\eta|< 1$.  For the ``toward'' region the 
contribution from the outgoing jets plus 
final state-radiation dominates for $p_T({\rm jet\# 1})$ values above about 
$5\gev$, whereas for the ``away'' region 
this does not occur until around $20\gev$.

\begin{figure}[htbp]
    \centerline{\psfig{file=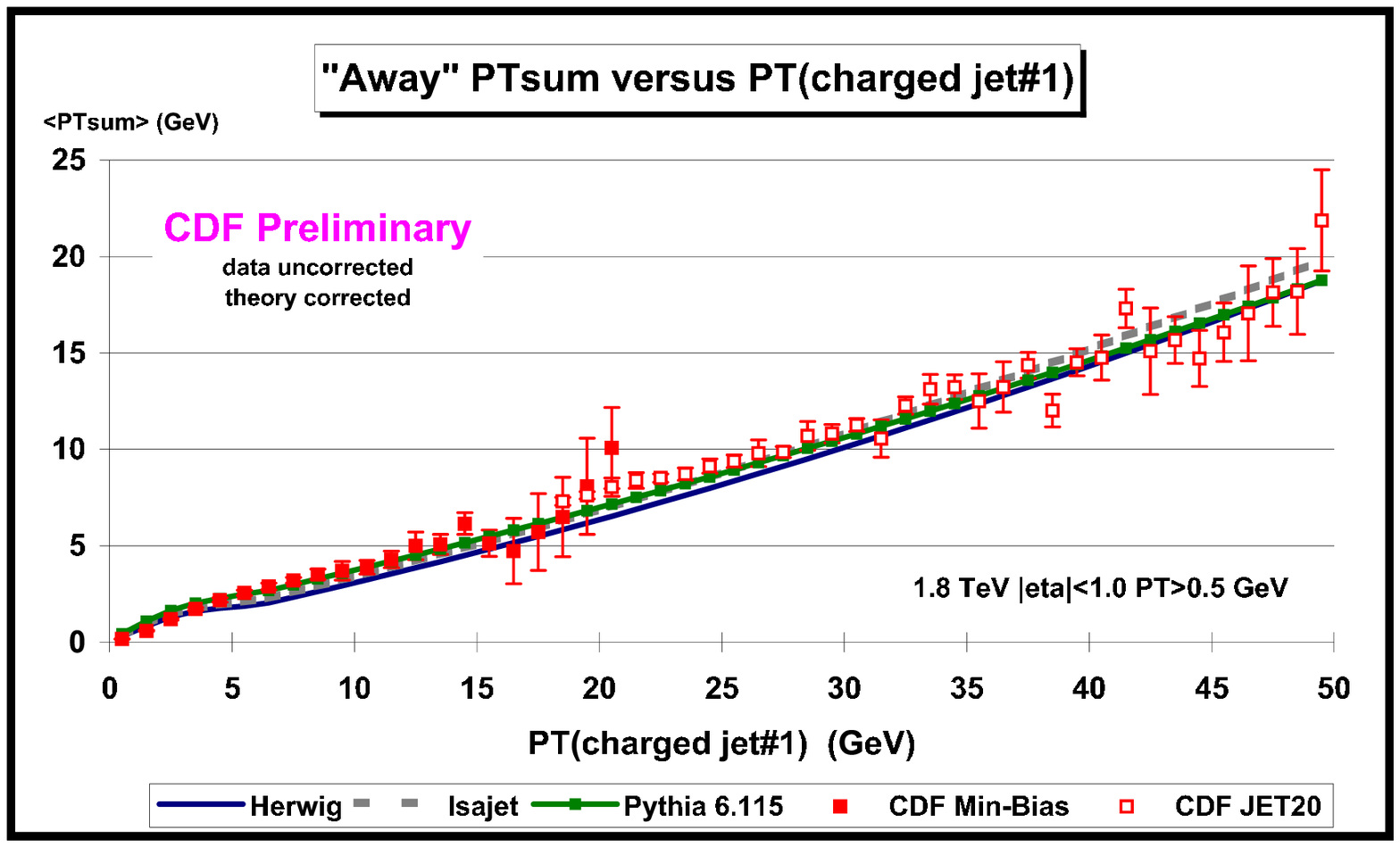,width=8cm}}
\caption{Data from Fig.~\ref{rdf_fig9} 
on the average {\it scalar} $p_T$ sum of  charged particles 
(($p_T > 0.5\gev$ and $|\eta| < 1$) as 
a function of $p_T({\rm jet\#1})$  ({\it leading charged jet}) for the 
``away'' region defined in 
Fig.~\ref{rdf_fig7} compared with 
the QCD ``hard scattering'' Monte-Carlo predictions of \herwig~5.9, 
\isajet~7.32, and \pythia~6.115. The 
errors on the ({\it uncorrected}) data include both statistical and 
correlated systematic uncertainties. 
The theory curves are corrected for the track finding efficiency and 
have an error ({\it statistical plus systematic}) of around $5\%$.
}
\label{rdf_fig14}
\end{figure}

\begin{figure}[htbp]
    \centerline{\psfig{file=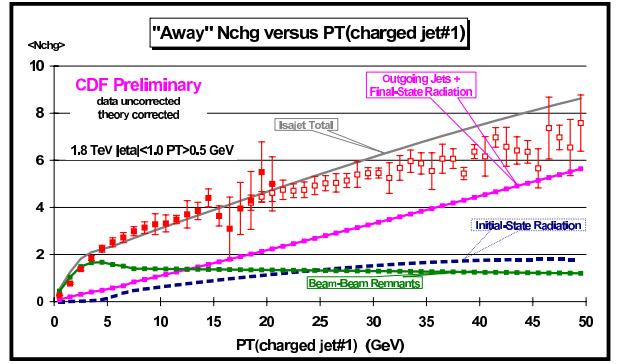,width=8cm}}
\caption{Data from Fig.~\ref{rdf_fig8} 
on the average number of  charged particles ($p_T > 0.5\gev$ and 
$|\eta| < 1$) as a function of 
$p_T({\rm jet\#1})$ ({\it leading charged jet}) for the ``away'' region 
defined in Fig.~\ref{rdf_fig7} 
compared with the QCD 
``hard scattering'' Monte-Carlo predictions of \isajet~7.32. The 
predictions of \isajet are divided into three 
categories: charged particles that arise from the break-up of the 
beam and target ({\it beam-beam remnants}), 
charged particles that arise from initial-state radiation, and charged 
particles that result from the outgoing jets plus 
final-state radiation (see Fig.~\ref{rdf_fig1}). 
The errors on the ({\it uncorrected}) data include both statistical and 
correlated 
systematic uncertainties. The theory curves are corrected for the 
track finding efficiency and have an 
error ({\it statistical plus systematic}) of around $5\%$.
}
\label{rdf_fig15}
\end{figure}

Both the ``toward'' and ``away'' regions are described fairly well by 
the QCD ``hard scattering'' 
Monte-Carlo models.  These regions are dominated by the outgoing 
``hard scattering'' jets and as we saw 
in Section C the Monte-Carlo models describe the leading outgoing 
jets fairly accurately.  We will now 
study the ``transverse'' region which is dominated by the underlying event.  

\subsection{The ``Transverse'' Region and the Underlying Event}

Fig.~\ref{rdf_fig8} shows that there is a lot of activity in the 
``transverse'' region.  
If we suppose that the ``transverse'' multiplicity 
is uniform in azimuthal angle $\phi$ and pseudo-rapidity $\eta$, 
the observed $2.3$ charged particles at $p_T({\rm 
jet\# 1}) = 20\gev$ translates to $3.8$ charged particles per unit 
pseudo-rapidity with $p_T > 0.5\gev$ (multiply by 
$3$ to get $360^\circ$, divide by $2$ for the two units of 
pseudo-rapidity, multiply by $1.09$ to correct for the track 
finding efficiency).  We know that if we include all $p_T$ that there 
are roughly $4$ charged particles per unit 
rapidity in a ``soft'' $p \bar p$ collision at $1.8\tev$, 
and the data show that in the underlying event of a ``hard 
scattering'' there are about $3.8$ charged particles per unit rapidity 
in the region $p_T > 0.5\gev$!  If one includes 
all $p_T$ values then the underlying event has a charge particle 
density that is at least a factor of two larger than the 
$4$ charged particles per unit rapidity seen in ``soft'' 
$p \bar p$ collisions at this energy.  
As can be seen in Fig.~\ref{rdf_fig8}, 
the charged particle density in the ``transverse'' region is a function 
of $p_T({\rm jet\# 1})$ and 
rises very rapidity at low  $p_T({\rm jet\# 1})$ values.  The 
``transverse'' charged multiplicity doubles 
in going from $p_T({\rm jet\# 1}) = 1.5\gev$ to $p_T({\rm jet\# 1}) 
= 2.5\gev$ and then forms an approximately constant 
``plateau'' for $p_T({\rm jet\# 1}) > 6\gev$.

\begin{figure}[htbp]
    \centerline{\psfig{file=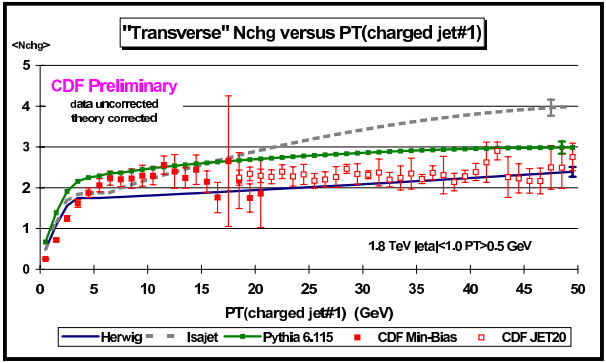,width=8cm}}
\caption{Data from Fig.~\ref{rdf_fig8}
 on the average number of  charged particles ($p_T > 0.5\gev$ and 
$|\eta| < 1$) as a function of 
$p_T({\rm jet\#1})$ ({\it leading charged jet}) for the ``transverse'' 
region defined in Fig.~\ref{rdf_fig7} 
compared with the 
QCD ``hard scattering'' Monte-Carlo predictions of \herwig~5.9, 
\isajet~7.32, and \pythia~6.115. The errors 
on the ({\it uncorrected}) data include both statistical and correlated 
systematic uncertainties. The theory curves are 
corrected for the track finding efficiency and have an error 
({\it statistical plus systematic}) of around $5\%$.
}
\label{rdf_fig16}
\end{figure}

\begin{figure}[htbp]
    \centerline{\psfig{file=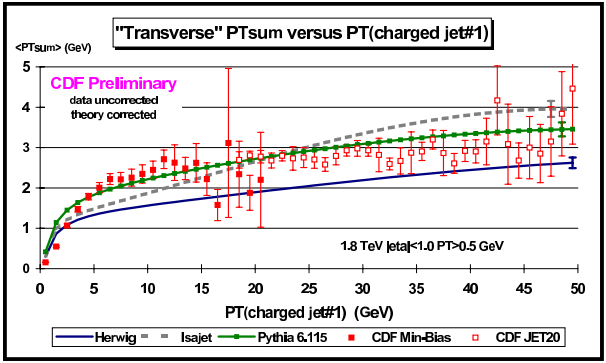,width=8cm}}
\caption{Data from Fig.~\ref{rdf_fig9} on the average 
{\it scalar} $p_T$ sum of  charged particles ($p_T > 0.5\gev$ and 
$|\eta| < 1$) as a 
function of $p_T({\rm jet\#1})$ ({\it leading charged jet}) for the 
``transverse'' region defined in Fig.~\ref{rdf_fig7} 
compared 
with the QCD ``hard scattering'' Monte-Carlo predictions of 
\herwig~5.9, \isajet~7.32, and \pythia~6.115. The 
errors on the ({\it uncorrected}) data include both statistical and 
correlated systematic uncertainties. The theory 
curves are corrected for the track finding efficiency and have an 
error ({\it statistical plus systematic}) of around $5\%$.
}
\label{rdf_fig17}
\end{figure}

\begin{figure}[htbp]
    \centerline{\psfig{file=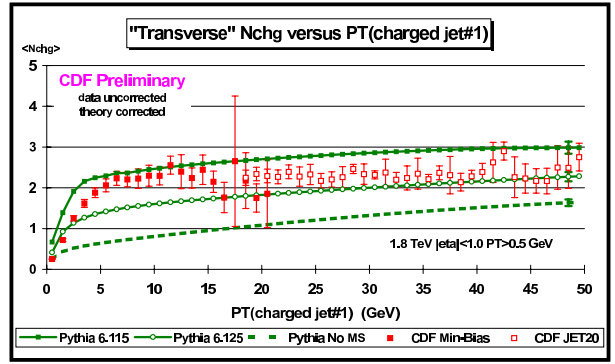,width=8cm}}
\caption{Data from Fig.~\ref{rdf_fig18} 
on the average number of  charged particles ($p_T > 0.5\gev$ and 
$|\eta| < 1$) as a function of 
$p_T({\rm jet\#1})$ ({\it leading charged jet}) for the ``transverse'' 
region defined in Fig.~\ref{rdf_fig7} compared with the 
QCD ``hard scattering'' Monte-Carlo predictions of \pythia~6.115, 
\pythia~6.125, and \pythia~with no 
multiple parton scattering (No MS). The errors on the ({\it 
uncorrected}) data include both statistical and correlated 
systematic uncertainties. The theory curves are corrected for the 
track finding efficiency and have an error ({\it 
statistical plus systematic}) of around $5\%$.
}
\label{rdf_fig18}
\end{figure}

\begin{figure}[htbp]
    \centerline{\psfig{file=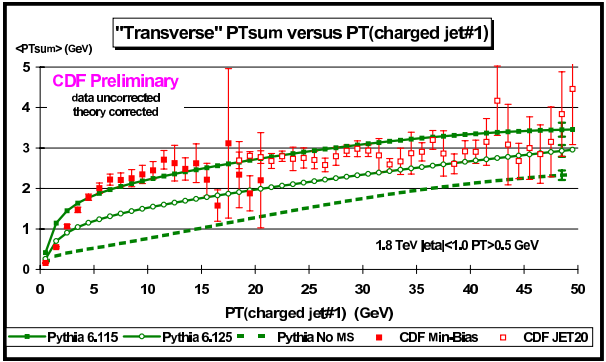,width=8cm}}
\caption{Data from Fig.~\ref{rdf_fig9} 
on the average {\it scalar} $p_T$ sum of  charged particles 
($p_T > 0.5\gev$ and $|\eta| < 1$) as a 
function of $p_T({\rm jet\#1})$ ({\it leading charged jet}) for the 
``transverse'' region defined in Fig.~\ref{rdf_fig7} 
compared 
with the QCD ``hard scattering'' Monte-Carlo predictions of 
\pythia~6.115, \pythia~6.125, and \pythia~with no 
multiple parton scattering (No MS). The errors on the ({\it
uncorrected}) data include both statistical and correlated 
systematic uncertainties. The theory curves are corrected for the 
track finding efficiency and have an error ({\it 
statistical plus systematic}) of around $5\%$.
}
\label{rdf_fig19}
\end{figure}

Fig.~\ref{rdf_fig16} and Fig.~\ref{rdf_fig17} 
compare the ``transverse'' $\langle N_{chg}\rangle$ 
and the ``transverse'' $\langle {p_T}_{sum} \rangle$, respectively, with 
the QCD ``hard scattering'' Monte-Carlo predictions of \herwig~5.9, 
\isajet~7.32, and \pythia~6.115.  Fig.~\ref{rdf_fig18}
and Fig.~\ref{rdf_fig19} compare the ``transverse'' 
$\langle N_{chg} \rangle$ and the ``transverse'' $\langle PT_{sum} 
\rangle$, respectively, with 
three versions of \pythia (6.115, 6.125, and no multiple scattering, 
see Table~\ref{rdf_table2}). \pythia~with no 
multiple parton scattering does not have enough activity in the 
underlying event.  \pythia~6.115 fits the 
``transverse'' $\langle N_{chg} \rangle$ the best, but overshoots 
slightly the ``toward'' $\langle N_{chg} \rangle$ in Fig.~\ref{rdf_fig11}.  
\isajet~has a lot 
of activity in the underlying event, but gives the wrong 
$p_T({\rm jet\# 1})$ dependence.  Instead of a ``plateau,'' 
\isajet~predicts a rising ``transverse'' $\langle N_{chg} \rangle$ 
and gives too much activity at large $p_T({\rm jet\# 1})$ 
values.  \herwig~does not have enough ``transverse'' $\langle PT_{sum} 
\rangle$.

\begin{figure}[htbp]
    \centerline{\psfig{file=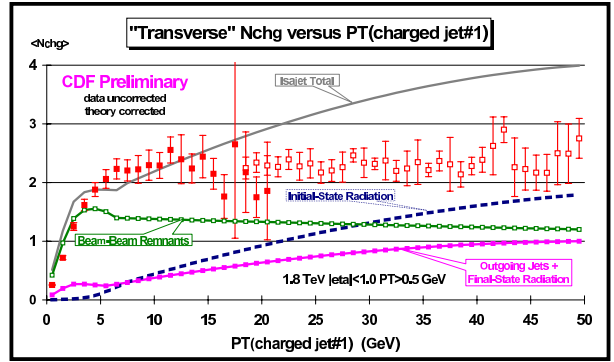,width=8cm}}
\caption{Data from Fig.~\ref{rdf_fig8} 
on the average number of  charged particles ($p_T > 0.5\gev$ and 
$|\eta| < 1$) as a function of 
$p_T({\rm jet\#1})$ ({\it leading charged jet}) for the ``transverse''
region defined in Fig.~\ref{rdf_fig7} compared with the 
QCD ``hard scattering'' Monte-Carlo predictions of \isajet~7.32. 
The predictions of \isajet are divided into three 
categories: charged particles that arise from the break-up of the beam 
and target ({\it beam-beam remnants}), 
charged particles that arise from initial-state radiation, and charged 
particles that result from the outgoing jets plus 
final-state radiation (see Fig.~\ref{rdf_fig1}). 
The errors on the ({\it uncorrected}) data include both statistical and 
correlated 
systematic uncertainties. The theory curves are corrected for the 
track finding efficiency and have an error ({\it 
statistical plus systematic}) of around $5\%$.
}
\label{rdf_fig20}
\end{figure}

We expect the ``transverse'' region to be composed predominately from 
particles that arise from the break-up of the 
beam and target and from initial-state radiation. This is clearly the 
case as can be seen in Fig.~\ref{rdf_fig20} where the 
predictions of \isajet~for the ``transverse'' region are divided into 
three categories: beam-beam remnants, initial-
state radiation, and outgoing jets plus final-state radiation.  It is 
interesting to see that it is the beam-beam remnants 
that are producing the approximately constant ``plateau''.  The 
contributions from initial-state radiation and from the 
outgoing hard scattering jets both increase as $p_T({\rm jet\# 1})$ 
increases.  In fact, for \isajet~it is the sharp rise 
in the initial-state radiation component that is causing the 
disagreement with the data for $p_T({\rm jet\# 1}) > 20\gev$.

\begin{figure}[htbp]
    \centerline{\psfig{file=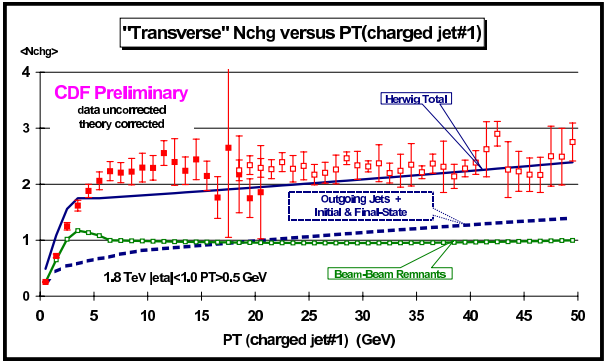,width=8cm}}
\caption{Data from Fig.~\ref{rdf_fig8} 
on the average number of  charged particles ($p_T > 0.5\gev$ and 
$|\eta| < 1$) as a function of 
$p_T({\rm jet\#1})$ ({\it leading charged jet}) for the ``transverse'' 
region defined in Fig.~\ref{rdf_fig7} compared with the 
QCD ``hard scattering'' Monte-Carlo predictions of \herwig~5.9. The 
predictions of \herwig~are divided into 
two categories: charged particles that arise from the break-up of the 
beam and target ({\it beam-beam remnants}), 
and charged particles that result from the outgoing jets plus initial 
and final-state radiation ({\it hard scattering 
component}) (see Fig.~\ref{rdf_fig1}). The errors on the ({\it
uncorrected}) data include both statistical and correlated 
systematic uncertainties. The theory curves are corrected for the 
track finding efficiency and have an error ({\it 
statistical plus systematic}) of around $5\%$.
}
\label{rdf_fig21}
\end{figure}

\begin{figure}[htbp]
    \centerline{\psfig{file=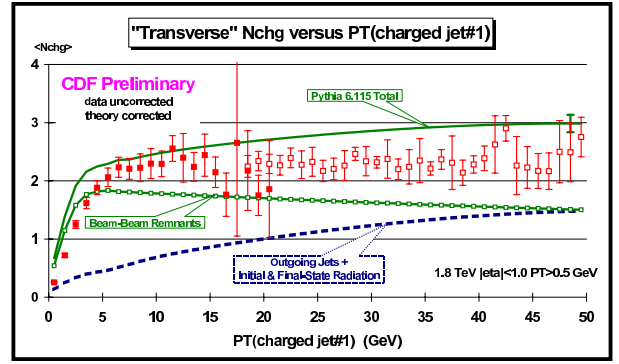,width=8cm}}
\caption{Data from Fig.~\ref{rdf_fig8} 
on the average number of  charged particles ($p_T > 0.5\gev$ and 
$|\eta| < 1$) as a function of 
$p_T({\rm jet\#1})$ ({\it leading charged jet}) for the ``transverse'' 
region defined in Fig.~\ref{rdf_fig7} 
compared with the 
QCD ``hard scattering'' Monte-Carlo predictions of \pythia~6.115. 
The predictions of \pythia~are divided into 
two categories: charged particles that arise from the break-up of 
the beam and target ({\it beam-beam remnants}), 
and charged particles that result from the outgoing jets plus initial 
and final-state radiation ({\it hard scattering 
component}). For \pythia, the beam-beam remnants include contributions 
from multiple parton scattering (see 
Fig.~\ref{rdf_fig2}). The errors on the ({\it uncorrected}) data include 
both statistical and correlated systematic uncertainties. 
The theory curves are corrected for the track finding efficiency and 
have an error ({\it statistical plus systematic}) of 
around $5\%$.
}
\label{rdf_fig22}
\end{figure}

\begin{figure}[htbp]
    \centerline{\psfig{file=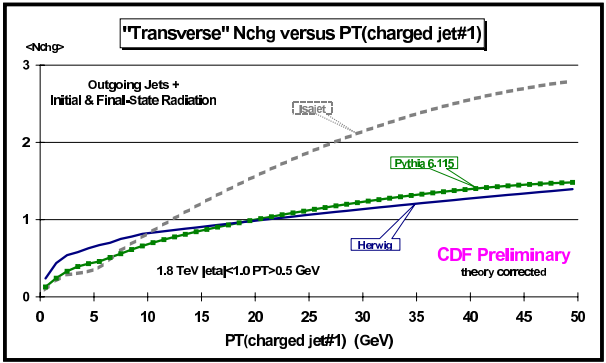,width=8cm}}
\caption{QCD ``hard scattering'' Monte-Carlo predictions from 
\herwig~5.9, \isajet~7.32, and \pythia~6.115  of the 
average number of  charged particles ($p_T > 0.5\gev$ and 
$|\eta| < 1$) as a function of $p_T({\rm jet\#1})$ ({\it 
leading charged jet}) for the ``transverse'' region defined in 
Fig.~\ref{rdf_fig7} arising from the outgoing jets plus initial and 
finial-state radiation ({\it hard scattering component}).  The curves are 
corrected for the track finding efficiency and 
have an error ({\it statistical plus systematic}) of around $5\%$.
}
\label{rdf_fig23}
\end{figure}

As we explained in Section B, for \herwig~and \pythia~it makes no 
sense to distinguish between particles that 
arise from initial-state radiation from those that arise from 
final-state radiation, but one can separate the ``hard 
scattering component'' from the beam-beam remnants. For \pythia~the 
beam-beam remnants include contributions 
from multiple parton scattering as illustrated in Fig.~\ref{rdf_fig2}. 
Fig.~\ref{rdf_fig21} and Fig.~\ref{rdf_fig22} 
compare the ``transverse'' 
$\langle N_{chg} \rangle$ with the QCD ``hard scattering'' Monte-Carlo 
predictions of \herwig~5.9 and \pythia~6.115, 
respectively.  Here the predictions are divided into two categories: 
charged particles that arise from the break-up of 
the beam and target ({\it beam-beam remnants}), and charged particles
that result from the outgoing jets plus initial 
and final-state radiation ({\it hard scattering component}).  As was the 
case with \isajet~the beam-beam remnants 
form the approximately constant ``plateau'' and the hard scattering 
component increase as $p_T({\rm jet\# 1})$ 
increases.  However, the hard scattering component of \herwig~and 
\pythia~does not rise nearly as fast as the 
hard scattering component of \isajet.  This can be seen clearly in 
Fig.~\ref{rdf_fig23} where we compare directly the hard 
scattering component ({\it outgoing jets plus initial and final-state
radiation}) of the ``transverse'' 
$\langle N_{chg}\rangle$ 
from \isajet~7.32, \herwig~5.9, and \pythia~6.115.  \pythia~and 
\herwig~are similar and rise gently as 
$p_T({\rm jet\# 1})$ increases, whereas \isajet produces a much 
sharper increase as $p_T({\rm jet\# 1})$ increases.  

There are two reasons why the hard scattering component of 
\isajet~is different from \herwig~and \pythia.  The 
first is due to different fragmentation schemes.  \isajet~uses 
independent fragmentation, which produces too many 
soft hadrons when partons begin to overlap.  The second difference 
arises from the way the QCD Monte-Carlo 
produce ``parton showers''.  \isajet~uses a leading-log picture in 
which the partons within the shower are ordered 
according to their invariant mass.  Kinematics requires that the 
invariant mass of daughter partons be less than the 
invariant mass of the parent.  \herwig~and \pythia~modify the 
leading-log picture to include ``color coherence 
effects'' which leads to ``angle ordering'' within the parton shower.  
Angle ordering produces less high $p_T$ 
radiation within a parton shower which is what is seen in 
Fig.~\ref{rdf_fig23}.  Without further study, we do not know how much 
of the difference seen in Fig.~\ref{rdf_fig23} is due to the 
different fragmentation schemes and how much is due to the color 
coherence effects.

\begin{figure}[htbp]
    \centerline{\psfig{file=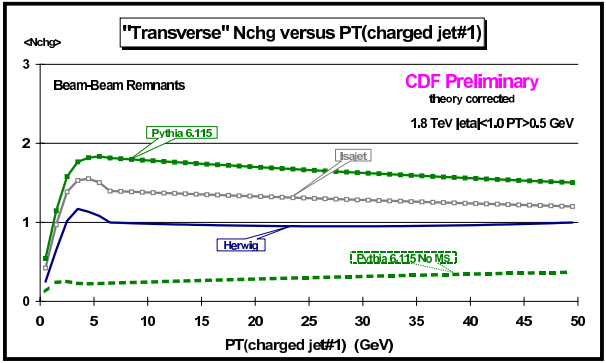,width=8cm}}
\caption{
QCD ``hard scattering'' Monte-Carlo predictions from \herwig~5.9, 
\isajet~7.32, \pythia~6.115, and \pythia~
with no multiple parton scattering (No MS) of the average number of  
charged particles ($p_T > 0.5\gev$ and $|\eta| 
< 1$) as a function of $p_T({\rm jet\#1})$  ({\it leading charged jet}) 
for the ``transverse'' region defined in Fig.~7  
arising from the break-up of the beam and target ({\it beam-beam 
remnants}). For \pythia~the beam-beam remnants 
include contributions from multiple parton scattering (see 
Fig.~\ref{rdf_fig2}). The curves are corrected for the track finding 
efficiency and have an error ({\it statistical plus systematic}) of around $5\%$.
}
\label{rdf_fig24}
\end{figure}

The beam-beam remnant contribution to the  ``transverse'' 
$\langle N_{chg} \rangle$ is different for each of the QCD Monte-
Carlo models. This can be seen in Fig.~\ref{rdf_fig24} where we 
compare directly the beam-beam remnant component of the 
``transverse'' $\langle N_{chg} \rangle$ from \isajet~7.32, 
\herwig~5.9, \pythia~6.115, and \pythia~with no multiple 
parton interactions.  Since we are considering only charged particles 
with $p_T > 0.5\gev$, the height of the 
``plateaus'' in Fig.~\ref{rdf_fig24} is related to the $p_T$ 
distribution of the beam-beam remnant contributions.  A 
steeper $p_T$ distribution means less particles with $p_T > 0.5\gev$.  
\pythia~uses multiple parton scattering to 
enhance the underlying event and we have included these contributions 
in the beam-beam remnants.  For \pythia~the height of the ``plateau'' 
in Fig.~\ref{rdf_fig24} can be adjusted by adjusting the amount of 
multiple parton scattering.  
\herwig~and \isajet~do not include multiple parton scattering.  For 
\herwig~and \isajet~the height of the 
``plateau'' can be adjusted by changing the $p_T$ distribution of 
the beam-beam remnants. 

\subsection{Summary and Conclusions}

For $p \bar p$ collisions at $1.8\tev$ min-bias does not necessarily 
imply ``soft'' physics.  There is a lot of 
QCD ``hard scattering'' in the Min-Bias data. We have studied both 
``local'' leading jet observables and ``global'' 
observables where to fit the data the QCD Monte-Carlo models have to 
correctly describe the entire event structure. 
Our summary and conclusions are as follows.

\vskip 0.1in
\centerline{\it The Evolution of Charge Particle Jets}

Charged particle jets are ``born'' somewhere around $p_T({\rm jet})$ 
of about $2\gev$ with, on the average, about 
$2$ charged particles and grow to, on the average,  about $10$ charged 
particles at $50\gev$.  The QCD ``hard 
scattering'' models describe quite well ({\it although not perfectly}) 
``local'' leading jet observables such as the 
multiplicity distribution of charged particles within the leading 
jet, the ``size'' of the leading jet, the radial flow of 
charged particles and $p_T$ around the leading jet direction, and the 
momentum distribution of 
charged particles within the leading jet.  In fact, the QCD ``hard'' 
scattering Monte-Carlo models agree as well with 
$2\gev$ charged particle jets as they do with $50\gev$ charged 
particle jets!  The charge particle jets in the Min-Bias 
data are simply the extrapolation ({\it down to small $p_T$}) of the 
high $p_T$ jets observed in the 
JET20 data.  For a fixed $p_T({\rm hard})$, the QCD ``hard'' 
scattering cross section grows with increasing collider 
energy.  As the center-of-mass energy of a $p \bar p$ collision 
grows, ``hard'' scattering becomes a larger and 
larger fraction of the total inelastic cross section.  At $1.8\tev$ 
``hard scattering'' makes up a sizable part of the 
``hard core'' inelastic cross section and a lot of min-bias events 
have $2\tev$ or $3\gev$ jets.

\vskip 0.1in
\centerline{\it The ``Underlying Event''}

A hard scattering collider event consists of large $p_T$ outgoing 
hadrons that originate from the 
large $p_T$ partons ({\it outgoing jets}) and also hadrons that originate 
from the break-up of the 
proton and antiproton ({\it beam-beam remnants}).  The ``underlying event'' 
is formed from the beam-beam 
remnants, initial-state radiation, and possibly from multiple parton 
interactions.  Our data show that the charged 
particle multiplicity and  scalar $p_T$ sum in the ``underlying event'' 
grows very rapidly with the transverse 
momentum of the leading charged particle jet and then forms an 
approximately constant ``plateau'' for $p_T({\rm 
jet\# 1}) > 6\gev$.  The height of this``plateau'' is at least twice 
that observed in ``soft'' collisions at the same corresponding energy. 
 
None of the QCD Monte-Carlo models we examined correctly describe all 
the properties of the underlying event 
seen in the data.  \herwig~5.9 and \pythia~6.125 do not have enough 
activity in the underlying event.  \pythia~
6.115 has about the right amount of activity in the underlying event, 
but as a result produces too much overall 
charged multiplicity.  \isajet~7.32 has a lot of activity in the 
underlying event, but with the wrong dependence on 
$p_T({\rm jet\# 1})$.  Because \isajet~uses independent fragmentation 
and \herwig~and \pythia~do not, there 
are clear differences in the hard scattering component ({\it mostly 
initial-state radiation}) of the underlying event 
between \isajet~and the other two Monte-Carlo models. Here the data 
strongly favor \herwig~and \pythia~over \isajet. 

The beam-beam remnant component of both \isajet~7.32 and \herwig~5.9 
has the wrong $p_T$ dependence. 
\isajet~and \herwig~both predict too steep of a $p_T$ distribution.  
\pythia~does a better job, but is still slightly 
too steep. It is, of course, understandable that the Monte-Carlo models 
might be somewhat off on the 
parameterization of the beam-beam remnants.  This component cannot be 
calculated from perturbation theory and 
must be determined from data.  With what we have learned from the data 
presented here, the beam-beam remnant 
component of the QCD ``hard scattering'' Monte-Carlo models can be 
tuned to better describe the overall event in $p \bar p$ collisions.



\end{document}